\documentclass[twoside,12pt]{article}
\usepackage{epsfig}
\usepackage{amsmath, amssymb}

\usepackage{color}

\definecolor{gray}{rgb}{0.5,0.5,0.5}

\providecommand{\keywords}[1]{\textbf{\textit{Keywords:}} #1}

\newcommand{\be}{\begin{equation}}
\newcommand{\eeq}{\end{equation}}
\newcommand{\bea}{\begin{eqnarray}}
\newcommand{\eea}{\end{eqnarray}}
\newcommand{\nn}{\nonumber}
\input symbols

\begin{document}




\title{Neutrino oscillations: the rise of the PMNS paradigm}

\author{C. Giganti$^1$, S.\ Lavignac$^2$, M.\ Zito$^3$ \\
\\
$^1$ LPNHE, CNRS/IN2P3, UPMC, Universit\'{e} Paris Diderot, Paris 75252, France\\
$^2$Institut de Physique Th\'{e}orique,
Universit\'e Paris Saclay, CNRS, CEA,\\
Gif-sur-Yvette, France\footnote{Unit\'e Mixte de Recherche
du CEA/DRF et du CNRS (UMR 3681).}\\
$^3$IRFU/SPP, CEA, Universit\'e Paris-Saclay, F-91191 Gif-sur-Yvette, France}

\maketitle
\begin{abstract}

Since the discovery of neutrino oscillations, the experimental progress in the last two decades has been very fast, with the precision measurements of the neutrino squared-mass differences and of the mixing angles, including the last unknown mixing angle $\theta_{13}$.

Today a very large set of oscillation results obtained with a variety of experimental configurations and techniques can be interpreted in the framework of three active massive neutrinos, whose mass and flavour eigenstates are related by a 3 $\times$ 3 unitary mixing matrix, the Pontecorvo-Maki-Nakagawa-Sakata (PMNS) matrix, parameterized by three mixing angles $\theta_{12}$, $\theta_{23}$, $\theta_{13}$ and a CP-violating phase \dcp.
The additional parameters governing neutrino oscillations are the squared-mass differences $\Delta m^2_{ji}=m^2_j-m^2_i$, where $m_i$ is the mass of the $i$th neutrino mass eigenstate. This review covers the rise of the PMNS three-neutrino mixing paradigm and the current status of the experimental determination of its parameters.

The next years will continue to see a rich program of experimental endeavour coming to fruition and addressing the three missing pieces of the puzzle, namely the determination of the octant and precise value of the mixing angle $\theta_{23}$, the unveiling of the neutrino mass ordering (whether $m_1 < m_2 < m_3$ or $m_3 < m_1 < m_2$) and the measurement of the CP-violating phase \dcp.

\end{abstract}









\tableofcontents
\section{Introduction}
\label{sec:intro}

It is remarkable that after a very long period of theoretical and experimental work, sometimes marked by heated controversies and debates, the original intuition of Bruno Pontecorvo in 1957~\cite{pontecorvo1, pontecorvo2} was at last confirmed in the few years between 1998 and 2002 with the beautiful discoveries of atmospheric and solar neutrino oscillations by the Super-Kamioka Neutrino Detection Experiment (Super-Kamiokande)~\cite{Fukuda:1998mi} and the Sudbury Neutrino Observatory (SNO)~\cite{sno2002}, shortly thereafter confirmed by the Kamioka Liquid Scintillator Antineutrino Detector (KamLAND) experiment~\cite{kamland2002}. These discoveries have been recognized by the 2015 Nobel Prize in Physics awarded to Takaaki Kajita and Arthur Mc Donald.

Since this milestone, experimental progress has been very fast, with the first significant indications of a nonzero value of the last unknown mixing angle $\theta_{13}$ by the long-baseline accelerator neutrino experiment Tokai-to-Kamioka (T2K)~\cite{t2k2011}, and its discovery by the reactor experiments Daya Bay~\cite{dayabay2012}, Reactor Experiment for Neutrino Oscillations
(RENO)~\cite{reno2012} and Double Chooz~\cite{dc2012}. The discovery of $\nu_\mu \rightarrow \nu_e$ appearance by T2K
in 2013~\cite{Abe:2013hdq}, later confirmed by the NuMI Off-Axis $\nu_e$ Appearance (\nova) experiment~\cite{Adamson:2017gxd}, plays a special role for future developments since it has demonstrated neutrino appearance in a direct way and opened the way towards probing three-flavour effects.
While experiments with natural sources have been at the forefront of the discovery of neutrino oscillations, recent precision measurements have been mainly carried out with man-made sources, like nuclear reactors and neutrino beams. 

On the theory side, the first ideas of oscillations date back to 1957, when Pontecorvo,
inspired by kaon-antikaon oscillations, suggested that neutrino-antineutrino oscillations
could take place if lepton number is violated~\cite{pontecorvo1, pontecorvo2}. This phenomenon,
which we would now describe as active-sterile neutrino oscillations, required the neutrino
to be massive, in contradiction with the then common belief that it was massless.
The idea of neutrino flavour mixing emerged only later, around the time where the muon
neutrino was discovered, in a paper by Maki, Nakagawa and Sakata~\cite{Maki:1962mu}.
Subsequently, neutrino flavour oscillations were proposed by Pontecorvo in 1967~\cite{Pontecorvo:1967fh}
and presented in the modern formalism by Gribov and Pontecorvo in 1969~\cite{Gribov:1968kq}.

Today a very large set of oscillation results obtained with an amazing variety of experimental configurations and techniques
can be interpreted  in the framework of three active massive neutrinos, whose mass and flavour eigenstates are related
by a 3 $\times$ 3 unitary mixing matrix, the Pontecorvo-Maki-Nakagawa-Sakata (PMNS) matrix, parameterized by three mixing angles $\theta_{12}$, $\theta_{23}$, $\theta_{13}$ and a CP-violating phase \dcp\ (the two additional phases present in the Majorana case do not affect oscillations). The additional parameters governing neutrino oscillations are the squared-mass differences $\Delta m^2_{ji}=m^2_j-m^2_i$, where $m_i$ is the mass of the $i$th neutrino mass eigenstate.
This description of the oscillation data in terms of three-neutrino mixing, which we will refer to as the PMNS paradigm,
has been successfully confirmed by the experimental progress made over the past two decades.
The next years will continue to see a rich program of experimental endeavour coming to fruition and addressing the three missing pieces of the puzzle, namely the determination of the octant and precise value of the mixing angle $\theta_{23}$, the unveiling of the neutrino mass ordering and the measurement of the CP-violating phase $\delta_{CP}$.

The study of neutrinos, and of neutrino oscillations in particular, plays a special role in our understanding of particle physics and explores the frontier of our knowledge in this domain. Indeed, the nonzero neutrino masses, whose only experimental evidence comes from neutrino oscillations, is our first positive indication of physics beyond the Standard Model~\cite{mohapatra}. Moreover, the structure of the lepton mixing matrix with three sizeable mixing angles is radically different from the analogous Cabibbo-Kobayashi-Maskawa (CKM) mixing matrix in the quark sector~\cite{cabibbo, kobayashi}, and this in itself represents a deep question that needs to be addressed and understood.

Interestingly, the large mixing angles of the PMNS matrix open the possibility for significant CP violation in the lepton sector.
These studies might help us understand the origin of the baryon asymmetry of the universe,
one of the most convincing interpretations of which is provided by the leptogenesis mechanism~\cite{Fukugita:1986hr}
(for a review of leptogenesis, see e.g. Ref.~\cite{leptoreview}).
In this scenario, the same hypothetical heavy Majorana neutrinos are responsible for the
baryon asymmetry of the universe and for the small neutrino masses via the so-called seesaw
mechanism~\cite{Minkowski:1977sc,GellMann:1980vs,Yanagida:1979as,Glashow:1979nm,Mohapatra:1979ia}.
 
This review aims to cover the recent progress in the measurements related to neutrino oscillations,
as well as the future experimental program aiming at the determination of the yet unknown
parameters (the $\theta_{23}$ octant, the neutrino mass ordering and the CP-violating phase \dcp).
It is organized as follows. In Section~\ref{sec:th}, we introduce the formalism of massive neutrino oscillations,
both in vacuum and in matter.
In Sections~3, 4 and~5,
we give a summary of the experimental evidence for neutrino oscillations in the 1-2, 2-3 and 1-3 sectors,
governed by the mixing angles $\theta_{12}$, $\theta_{23}$ and $\theta_{13}$, respectively.
We then turn in Section~6 to the new experimental results that can only
be interpreted as genuine three-flavour effects, and especially to the study of CP-violating effects
that has been initiated by the T2K and \nova experiments. We then briefly summarize in Section~7
our current level of understanding of the various parameters of the PMNS framework and the role played by global fits.
After briefly mentioning the remaining anomalies that cannot be interpreted within the PMNS framework
in Section~8, we give in Section~9 an overview of the future experimental program and discuss how
it can answer the open questions. 

Given the broadness and diversity of the field of neutrino oscillations, it is not possible
to address all subtleties and details relative to the theoretical developments,
experimental methods and results within the scope of this review. We therefore refer the interested reader to the existing books
on the subject, out of which we quote only a few ones~\cite{Fukugita:2003en,Mohapatra:2004,book-giunti,Bilenky:2010,book-zuber,Barger:2012pxa,Valle:2015pba,Suekane:2015yta},
as well as to the Particle Data Group (PDG) review on neutrino mass, mixing and oscillations~\cite{pdg}.
There are also numerous recent reviews on more specialized subjects, such as neutrino propagation
in matter~\cite{Blennow:2013rca} (see also the older Ref.~\cite{Kuo:1989qe}),
solar neutrinos~\cite{ianni},
\thint\ measurements with reactor experiments~\cite{lachenmaier},
the question of mass ordering and its determination by future experiments~\cite{xian},
experimental anomalies and sterile neutrinos~\cite{Abazajian:2012ys,Gariazzo:2015rra}.
Finally, a pedagogical account of the history of neutrino physics, including both
theoretical and experimental developments, can be found in Ref.~\cite{Bilenky:2012qb}.

\section{Theory and phenomenology of neutrino oscillations}
\label{sec:th}

\subsection{Flavour mixing in the lepton sector    %
\label{subsec:mixing}}                                           %

Neutrino oscillations in vacuum~\cite{pontecorvo1,pontecorvo2,Pontecorvo:1967fh}
are a quantum-mechanical phenomenon that is made possible by the existence
of non-degenerate neutrino masses and lepton flavour mixing~\cite{Maki:1962mu,Pontecorvo:1967fh}.
As for quarks, the origin of flavour mixing in the lepton sector lies in the mismatch
between the basis of (weak) gauge eigenstates
and the basis of mass eigenstates,
namely, the fact that the neutrino mass matrix is not diagonal when written in the flavour basis,
defined as the weak eigenstate basis corresponding to the charged lepton mass eigenstates
$e$, $\mu$ and $\tau$.
The unitary transformation relating the flavour to the mass eigenstate left-handed neutrino fields
is the {\it lepton mixing matrix}, known as the PMNS (Pontecorvo-Maki-Nakagawa-Sakata) matrix:
\be
  \left(\!\! \begin{array}{c} \nu_e(x) \\ \nu_\mu(x) \\ \nu_\tau(x) \end{array}\!\! \right)_{\!\!L}
  =\, U\, \left(\!\! \begin{array}{c} \nu_1(x) \\ \nu_2(x) \\ \nu_3(x) \end{array}\!\! \right)_{\!\!L}
  =\, \left(\! \begin{array}{ccc} U_{e1} & U_{e2} & U_{e3} \\ U_{\mu1} & U_{\mu2} & U_{\mu3} \\ U_{\tau 1}
    & U_{\tau2} & U_{\tau3} \end{array}\! \right)\! \left(\!\!
    \begin{array}{c} \nu_1(x) \\ \nu_2(x) \\ \nu_3(x) \end{array}\!\! \right)_{\!\!L}\, .
\label{eq:flavour_mass_relation_long}
\eeq
In Eq.~(\ref{eq:flavour_mass_relation_long}),
$\nu_{e L} (x)$, $\nu_{\mu L} (x)$ and $\nu_{\tau L} (x)$ are the fields describing
the left-handed {\it flavour eigenstate} neutrinos,
defined as the neutrinos that couple via charged weak current to the electron, the muon
and the tau, respectively,
and $\nu_{1 L} (x)$, $\nu_{2 L} (x)$ and $\nu_{3 L} (x)$ describe the left-handed
mass eigenstate neutrinos with masses $m_1$, $m_2$ and $m_3$.
In shorthand notations, the relation~(\ref{eq:flavour_mass_relation_long}) reduces to
\be
  \nu_{\alpha L} (x)\, =\, \sum_i U_{\alpha i} \nu_{i L} (x)\, ,
\label{eq:flavour_mass_relation}
\eeq
where $\alpha = e, \mu, \tau$ and $i=1,2,3$.
For the sake of notational simplicity, we shall omit the subscript ``$L$'' and the $x$ dependence
of the fields in the following.
As a consequence of flavour mixing,
the neutrino that couples to a charged lepton of a given flavour (an electron,
a muon or a tau) via the weak charged current (CC) is not a mass eigenstate,
but a {\it coherent superposition of mass eigenstates}:
\be
  {\cal L}_{\rm CC}\, =\, \frac{g}{\sqrt{2}}\ W^-_\mu\!\!
    \sum_{\alpha = e, \mu, \tau}\!\! \bar \ell_{\alpha L} \gamma^\mu \nu_{\alpha L} + \mbox{h.c.}\,
  =\, \frac{g}{\sqrt{2}}\ W^-_\mu\!\! \sum_{\alpha = e, \mu, \tau}\!\! \bar \ell_{\alpha L} \gamma^\mu\!\!
    \sum_{i = 1,2,3}\!\! U_{\alpha i}\, \nu_{i L} + \mbox{h.c.}\, .
\label{eq:L_CC}
\eeq
It is this coherence that makes neutrino oscillations possible.

Being associated with a change of basis, the PMNS matrix is a unitary matrix. Like the CKM matrix,
it satisfies unitary relations, derived from $U U^\dagger = U^\dagger U = \mathbf{1}$:
\be
  \sum_i\, U_{\alpha i} U^*_{\beta i}\, =\, \delta_{\alpha \beta} \quad \left( \alpha, \beta = e, \mu, \tau \right) ,
  \qquad  \sum_\alpha\, U^*_{\alpha i} U_{\alpha j}\, =\, \delta_{ij} \quad \left( i, j = 1,2,3 \right) .
\eeq
Like any $3 \times 3$ unitary matrix, $U$ can be parametrized by 3 mixing angles and 6 phases.
However, not all of these phases are physical, since lepton fields can be rephased
to absorb some of them. Namely, if neutrinos are Dirac fermions, one can rephase
both the charged lepton and the neutrino fields, $\ell_\alpha (x) \to e^{i\phi_\alpha}\, \ell_\alpha (x)$
and $\nu_i (x) \to e^{i\phi_i}\, \nu_i (x)$, where $\ell_\alpha (x)$ and $\nu_i (x)$ denote
the 4-component Dirac fields (i.e. the phases of the left-handed and right-handed lepton fields
are shifted by the same amount, in order not to affect the mass terms
$- \sum_\alpha m_{\ell_\alpha} \bar \ell_{\alpha R} \ell_{\alpha L} - \sum_i m_i  \bar \nu_{i R} \nu_{i L} + \mbox{h.c.}$).
This leaves the charged current term~(\ref{eq:L_CC}) invariant, provided that one redefines
the PMNS matrix in the following way:
\be
  U_{\alpha i}\, \to\, e^{i (\phi_\alpha - \phi_i)}\, U_{\alpha i}\, .
\eeq
Since there are 5 independent phase differences $\phi_\alpha - \phi_i$, one can
remove 5 phases from the PMNS matrix, leaving only one physical CP-violating
phase, as in the CKM matrix. If neutrinos are Majorana fermions, however, it is
not possible to rephase the left-handed neutrino fields, because this would make
their masses complex. Indeed, Majorana mass terms are of the form
$- \frac{1}{2}\, m_i\, \nu^T_{i L} C \nu_{i L} + \mbox{h.c.}$, where $C$ is the charge
conjugation matrix satisfying $C \gamma_\mu C^{-1} = - \gamma^T_\mu$
(for a review on Majorana neutrinos, see e.g. Ref.~\cite{Petcov:2013poa}).
Thus only the charged lepton fields can be rephased, leading to
\be
  U_{\alpha i}\, \to\, e^{i \phi_\alpha}\, U_{\alpha i}\, .
\eeq
One is therefore left with 3 physical CP-violating phases in the Majorana case,
instead of a single one in the Dirac case\footnote{This parameter counting can be
generalized to an arbitrary number $N$ of lepton flavours.
One finds, in the Dirac case, $N(N-1)/2$ mixing angles and $(N-1)(N-2)/2$ phases,
and $N-1$ additional phases in the Majorana case~\cite{Bilenky:1980cx,Schechter:1980gr,Doi:1980yb}.
Thus, at variance with the quark sector, CP violation is possible already with 2 generations
of leptons if neutrinos are Majorana fermions, although CP violation in oscillations requires
at least 3 generations (see Section~\ref{subsec:CPV}).}.

Based on this parameter counting, the PMNS matrix can be written as the product of three rotations
through angles $\theta_{23}$, $\theta_{13}$ and $\theta_{12}$,  where the second
(unitary) rotation depends on a phase $\delta_{\rm CP}$, and of a diagonal matrix of phases $P$:
\bea
  U\! & =\! & \left( \begin{array}{ccc} 1 & 0 & 0 \\ 0 & c_{23} & s_{23} \\ 0 & - s_{23} & c_{23} \end{array} \right)
    \left( \begin{array}{ccc} c_{13} & 0 & s_{13} e^{-i \delta_{\rm CP}} \\ 0 & 1 & 0 \\
    - s_{13} e^{i \delta_{\rm CP}} & 0 & c_{13} \end{array} \right)
    \left( \begin{array}{ccc} c_{12} & s_{12} & 0 \\ - s_{12} & c_{12} & 0 \\ 0 & 0 & 1 \end{array} \right) P  \nn \\
  & =\! & \left( \begin{array}{ccc} c_{12} c_{13} & s_{12} c_{13} & s_{13} e^{-i \delta_{\rm CP}} \\
    - s_{12} c_{23} - c_{12} s_{13} s_{23} e^{i \delta_{\rm CP}} &
    c_{12} c_{23} - s_{12} s_{13} s_{23} e^{i \delta_{\rm CP}} & c_{13} s_{23} \\
    s_{12} s_{23} - c_{12} s_{13} c_{23} e^{i \delta_{\rm CP}} &
    - c_{12} s_{23} - s_{12} s_{13} c_{23} e^{i \delta_{\rm CP}} & c_{13} c_{23} 
   \end{array} \right) P\ .
\label{eq:PMNS}
\eea
In Eq.~(\ref{eq:PMNS}), $c_{ij} \equiv \cos \theta_{ij}$, $s_{ij} \equiv \sin \theta_{ij}$ and $P$
is either the unit matrix $\mathbf{1}$ in the Dirac case, or a diagonal matrix containing
the two phases associated with the Majorana nature of neutrinos in the Majorana case.
Without loss of generality, one can take $\theta_{ij} \in \left[ 0, \frac{\pi}{2} \right]$ and
$\delta_{\rm CP} \in \left[ 0, 2 \pi \right[$.
This parametrization has now become standard, except for the explicit form of the matrix $P$
in the Majorana case, for which different conventions can be found in the literature, e.g.
\be
  P_{\rm Majorana}\, =\, \left( \begin{array}{ccc} e^{i \alpha_1} & 0 & 0 \\ 0 & e^{i \alpha_2} & 0 \\ 0 & 0 & 1 \end{array} \right) ,
    \quad
  \left( \begin{array}{ccc} 1 & 0 & 0 \\ 0 & e^{i \phi_2} & 0 \\ 0 & 0 & e^{i (\phi_3 + \delta_{\rm CP})} \end{array} \right) ,
    \quad
  \left( \begin{array}{ccc} e^{i \rho} & 0 & 0 \\ 0 & 1 & 0 \\ 0 & 0 & e^{i \sigma} \end{array} \right) .
\eeq
All these choices are related by rephasings of the charged lepton fields and are
therefore physically equivalent.
The phase $\delta_{\rm CP}$ is often called the ``Dirac phase'' of the PMNS matrix, while
the phases contained in $P$, which can be restricted to the range $\left[ 0, \pi \right[$
without loss of generality, are called ``Majorana phases''. 
We will not be concerned with Majorana phases in this review because, as we are
going to see, they do not enter oscillation probabilities.
They appear only in lepton number violating processes like neutrinoless double beta decay,
in which the Majorana nature of neutrinos plays a crucial role (for reviews of neutrinoless
double beta decay, see e.g. Refs.~\cite{Vergados:2012xy,DellOro:2016tmg} for the theoretical
aspects and Ref.~\cite{Gomez-Cadenas:2015twa} for the experimental aspects).
The phase $\delta_{\rm CP}$, on the contrary, is relevant to neutrino oscillations and gives
rise to an asymmetry between neutrino and antineutrino oscillations in vacuum,
as will be discussed in Section~\ref{subsec:CPV}.

\subsection{Neutrino oscillations in vacuum    %
\label{subsec:vacuum}}                                   %

Schematically, an (idealized) oscillation experiment involves three steps.
The first one is the production of a pure flavour state from a charged current process,
e.g. a $\nu_\mu$ from a charged pion decay $\pi^+ \to \mu^+ \nu_\mu$. This flavour eigenstate
is a coherent superposition of mass eigenstates determined by the PMNS matrix\footnote{Note
that the relation between the flavour and mass eigenstate neutrino {\it states} involves the
complex conjugate of the PMNS matrix, as opposed to the PMNS matrix itself for neutrino
{\it fields}, Eq.~(\ref{eq:flavour_mass_relation}). This is because the quantum neutrino field
$\nu_\alpha(x)$ annihilates a neutrino of flavour $\alpha$, while the neutrino state
$\left| \nu_\alpha (\vec{p}) \right>$ is obtained by acting with the creation operator
$a^\dagger_\alpha (\vec{p})$ on the vacuum. For antineutrinos, one has
$\bar \nu_\alpha(x) = \sum_i U^*_{\alpha i} \bar \nu_i(x)$ and
$|\bar \nu_\alpha \rangle = \sum_i U_{\alpha i} |\bar \nu_i\rangle$.}:
\be
  |\nu (t=0) \rangle\, =\, |\nu_\alpha \rangle\, =\, \sum_i U^*_{\alpha i} |\nu_i\rangle\, .
\eeq
The second step is the propagation of the neutrino. Each mass eigenstate,
being en eigenstate of the Hamiltonian in vacuum, evolves with its own phase factor
$e^{-i E_i t}$, where $E_i = \sqrt{p^2 + m^2_i}$ is the energy of the $i$-th mass eigenstate
(in the standard convention $\hbar = c = 1$).
This modifies the coherent superposition, which is no longer a pure flavour eigenstate:
\be
  |\nu (t) \rangle\, =\, \sum_i U^*_{\alpha i}\, e^{-i E_i t}\, |\nu_i \rangle\,
    =\, \sum_i U^*_{\alpha i}\, e^{-i E_i t} \sum_\beta U_{\beta i} |\nu_\beta \rangle\, .
\eeq
The last step is the detection of a specific flavour via a charged current interaction.
The probability amplitude for the flavour $\alpha$ neutrino to have oscillated into
a different flavour $\beta$ at the time $t$ is given by $\langle \nu_\beta |\nu (t) \rangle$,
yielding an oscillation probability
\be
  P (\nu_\alpha \to \nu_\beta)\, =\, \left| \langle \nu_\beta |\nu (t) \rangle \right|^2\,
    =\, \left| \sum_i U_{\beta i} U^*_{\alpha i}\, e^{-i E_i t} \right|^2\, .
\eeq
Since neutrinos are ultra-relativistic in all practical experimental conditions, one
can expand $E_i = \sqrt{p^2 + m^2_i}\, \simeq\, p + m^2_i/(2E)$ (using $E \simeq p$).
The neutrino oscillation formula then reads\footnote{In deriving the oscillation
formula~(\ref{eq:oscillation_probability}), we made several simplifying assumptions:
the propagating mass eigenstates were described by plane waves with well-defined
momenta $\vec{p}_i$, which were further assumed to be equal ($\vec{p}_i = \vec{p}$).
The proper treatment of neutrino oscillations should be done in the wave-packet
formalism, or even in the framework of quantum field theory. However, under appropriate
coherence conditions, these approaches lead to the same oscillation probability
as the standard derivation presented here. See e.g. Ref.~\cite{Akhmedov:2009rb} for a discussion.}
\bea
  P (\nu_\alpha \to \nu_\beta)\ = &\! \delta_{\alpha \beta}\!\! &
    -\ 4 \sum_{i < j}\, \mbox{Re} \left[ U_{\alpha i} U^*_{\beta i} U^*_{\alpha j} U_{\beta j} \right]
    \sin^2 \left( \frac{\Delta m^2_{ji} L}{4 E} \right)  \nn \\
    && +\ 2 \sum_{i < j}\, \mbox{Im} \left[ U_{\alpha i} U^*_{\beta i} U^*_{\alpha j} U_{\beta j} \right]
    \sin \left( \frac{\Delta m^2_{ji} L}{2 E} \right) ,
\label{eq:oscillation_probability}
\eea
in which $\Delta m^2_{ji} \equiv m^2_j - m^2_i$ and $L \simeq c t$ is the distance
travelled by the neutrino. For antineutrino oscillations, one must replace $U$ by
$U^*$ in Eq.~(\ref{eq:oscillation_probability}), which amounts to change the sign
of the last term.

From Eq.~(\ref{eq:oscillation_probability}), a few obvious comments can be made
about the properties of neutrino oscillations. First of all, oscillations require neutrinos
to have non-degenerate masses ($\Delta m^2_{ji} \neq 0$) and non-trivial flavour
mixing ($U \neq \mathbf{1}$). The oscillation probability $P (\nu_\alpha \to \nu_\beta)$
depends on the three mixing angles $\theta_{12}$, $\theta_{23}$, $\theta_{13}$
and on two independent squared-mass differences, which can be chosen to be
$\Delta m^2_{21}$ and $\Delta m^2_{31}$ (then $\Delta m^2_{32}$ is determined
by $\Delta m^2_{32} = \Delta m^2_{31} - \Delta m^2_{21}$).
Oscillations also depend on the ``Dirac'' CP-violating phase $\delta_{\rm CP}$, but not
on the ``Majorana phases'', as can be seen from the fact that the PMNS matrix
entries appear in Eq.~(\ref{eq:oscillation_probability}) only
in the combinations $U_{\alpha i} U^*_{\beta i}$, to which the phases contained
in the diagonal matrix $P$ in Eq.~(\ref{eq:PMNS}) do not contribute.
Therefore, Dirac and Majorana
neutrinos have the same oscillation probabilities. This fact can be understood
on more general grounds, since oscillations
conserve total lepton number,
while the Majorana nature of neutrinos manifests itself in processes that violate
lepton number, like neutrinoless double beta decay.
Another consequence of formula~(\ref{eq:oscillation_probability}) is that
CP violation (namely, the fact that
$P (\bar \nu_\alpha \to \bar \nu_\beta) \neq P (\nu_\alpha \to \nu_\beta)$)
is possible only in appearance channels ($\beta \neq \alpha$), not in disappearance
channels ($\beta = \alpha$). Indeed, the survival (i.e. non-oscillation) probability
is the same for neutrinos and antineutrinos:
\bea
  P (\nu_\alpha \to \nu_\alpha)\, =\, 1 - 4 \sum_{i < j}\, \left| U_{\alpha i} U_{\alpha j} \right|^2
    \sin^2 \left( \frac{\Delta m^2_{ji} L}{4 E} \right) =\, P (\bar \nu_\alpha \to \bar \nu_\alpha) \, ,
\eea
because the combination $U_{\alpha i} U^*_{\beta i} U^*_{\alpha j} U_{\beta j}$
is real for $\alpha = \beta$.

While a detailed description of neutrino oscillations, including subleading and
CP-violation effects, necessitates the use of the full three-flavour formula,
it is a good approximation in many experimental situations to neglect the subleading
terms in Eq.~(\ref{eq:oscillation_probability}). One is then left with effective
two-flavour oscillations, governed by a single $\Delta m^2$ and a single mixing
angle $\theta$:
\be
  P (\nu_\alpha \to \nu_\beta)\, =\, P (\bar \nu_\alpha \to \bar \nu_\beta)\, =\,
    \sin^2 2 \theta\, \sin^2 \left( \frac{\Delta m^2 L}{4E} \right) \qquad (\beta \neq \alpha)\, ,
\label{eq:2f_oscillations}
\eeq
(we note in passing that oscillations do not violate the  CP symmetry in the two-flavour case).
In this case, the amplitude of oscillations is simply given by $\sin^2 2 \theta$,
and the oscillation length is proportional to the neutrino energy $E$ and inversely
proportional to $\Delta m^2$:
\be
  L_{\rm osc.} \mbox{(km)}\, =\, 2.48\, E \mbox{(GeV)} / \Delta m^2 (\mbox{eV}^2)\, , \qquad
  P (\nu_\alpha \to \nu_\beta)\, =\, \sin^2 2 \theta\, \sin^2 \left( \pi L / L_{\rm osc.} \right) .
\eeq
It should be stressed that the oscillation probability~(\ref{eq:2f_oscillations})
is sensitive neither to the sign of $\Delta m^2$ nor to the {\it octant} of $\theta$,
i.e. to whether $\theta \in\ ] 0, \frac{\pi}{4} [$ (first octant) or $\theta \in\ ] \frac{\pi}{4} , \frac{\pi}{2} [$
(second octant), $\theta = \frac{\pi}{4}$ corresponding to the maximal mixing angle.
This two-flavour formula has been used for decades, before oscillation experiments reached
a level of precision that made them sensitive to subdominant effects.
In particular, the oscillations of solar and atmospheric neutrinos turned out
to be well described in the framework of two-flavour oscillations (with the inclusion
of matter effects in the case of solar neutrinos, see Section~\ref{subsec:matter}) with parameters
($\Delta m^2_{\rm sol}$, $\theta_{\rm sol}$) and ($\Delta m^2_{\rm atm}$, $\theta_{\rm atm}$),
respectively, such that $\Delta m^2_{\rm sol} \ll \Delta m^2_{\rm atm}$ and both mixing
angles $\theta_{\rm sol}$ and $\theta_{\rm atm}$ are large. In order to interpret these
results in the framework of three-flavour oscillations, the following conventions
were adopted: {\it (i)} $\Delta m^2_{\rm sol}$ is identified with the squared-mass splitting
between $\nu_1$ and $\nu_2$; {\it (ii)} these mass eigenstates are labelled
in such a way that $m_2 > m_1$, i.e. $\Delta m^2_{21} = \Delta m^2_{\rm sol} > 0$.
Then $\Delta m^2_{\rm atm}$ must be identified with $|\Delta m^2_{31}|$ or $|\Delta m^2_{32}|$.
Since $\Delta m^2_{\rm sol} \ll \Delta m^2_{\rm atm}$, this implies
\be
  \Delta m^2_{\rm sol} = \Delta m^2_{21}\, \ll\, |\Delta m^2_{31}| \simeq |\Delta m^2_{32}| \simeq \Delta m^2_{\rm atm}\, .
\label{eq:Delta_m2}
\eeq
We are left with two possibilities for the mass spectrum: either $m_1 < m_2 < m_3$,
which is referred to as the {\it normal mass ordering} or {\it normal hierarchy}, characterized by
$\Delta m^2_{31} > 0$; or $m_3 < m_1 < m_2$, which is known as the {\it inverted mass ordering}
or {\it inverted hierarchy}, characterized by $\Delta m^2_{31} < 0$. The mass ordering
$m_1 < m_3 < m_2$ is excluded, as it is not consistent with Eq.~(\ref{eq:Delta_m2}).

Experiments characterized by a baseline $L$ and a beam energy $E$ such that $\Delta m^2_{21} L / E \ll 1$
can be described to a good approximation by setting $\Delta m^2_{21} = 0$ in the
three-flavour formula~(\ref{eq:oscillation_probability}), so that one is left with a single
oscillation ``frequency'' $\Delta m^2_{31} = \Delta m^2_{32}$. This approximation is valid
because all mixing angles, with the exception of $\theta_{13}$ which is a bit smaller,
are of comparable magnitude. The oscillation (appearance)
probability becomes~\cite{DeRujula:1979brg,Barger:1980hs}
\be
  P (\nu_\alpha \to \nu_\beta)\, =\, \sin^2 2 \theta^{\rm eff}_{\alpha \beta}\,
    \sin^2 \left( \frac{\Delta m^2_{31} L}{4E} \right) ,  \qquad
    \sin^2 2 \theta^{\rm eff}_{\alpha \beta}\, \equiv\, 4 \left| U_{\alpha 3} U_{\beta 3} \right|^2 \qquad (\beta \neq \alpha)\, ,
\eeq
while for the non-oscillation (disappearance) probability:
\be
  P (\nu_\alpha \to \nu_\alpha)\, =\, 1 - \sin^2 2 \theta^{\rm eff}_{\alpha \alpha}\,
    \sin^2 \left( \frac{\Delta m^2_{31} L}{4E} \right) ,  \qquad  \sin^2 2 \theta^{\rm eff}_{\alpha \alpha}\,
    \equiv\, 4 \left| U_{\alpha 3} \right|^2 \left( 1 - \left| U_{\alpha 3} \right|^2 \right) .
\eeq
These formulae describe the dominant oscillations in atmospheric neutrinos,
long-baseline accelerator neutrino experiments and short-baseline reactor experiments.
For instance, the probability of muon neutrino disappearance is
\bea
  P (\nu_\mu \to \nu_\mu)\!\! & =\!\! &  1 - \left( \cos^2 \theta_{13} \sin^2 2 \theta_{23}
    + \sin^4 \theta_{23} \sin^2 2 \theta_{13}  \right) \sin^2 \left( \frac{\Delta m^2_{31} L}{4E} \right)  \\
  & \simeq\!\! & 1 - \sin^2 2 \theta_{23}\, \sin^2 \left( \frac{\Delta m^2_{31} L}{4E} \right) ,
\label{eq:numudisappApp}
\eea
where terms proportional to $\sin^2 \theta_{13}$ were neglected in the second line.
This justifies the widely-used terminology ``atmospheric mixing angle'' for $\theta_{23}$
and ``atmospheric $\Delta m^2$'' for $\Delta m^2_{31}$ (or $\Delta m^2_{32}$).
For electron and tau neutrino appearance in a muon neutrino beam, one has
\bea
  P (\nu_\mu \to \nu_e)\!\! & =\!\! & \sin^2 \theta_{23} \sin^2 2 \theta_{13}\, \sin^2 \left( \frac{\Delta m^2_{31} L}{4E} \right) ,
    \label{eq:nu_mu_nu_e}  \\
  P (\nu_\mu \to \nu_\tau)\!\! & =\!\! & \cos^4 \theta_{13} \sin^2 2 \theta_{23}\, \sin^2 \left( \frac{\Delta m^2_{31} L}{4E} \right) ,
    \label{eq:nu_mu_nu_tau}
\eea
while for short-baseline disappearance of reactor antineutrinos:
\be
  P (\bar \nu_e \to \bar \nu_e)\, =\, 1 - \sin^2 2 \theta_{13}\, \sin^2 \left( \frac{\Delta m^2_{31} L}{4E} \right) .
\label{eq:nu_e_disapp}
\eeq
It should be kept in mind that the above expressions receive corrections from three-flavour effects,
which must be taken into account to comply with the precision of present-day experiments.
Note in passing that although governed by a single oscillation frequency,
the probabilities~(\ref{eq:nu_mu_nu_e}) and~(\ref{eq:nu_mu_nu_tau}) are not simple
two-flavour formulae, as they depend on the two mixing angles $\theta_{13}$ and $\theta_{23}$.
In particular, Eq.~(\ref{eq:nu_mu_nu_e}) is sensitive to the octant of $\theta_{23}$.

When instead $\Delta m^2_{31} L / E \gg 1$ and $\Delta m^2_{21} L / E \gtrsim 1$,
$\Delta m^2_{31}$-driven oscillations are averaged and oscillations of electron neutrinos
are dominated by $\Delta m^2_{21}$ rather than $\Delta m^2_{31}$.
Neglecting terms suppressed by $\sin^2 \theta_{13}$, one obtains
\be
\label{eq:nuereactor}
  P (\nu_e \to \nu_e)\, =\, P (\bar \nu_e \to \bar \nu_e)\, \simeq\,
    1 - \sin^2 2 \theta_{12}\, \sin^2 \left( \frac{\Delta m^2_{21} L}{4E} \right) .
\eeq
This formula applies to the long-baseline reactor neutrino experiment KamLAND and to
low-energy solar neutrinos (with the oscillating term averaged), for which matter effects
are subdominant compared with vacuum oscillations. This justifies the popular terminology
``solar mixing angle'' for $\theta_{12}$ and ``solar $\Delta m^2$'' for $\Delta m^2_{21}$.
Restoring $\theta_{13}$, one obtains the more accurate expression
\be
  P (\nu_e \to \nu_e)\, =\, P (\bar \nu_e \to \bar \nu_e)\, =\, \sin^4 \theta_{13}
    + \cos^4 \theta_{13} \left( 1 - \sin^2 2 \theta_{12}\, \sin^2 \left( \frac{\Delta m^2_{21} L}{4E} \right) \right) .
\label{eq:nue_nue_2f_improved}
\eeq
%

\subsection{CP violation in neutrino oscillations and three-flavour effects   %
\label{subsec:CPV}}                                                                                     %

As we have seen in Section~\ref{subsec:vacuum}, oscillations depend on the phase
$\delta_{\rm CP}$ of the PMNS matrix, making it possible to observe the violation of the CP
symmetry in neutrino oscillations~\cite{Cabibbo:1977nk} -- namely, the fact that neutrinos
and antineutrinos oscillate with different probabilities in vacuum. Before discussing this possibility
in detail, let us summarize the action of the different discrete symmetries\footnote{It should
be stressed that $P (\bar \nu_\alpha \to \bar \nu_\beta)$
is the image of $P (\nu_\alpha \to \nu_\beta)$ by CP, not by the charge conjugation C.
Indeed, $\nu_\alpha$ and $\nu_\beta$ are left-handed neutrinos, whose antiparticles
$\bar \nu_\alpha$ and $\bar \nu_\beta$ are right-handed antineutrinos, i.e. the
CP conjugates of $\nu_\alpha$ and $\nu_\beta$. Their charge conjugates would be
hypothetical left-handed antineutrinos, which do not couple to the $W$ and $Z$ bosons
and are not produced in weak processes.} on oscillation
probabilities (we refer the reader to Ref.~\cite{Akhmedov:2004ve}
for a detailed discussion of the action of CP and T on oscillations, both in vacuum and in matter):
\bea
  P (\nu_\alpha \to \nu_\beta) & \xrightarrow{\mbox{  CP  }} & P (\bar \nu_\alpha \to \bar \nu_\beta)\ ,  \\
  & \xrightarrow{\mbox{\phantom{P}T\phantom{P}}} & P (\nu_\beta \to \nu_\alpha)\ ,  \\
  & \xrightarrow{\mbox{CPT}} & P (\bar \nu_\beta \to \bar \nu_\alpha)\ .
\eea
Thus, if CPT conservation is assumed, $P (\nu_\alpha \to \nu_\beta) = P (\bar \nu_\beta \to \bar \nu_\alpha)$,
implying that the CP and T asymmetries in neutrino oscillations are equal:
\be
  A_{\alpha \beta}\, \equiv\, \frac{P (\nu_\alpha \to \nu_\beta) - P (\bar \nu_\alpha \to \bar \nu_\beta)}
    {P (\nu_\alpha \to \nu_\beta) + P (\bar \nu_\alpha \to \bar \nu_\beta)}\,
  =\, \frac{P (\nu_\alpha \to \nu_\beta) - P (\nu_\beta \to \nu_\alpha)}
    {P (\nu_\alpha \to \nu_\beta) + P (\nu_\beta \to \nu_\alpha)}\ .
\eeq
Another implication of CPT conservation is $P (\nu_\alpha \to \nu_\alpha) = P (\bar \nu_\alpha \to \bar \nu_\alpha)$,
i.e. there is no CP violation in disappearance experiments, a fact we have
already deduced from the general three-flavour oscillation formula.

In order to study the effect of CP violation (CPV) in neutrino oscillations, it is convenient
to introduce the quantity
$\Delta P_{\alpha \beta} \equiv P (\nu_\alpha \to \nu_\beta) - P (\bar \nu_\alpha \to \bar \nu_\beta)$,
which is nothing but twice the CP-odd part in the three-flavour oscillation
probability~(\ref{eq:oscillation_probability}). One can show that it can be expressed
as~\cite{Bilenky:1980cx,Barger:1980jm}
\be
  \Delta P_{\alpha \beta}\, =\, \pm 16 J \sin \left( \frac{\Delta m^2_{21} L}{4 E} \right)
    \sin \left( \frac{\Delta m^2_{31} L}{4 E} \right) \sin \left( \frac{\Delta m^2_{32} L}{4 E} \right) ,\ \ \quad
    J\, \equiv\, \mbox{Im} \left[ U_{e 1} U^*_{\mu 1} U^*_{e 2} U_{\mu 2} \right] ,
\label{eq:CP_asymmety}
\eeq
with a $+$ sign when $(\alpha, \beta, \gamma)$ (with $\gamma \neq \alpha, \beta$)
is an even permutation of $(e, \mu, \tau)$, and a $-$ sign when it is an odd permutation.
The quantity $J$ in Eq.~(\ref{eq:CP_asymmety}) is a measure of CP violation from the
``Dirac'' phase of the PMNS matrix and is called {\it Jarlskog invariant}~\!\footnote{The name
``invariant'' refers to the fact that $J$ does not depend on the phase convention of the PMNS
matrix, i.e. it is invariant under rephasings of the lepton fields. Historically, the Jarlskog
invariant has been introduced to parametrize CP violation in the quark sector~\cite{Jarlskog:1985cw};
the invariant $J$ in Eq.~(\ref{eq:CP_asymmety}) is its generalization to the lepton sector.}.
Using the standard parametrization of the PMNS matrix, one can write
\be
  J\, =\, \frac{1}{8} \cos \theta_{13} \sin 2 \theta_{12} \sin 2 \theta_{13} \sin 2 \theta_{23} \sin \delta_{\rm CP}\, .
\label{eq:Jarlskog}
\eeq
Therefore, a necessary condition for CP violation in neutrino oscillations
is that all three mixing angles $\theta_{ij}$ are nonzero and that the phase
$\delta_{\rm CP}$ is different from $0$ and $\pi$.
Furthermore, one can see from Eq.~(\ref{eq:CP_asymmety}) that $\Delta m^2_{21}$,
$\Delta m^2_{31}$ and $\Delta m^2_{32}$ must be non vanishing, i.e. all neutrinos masses
should be different:
\bea
  \mbox{\it Conditions for CPV in oscillations :} \qquad
  \theta_{ij} \neq 0\, , \ \delta_{\rm CP} \neq 0, \pi\, , \ m_1 \neq m_2\, , \ m_2 \neq m_3\, , \ m_3 \neq m_1\, .
\eea
These criteria parallel the ones for CP violation in the quark sector.
They are of rather academic interest now that all $\theta_{ij}$'s and
$\Delta m^2_{ji}$'s have been measured, but for a long time the only
experimental information we had on $\theta_{13}$ was an upper bound
(see Section 5.1 for a discussion of early limits and first indications).
A value of $\sin^2 2 \theta_{13}$ below $10^{-4}$ 
would have made it very
difficult to observe CP violation in oscillation experiments, even if the CP-violating
phase $\delta_{\rm CP}$ were maximal.

The formula~(\ref{eq:CP_asymmety}) contains a lot of information. First of all, it tells us
that the CP-violating term in neutrino oscillations is universal, i.e. it does not depend
on the oscillation channel (up to a sign). This follows from the unitarity of the PMNS
matrix, which implies
\bea
  \mbox{Im} \left[ U_{e 1} U^*_{\mu 1} U^*_{e 2} U_{\mu 2} \right]
    = - \mbox{Im} \left[ U_{e 1} U^*_{\tau 1} U^*_{e 2} U_{\tau 2} \right]
    = \mbox{Im} \left[ U_{\mu 1} U^*_{\tau 1} U^*_{\mu 2} U_{\tau 2} \right] ,
\eea
from which $\Delta P_{e \mu}\, = - \Delta P_{e \tau}\, = \Delta P_{\mu \tau}$ follows.
Another important information contained in Eq.~(\ref{eq:CP_asymmety}) is that
the effect of CP violation is proportional to $\sin ( \Delta m^2_{21} L / 4 E )$,
i.e. it can be observed only in experiments that are sensitive to the subdominant
oscillations governed by $\Delta m^2_{21}$. This is why experiments searching
for CP violation involve long baselines (several hundreds of km),
intense neutrino beams and large detectors.
Typically, the experimental conditions are such that $\Delta m^2_{31} L / E \sim 1$
and $\Delta m^2_{21} L / E \ll 1$, hence Eq.~(\ref{eq:CP_asymmety}) can be simplified to,
at second order in the small parameters $\sin \left( \Delta m^2_{21} L / 4 E \right)$ and $\sin \theta_{13}$:
\be
  \Delta P_{\alpha \beta}\, \simeq\, \pm 16 J
    \sin \left( \frac{\Delta m^2_{21} L}{4 E} \right) \sin^2 \left( \frac{\Delta m^2_{31} L}{4 E} \right) .
\label{eq:CP_asymmety_approx}
\eeq
Long baselines however imply that neutrinos propagate in the Earth crust, so that
their oscillations are affected by their interactions with matter (see Subsection~\ref{subsec:constant}).
This in turn creates, in the $\nu_\mu$--$\nu_e$ channel relevant to long-baseline experiments,
an asymmetry between neutrino and antineutrino oscillations
whose sign is related to the type of mass hierarchy,
normal ($\Delta m^2_{31} > 0$) or inverted ($\Delta m^2_{31} < 0$).
It is therefore necessary to disentangle the effect of neutrino interactions with matter
from the one of CP violation in the experimental data.

Neglecting matter effects (which is a reasonable approximation for a long-baseline experiment
like T2K), one can expand the full $\nu_\mu \to \nu_e$ oscillation probability to second order
in the small quantities $\sin \left( \Delta m^2_{21} L / 4 E \right)$ and $\sin \theta_{13}$:
\bea
  P (\nu_\mu \to \nu_e)\!\! & \simeq\!\! & \sin^2 \theta_{23} \sin^2 2 \theta_{13}\, \sin^2 \left( \frac{\Delta m^2_{31} L}{4 E} \right)
    + \cos^2 \theta_{23} \sin^2 2 \theta_{12}\, \sin^2 \left( \frac{\Delta m^2_{21} L}{4 E} \right)  \nn \\
  && +\, \frac{1}{2} \cos \theta_{13} \sin 2 \theta_{12} \sin 2 \theta_{13} \sin 2 \theta_{23} \cos \delta_{\rm CP}\,
    \sin \left( \frac{\Delta m^2_{21} L}{4 E} \right) \sin \left( \frac{\Delta m^2_{31} L}{2 E} \right)  \nn \\
  && -\, \cos \theta_{13} \sin 2 \theta_{12} \sin 2 \theta_{13} \sin 2 \theta_{23} \sin \delta_{\rm CP}\,
    \sin \left( \frac{\Delta m^2_{21} L}{4 E} \right) \sin^2 \left( \frac{\Delta m^2_{31} L}{4 E} \right) .
\label{eq:numu_nue_3f}
\eea
The first term corresponds to the dominant, $\Delta m^2_{31}$-driven oscillations;
the second one to the $\Delta m^2_{21}$-driven oscillations; the third and fourth terms involve
both $\Delta m^2_{21}$ and $\Delta m^2_{31}$ and are CP-even and CP-odd, respectively.
While, as shown above, the CP-odd term is proportional to the Jarlskog invariant $J$,
the CP-even term is proportional to $J \cot \delta_{\rm CP} \propto \cos \delta_{\rm CP}$.
Note that the dominant oscillations are sensitive to the octant of $\theta_{23}$.
Eq.~(\ref{eq:numu_nue_3f}) can be rewritten in the more compact form
\be
  P (\nu_\mu \to \nu_e)\, \simeq\, A^2_{\rm atm} + A^2_{\rm sol} + 2 \cos \theta_{13} A_{\rm atm} A_{\rm sol}
    \cos \left( \frac{\Delta m^2_{31} L}{4 E} + \delta_{\rm CP} \right) ,
\label{eq:numu_nue_compact}
\eeq
where $A_{\rm atm} \equiv \sin \theta_{23} \sin 2 \theta_{13}\, \sin \left( \frac{\Delta m^2_{31} L}{4 E} \right)$
and $A_{\rm sol} \equiv \cos \theta_{23} \sin 2 \theta_{12}\, \sin \left( \frac{\Delta m^2_{21} L}{4 E} \right)$.
The corresponding formulae for $P (\bar \nu_\mu \to \bar \nu_e)$ can be derived by
switching the sign of the phase $\delta_{\rm CP}$ in the above expressions
(in Eq.~(\ref{eq:numu_nue_3f}), this amounts to change the sign of the last term).
One can quantify the amount of CP violation in the $\nu_\mu$--$\nu_e$ channel
with the CP asymmetry parameter
\be
  A_{\mu e}\, \equiv\, \frac{P (\nu_\mu \to \nu_e) - P (\bar \nu_\mu \to \bar \nu_e)}
    {P (\nu_\mu \to \nu_e) + P (\bar \nu_\mu \to \bar \nu_e)}\ \simeq\,
    -\, \frac{ \cos \theta_{23} \sin 2 \theta_{12}}{\sin \theta_{23} \sin \theta_{13}}\
    \sin \left( \frac{\Delta m^2_{21} L}{4 E} \right) \sin \delta_{\rm CP}\, .
\eeq
%

\subsection{Other three-flavour effects    %
\label{subsec:3-flavour}}                              %

With the increased precision of short-baseline reactor neutrino experiments, the experimental
uncertainty on the squared-mass difference governing $\bar \nu_e$ disappearance approaches
the difference between $\Delta m^2_{31}$ and $\Delta m^2_{32}$. In this context, the two-flavour
formula~(\ref{eq:nu_e_disapp}) is no longer appropriate and should be replaced by
the three-flavour survival probability
\bea
  P (\bar \nu_e \to \bar \nu_e)\!\! & =\!\! & 1
    - \cos^4 \theta_{13} \sin^2 2 \theta_{12} \sin^2 \left( \frac{\Delta m^2_{21} L}{4 E} \right)
    - \cos^2 \theta_{12} \sin^2 2 \theta_{13} \sin^2 \left( \frac{\Delta m^2_{31} L}{4 E} \right)  \nn \\
  && \phantom{1 } - \sin^2 \theta_{12} \sin^2 2 \theta_{13} \sin^2 \left( \frac{\Delta m^2_{32} L}{4 E} \right) .
\label{eq:nue_nue}
\eea
It has been shown in Ref.~\cite{nunokawa} that Eq.~(\ref{eq:nue_nue})
is very well approximated at short distances by the formula
\bea
  P (\bar \nu_e \to \bar \nu_e)\!\! & \simeq\!\! & 1
    - \cos^4 \theta_{13} \sin^2 2 \theta_{12} \sin^2 \left( \frac{\Delta m^2_{21} L}{4 E} \right)
    - \sin^2 2 \theta_{13} \sin^2 \left( \frac{\Delta m^2_{ee} L}{4 E} \right) ,
\label{eq:nue_nue_approx}
\eea
in which $\Delta m^2_{ee} \equiv \cos^2 \theta_{12} \Delta m^2_{31} + \sin^2 \theta_{12} \Delta m^2_{32}$
can be viewed as the ``$\nu_e$-weighted average'' of $\Delta m^2_{31}$ and $\Delta m^2_{32}$.
Namely, Eq.~(\ref{eq:nue_nue_approx}) is accurate to better than one part in $10^4$
for $L/E < 1\, \mbox{km/MeV}$~\cite{parkemee}. This makes it possible to determine
$|\Delta m^2_{31}|$ from the measurement of $|\Delta m^2_{ee}|$
(assuming a given mass ordering
and using the knowledge of $\theta_{12}$ and $\Delta m^2_{21}$ from other experiments),
with a better precision than through Eq.~(\ref{eq:nu_e_disapp}).
For more precise reactor neutrino experiments, a more accurate formula than Eq.~(\ref{eq:nue_nue_approx})
can be used to determine $|\Delta m^2_{ee}|$~\cite{parkemee}.

Long-baseline reactor neutrino experiments such that $\Delta m^2_{21} L / E \sim 1$
are sensitive to oscillations governed by $\Delta m^2_{21}$, while the subleading $\Delta m^2_{31}$-
and $\Delta m^2_{32}$-driven oscillations in Eq.~(\ref{eq:nue_nue}) are averaged
due to the limited energy resolution of the detector, leading to the survival
probability~(\ref{eq:nue_nue_2f_improved}). However, an improvement of the energy
resolution could in principle render these experiments sensitive to
the fast, small-amplitude oscillations governed by $\Delta m^2_{31}$ and $\Delta m^2_{32}$
that develop on top of the dominant
$\Delta m^2_{21}$-driven oscillations. It has been suggested that a precision measurement
of the energy spectrum of reactor antineutrinos at a far detector would make it possible
to distinguish between the normal and inverted mass hierarchies (see Section~\ref{sec:future}
for details). Indeed, the last two terms
in Eq.~(\ref{eq:nue_nue}) lead to different distorsions of the energy spectrum,
depending on whether the hierarchy is normal (in which case
$|\Delta m^2_{31}| > |\Delta m^2_{32}|$) or inverted
(in which case $|\Delta m^2_{32}| > |\Delta m^2_{31}|$).

\subsection{Neutrino propagation in matter    %
\label{subsec:matter}}                                         %

The interactions of neutrinos with electrons, protons and neutrons affect their propagation in matter,
leading to a variety of new phenomena, among which:
{\it (i)} oscillations in matter with modified parameters with respect to vacuum oscillations~\cite{Wolfenstein:1977ue,Barger:1980tf};
{\it (ii)} resonant amplification of oscillations in a medium of constant density~\cite{Mikheev:1986gs,Mikheev:1986wj};
{\it (iii)} adiabatic flavour conversion in a medium of varying density like the Sun~\cite{Mikheev:1986gs,Mikheev:1986wj};
{\it (iv)} parametric enhancement of oscillations in a medium with several layers of alternating
densities~\cite{Akhmedov:1988kd,Krastev:1989ix}, which can be experienced by neutrinos
passing through the core of the Earth~\cite{Petcov:1998su,Akhmedov:1998ui};
{\it (v)} collective effects from neutrino-neutrino interactions in core-collapse supernovae
or in the early universe~\cite{Pantaleone:1992eq,Samuel:1993uw};
{\it (vi)} non-standard matter effects if new, flavour-violating neutrino interactions are present~\cite{Wolfenstein:1977ue}.
In this review, we will essentially encounter matter effects of the type {\it (i)}--{\it (iii)},
on which we therefore focus in this section (see also Refs.~\cite{Blennow:2013rca,Kuo:1989qe}
for specialized reviews on the subject). We refer the reader to other reviews for
a discussion of the phenomena
that are not addressed here, e.g. to Refs.~\cite{pdg,Blennow:2013rca}
for item~{\it (iv)} and to Section~4 of Ref.~\cite{Mirizzi:2015eza} for item~{\it (v)}.

\subsubsection{Basic formalism                      %
\label{subsec:formalism}}                                %

Neutrino propagation in matter can be described by a Schr\"odinger-like equation:
\be
  i \frac{d}{dt} \left| \nu(t) \right>\, =\, H \left| \nu(t) \right> ,
\label{eq:Schroedinger}
\eeq
where $\left| \nu(t) \right>$ is the neutrino state vector at time $t$, and the Hamiltonian $H$
can be split into a free (kinetic energy) part $H_0$ describing neutrino propagation in vacuum
and a potential term $V$ induced by the interactions of neutrinos in the medium~\cite{Wolfenstein:1977ue}:
\be
  H\, =\, H_0 + V\, .
\eeq
It is convenient to write the evolution equation~(\ref{eq:Schroedinger}) in
the flavour eigenstate basis $\{\, \left| \nu_e \right>, \left| \nu_\mu \right>, \left| \nu_\tau \right> \}$:
\be
  i \frac{d}{dt}\, \nu_\beta(t)\, =\, \sum_\gamma H_{\beta \gamma}\, \nu_\gamma(t)\, , \qquad \qquad
    \beta, \gamma = e, \mu, \tau\, .
\label{eq:Schroedinger_components}
\eeq
In Eq.~(\ref{eq:Schroedinger_components}),
the $H_{\beta \gamma} \equiv \left< \nu_\beta \right| H \left| \nu_\gamma \right>$ are
the Hamiltonian matrix elements in the flavour basis,
and $\nu_\beta(t) \equiv \left< \nu_\beta | \nu(t) \right>$
is the projection of the neutrino state vector onto the basis vector $\left| \nu_\beta \right>$,
i.e. the probability amplitude to find
the neutrino in the flavour eigenstate $\left| \nu_\beta \right>$ at time $t$.
Thus, if the neutrino is produced at $t=0$ in the flavour eigenstate $\left| \nu_\alpha \right>$,
the oscillation probability
is given by $P(\nu_\alpha \to \nu_\beta; t) = |\nu_\beta(t)|^2$.

Let us first consider the vacuum Hamiltonian. In the flavour basis, it is given by\footnote{This expression
can be easily derived by noting that, in the mass eigenstate basis, the vacuum Hamiltonian is
diagonal with eigenvalues $E_i$:  $\left< \nu_i \right| H \left| \nu_j \right> = E_i \delta_{ij}$.
Using $\left|\nu_\beta \right> = \sum_i U^*_{\beta i} \left|\nu_i \right>$, one then arrives at
$H_{\beta \gamma} = \left< \nu_\beta \right| H \left| \nu_\gamma \right>
= \sum_{i,j} U_{\beta i} U^*_{\gamma j} \left< \nu_i \right| H \left| \nu_j \right>
= \sum_i U_{\beta i} U^*_{\gamma i} E_i$.}:
\be
  H_0\, =\, U\, \mbox{Diag}\, (E_1, E_2, E_3)\, U^\dagger ,  \qquad \qquad  E_i = \sqrt{p^2 + m^2_i}\ ,
\eeq
where $p$ is the modulus of the neutrino momentum,
$m_i$ the mass of the $i^{\rm th}$ neutrino mass eigenstate
($i = 1, 2, 3$) and $U$ the lepton mixing matrix.
Assuming ultrarelativistic neutrinos, one can expand
$E_i \simeq p + m^2_i/(2E)$ (in which $E \simeq p$)
and redefine $H_0 \to H_0 - p \mathbf{1}$ to obtain:
\be
  H_0\, =\, \frac{1}{2E}\ U\, \mbox{Diag}\, (m^2_1, m^2_2, m^2_3)\, U^\dagger\,
    =\, \frac{M^\dagger_\nu M_\nu}{2E}\ ,
\label{eq:H0}
\eeq
where $M_\nu$ is the neutrino mass matrix in the flavour basis.
Indeed, removing a term proportional to the unit matrix from $H$ only affects the overall
phase of the neutrino state vector\footnote{This is true even if this term is time dependent.
If $\nu_\beta(t)$ satisfies the evolution equation~(\ref{eq:Schroedinger_components})
with Hamiltonian $H$, then
$\nu'_\beta(t) = e^{i \int_0^t E_0(t')dt'}\, \nu_\beta(t)$ satisfies the same equation  
with the shifted Hamiltonian $H' = H - E_0(t) \mathbf{1}$. $H$ and $H'$ lead to the same
oscillation probabilities since $|\nu'_\beta(t)|^2 = |\nu_\beta(t)|^2$.}, which is unobservable.
In the two-flavour case, Eq.~(\ref{eq:H0}) reduces to (after subtracting another term
proportional to $\mathbf{1}$ from $H_0$):
\be
  H_0\, = \left(\!\! \begin{array}{cc}
    - \frac{\Delta m^2}{2E} \cos 2 \theta &  \frac{\Delta m^2}{4E} \sin 2 \theta  \\
    \frac{\Delta m^2}{4E} \sin 2 \theta & 0  \end{array}\!\! \right) ,
\eeq
where $\Delta m^2 = m^2_2 - m^2_1$, and $\theta$ is the angle that parametrizes the
$2 \times 2$ lepton mixing matrix (omitting a phase irrelevant to oscillations in the
Majorana case):
\be
  \left(\!\! \begin{array}{c} \left|\nu_\alpha \right> \\ \left|\nu_\beta \right> \end{array}\!\! \right)
    =\, \left(\!\! \begin{array}{cc} \cos \theta & \sin \theta \\ - \sin \theta & \cos \theta \end{array}\!\! \right)
    \left(\!\! \begin{array}{c} \left|\nu_1 \right> \\ \left|\nu_2 \right> \end{array}\!\! \right) .
\eeq

The matter potential $V$ is induced by coherent forward scatterings of neutrinos
on electrons and nucleons in the medium, which leave the neutrino momentum
unchanged and can therefore interfere with the propagation of the unscattered
neutrinos~\cite{Wolfenstein:1977ue}.
It receives a contribution from W boson exchange (charged current) that
is present only for electron neutrinos, as ordinary matter does not contain muons nor taus,
and another one from Z boson exchange (neutral current) that is identical for all neutrino flavours.
The matter potential is diagonal in the flavour basis:
\be
  V_{\alpha \beta}\, =\, V_\alpha\, \delta_{\alpha \beta}\,
    = \left( V_{{\rm CC}, \alpha} + V_{{\rm NC}, \alpha} \right) \delta_{\alpha \beta}\, ,
\eeq
where the charged current contribution $V_{{\rm CC}, \alpha}$ depends on the neutrino flavour $\alpha$,
while the neutral current contribution $V_{{\rm NC}, \alpha}$ is flavour universal 
(a detailed derivation of these potentials can be found in the review~\cite{GonzalezGarcia:2007ib}):
\be
  V_{{\rm CC}, \alpha}\, =\, \left\{\!\! \begin{array}{cl}
      \sqrt{2}\, G_F n_e(x) & \quad \alpha = e \\ 0 & \quad \alpha = \mu, \tau \end{array} \right. ,  \qquad
  V_{{\rm NC}, \alpha}\, =\,
      - \frac{G_F}{\sqrt{2}}\, n_n(x)  \quad \left( \alpha = e, \mu, \tau \right)\, .
\label{eq:V_CC_NC}
\eeq
In Eq.~(\ref{eq:V_CC_NC}), $G_F = 1.166 \times 10^{-5}\, \mbox{GeV}^{-2}$
is the Fermi constant and $n_e(x)$, $n_n(x)$ are the electron and neutron densities
in the medium\footnote{Note that the neutral current contribution to the matter potential
only depends on the neutron density: the proton and electron contributions cancel out
in $V_{\rm NC}$ due to the assumed neutrality of the medium, which implies $n_p = n_e$.
Eq.~(\ref{eq:V_CC_NC}) also assumes that the medium is unpolarized and made of
non-relativistic particles.},
which a priori depend on the spatial position $x$. For antineutrinos,
the matter potential has the opposite sign:
\be
  V_\alpha (\bar \nu)\, =\, - V_\alpha (\nu)\, .
\eeq
After subtraction of the universal neutral current contribution, the matter Hamiltonian for neutrinos is given by, in the flavour basis:
\be
  H_{\beta \gamma}\, =\, \frac{1}{2E}\, \sum_i U_{\beta i} U^*_{\gamma i} m^2_i
    + V_{{\rm CC}, \beta}\, \delta_{\beta \gamma}\, .
\label{eq:H_components}
\eeq
For antineutrinos, the following replacements should be made in Eq.~(\ref{eq:H_components}):
\be
  U\, \to\, U^*\, ,  \qquad  V\, \to\, - V\, .
\eeq

In matter, the propagation eigenstates are not the mass eigenstates $\left|\nu_i \right>$ as in vacuum,
but the eigenstates of the matter Hamiltonian $\left|\nu^m_i \right>$, called {\it matter eigenstates}. 
They are related to the flavour eigenstates
by the mixing matrix in matter $U_m$, which diagonalizes $H$:
\be
  H\, =\, U_m \left(\!\! \begin{array}{ccc} E^m_1 & 0 & 0 \\ 0 & E^m_2 & 0  \\
    0 & 0 & E^m_3 \end{array}\!\! \right) U^\dagger_m\, ,
  \qquad  \qquad  \left(\!\! \begin{array}{c} \left|\nu_e \right> \\ \left|\nu_\mu \right> \\ \left|\nu_\tau \right> \end{array}\!\! \right)
    =\, U^*_m \left(\!\! \begin{array}{c} \left|\nu^m_1 \right> \\ \left|\nu^m_2 \right> \\ \left|\nu^m_3 \right> \end{array}\!\! \right) .
\label{eq:H_diagonalization}
\eeq
The eigenvalues $E^m_i$ of the matter Hamiltonian $H$ are called the {\it energy levels in matter}.
The amplitude of probability to find the neutrino in the matter eigenstate $\left| \nu^m_i \right>$
at the time $t$, $\nu^m_i(t) = \left< \nu^m_i | \nu(t) \right>$, is related to the amplitude
of probability $\nu_\beta(t)$ to find it in the flavour eigenstate $\left| \nu_\beta \right>$
by the mixing matrix in matter:
\be
  \nu_\beta(t)\, =\, \sum_i (U_m)_{\beta i}\, \nu^m_i(t)\, .
\label{eq:nu^m_i-nu_alpha}
\eeq
Eq.~(\ref{eq:nu^m_i-nu_alpha}) is nothing but the generalization of the vacuum relation
$\nu_\beta(t)\, =\, \sum_i U_{\beta i}\, \nu_i(t)$.

In many physical contexts, such as neutrino propagation in the Sun, it is a good approximation
to work in the two-flavour framework.
In the flavour eigenstate basis $\{\, \left| \nu_e \right>, \left| \nu_\beta \right> \}$
(where $\beta = \mu, \tau$),
the matter Hamiltonian is given by:
\be
  H\, = \left(\!\! \begin{array}{cc}
    - \frac{\Delta m^2}{2E} \cos 2 \theta \pm \sqrt{2}\, G_F n_e &  \frac{\Delta m^2}{4E} \sin 2 \theta  \\
    \frac{\Delta m^2}{4E} \sin 2 \theta & 0  \end{array}\!\! \right) ,
\label{eq:H_2f}
\eeq
in which $\Delta m^2$ and $\theta$ are the relevant vacuum oscillation parameters,
and the $+$ sign (resp. the $-$ sign) is for neutrinos (resp. for antineutrinos).
This Hamiltonian is diagonalized by the mixing angle in matter $\theta_m$~\cite{Wolfenstein:1977ue}:
\be
  H\, =\, U_m \left(\!\! \begin{array}{cc} E^m_1 & 0 \\ 0 & E^m_2 \end{array}\!\! \right) U^\dagger_m\ , \qquad \qquad
  U_m\, = \left(\!\! \begin{array}{cc} \cos \theta_m & \sin \theta_m \\ - \sin \theta_m & \cos \theta_m \end{array}\!\! \right) ,
\eeq
where (with a $-$ sign for neutrinos and a $+$ sign for antineutrinos)
\bea
  && E^m_2 - E^m_1\, =\, \frac{\Delta m^2}{2E}\,
    \sqrt{\left( 1 \mp \frac{n_e}{n_{\rm res}} \right)^2 \cos^2 2 \theta + \sin^2 2 \theta}\ ,  \label{eq:Delta_Em}  \\
  && \sin 2 \theta_m\, =\, \frac{\sin 2 \theta}
    {\sqrt{\left( 1 \mp \frac{n_e}{n_{\rm res}} \right)^2 \cos^2 2 \theta + \sin^2 2 \theta}}\ ,  \label{eq:sin_theta_m}  \\
  && \cos 2 \theta_m\, =\, \frac{\left( 1 \mp \frac{n_e}{n_{\rm res}} \right) \cos 2 \theta}
    {\sqrt{\left( 1 \mp \frac{n_e}{n_{\rm res}} \right)^2 \cos^2 2 \theta + \sin^2 2 \theta}}\ ,  \label{eq:cos_theta_m}
\eea
in which we have introduced the {\it resonance density}
\be
  n_{\rm res}\, =\, \frac{\Delta m^2 \cos 2 \theta}{2 \sqrt{2}\, G_F E}\ .
\eeq
If $\Delta m^2 \cos 2 \theta > 0$ (resonance condition for neutrinos), the mixing angle
in matter $\theta_m$ is maximal when $n_e = n_{\rm res}$, irrespective of the (nonzero) value
of the vacuum mixing angle $\theta$~\cite{Mikheev:1986gs,Mikheev:1986wj}:
\be
  \sin^2 2 \theta_m\, =\, 1 \quad \mbox{for}\ n_e = n_{\rm res}  \qquad \qquad
    \mbox{(case $\Delta m^2 \cos 2 \theta > 0$)}\, .
\eeq
This is the well-known {\it MSW (Mikheev-Smirnov-Wolfenstein) resonance}. For antineutrinos,
the resonance condition is $\Delta m^2 \cos 2 \theta < 0$, and the resonance occurs
for $n_e = - n_{\rm res}$ (in this case, it is $- n_{\rm res}$ that is positive and can be
interpreted as a resonance density).

The physics of neutrino flavour transitions in matter depends on whether the matter density
is constant or not. We discuss both cases in turn below.

\subsubsection{Medium with constant matter density                %
\label{subsec:constant}}                                                             %

Let us first consider the case of a medium with constant matter density, $n_e(x) = n_e = const$.
In this case, the Hamiltonian remains constant during the propagation of the neutrinos;
hence the matter eigenstates $\left|\nu^m_i \right>$, the energy levels $E^m_i$ and the mixing
matrix $U_m$ do not depend on time.
Inserting Eq.~(\ref{eq:nu^m_i-nu_alpha}) into Eq.~(\ref{eq:Schroedinger_components}), one
then obtains $n$ decoupled evolution equations for the probability amplitudes $\nu^m_i(t)$:
\be
   i \frac{d}{dt}\, \nu^m_i(t)\, =\, E^m_i \nu^m_i(t)\, ,
\eeq
which are trivially solved by $\nu^m_i (t) = e^{-i E^m_i t}\, \nu^m_i(0)\, $. Using again
Eq.~(\ref{eq:nu^m_i-nu_alpha}), one arrives at the oscillation probability in matter:
\be
  P_m (\nu_\alpha \to \nu_\beta)\, =\, \left| \nu_\beta (t) \right|^2\, =\, \left| \sum_i (U_m)_{\beta i}\, \nu^m_i (t) \right|^2\,
    =\, \left| \sum_i (U_m)_{\beta i} (U_m)^*_{\alpha i}\, e^{-i E^m_i t} \right|^2\, .
\label{eq:Pm_alpha_beta}
\eeq
Thus neutrino oscillations in a medium of constant density are governed by the same formula
as vacuum oscillations, with the oscillation parameters in vacuum replaced by the
oscillation parameters in matter. More specifically, in the two-flavour case~\cite{Wolfenstein:1977ue}:
\be
  P_m (\nu_\alpha \to \nu_\beta)\, =\, \sin^2 2 \theta_m \sin^2 \frac{(E^m_2 - E^m_1) t}{2}
    \qquad \qquad (\alpha \neq \beta)\ ,
\eeq
from which one can define an oscillation length in matter:
\be
  L^m_{\rm osc.}\, =\, \frac{2 \pi}{|E^m_2 - E^m_1|}\ .
\eeq
The oscillation length is maximal at the resonance $n_e = n_{\rm res}$, where it is related
to the oscillation length in vacuum by
$L^m_{\rm osc.} = 4\pi E / (|\Delta m^2| \sin 2 \theta) = L_{\rm osc.} / \sin 2 \theta$.

Matter effects can have a spectacular impact on oscillations when the vacuum
mixing angle is small (this is the case only for $\theta_{13}$ in practice).
Assuming non-maximal mixing in vacuum,
one can identify three noticeable regimes:
\begin{itemize}
\item[{\it (i)}] low density ($n_e \ll |n_{\rm res}|$):
$\sin^2 2 \theta_m \simeq \sin^2 2 \theta \left( 1 \pm \frac{2 n_e}{n_{\rm res}}\, \cos^2 2 \theta \right)$,
where the $+$ sign is for neutrinos, and the $-$ sign for antineutrinos.
Vacuum oscillations dominate, with subleading matter effects;
\item[{\it (ii)}] close to the resonance ($n_e \simeq |n_{\rm res}|$), oscillations are enhanced
with respect to vacuum oscillations ($\sin^2 2 \theta_m \simeq 1$) if the resonance condition
is satisfied, while they are suppressed if it is not satisfied
($\sin^2 2 \theta_m \simeq \tan^2 2 \theta / (4 + \tan^2 2 \theta)$);
\item[{\it (iii)}] high density ($n_e \gg |n_{\rm res}|$):
$\sin^2 2 \theta_m \simeq \tan^2 2 \theta / (\frac{n_e}{n_{\rm res}})^2$.
Oscillations are suppressed by matter effects.
\end{itemize}

Due to their different interactions with matter, neutrinos and antineutrinos oscillate with unequal
probabilities when they travel through a medium, even in the absence of CP violation.
This effect
is maximal at the resonance, where depending on the sign of $\Delta m^2 \cos 2 \theta$
either neutrino or antineutrino oscillations are resonantly enhanced.
This property is used by experiments aiming at determining the mass hierarchy, i.e. whether
$\Delta m^2_{31} > 0$ or $\Delta m^2_{31} < 0$. In the first case (normal mass ordering),
neutrino oscillations are enhanced over antineutrino oscillations,
while the opposite is true in the second case (inverted mass ordering).

Earth matter effects must be taken into account in the analysis of solar and atmospheric
neutrino data\footnote{While the electron density is not constant in the Earth, it is a good
approximation to consider it as made of layers of constant density (the crust, the mantle,
the outer core and the inner core). For the study of neutrino oscillations in the Earth,
the two-layer (mantle-core) approximation is often sufficient~\cite{pdg}.}.
Indeed, at night, neutrinos emitted by the Sun travel through the Earth
before reaching the detector, leading to the so-called {\it Earth regeneration effect}~\cite{Smirnov:1986ij}:
part of the solar neutrinos that have oscillated into muon or tau
neutrinos are converted back to electron neutrinos in the Earth. As a result, one can
observe a day-night asymmetry in the solar neutrino data (which is numerically small
in practice, see Section~\ref{sec:solar}). The origin of this effect
is simply that oscillation parameters in matter differ from oscillation parameters in vacuum.
In particular, high-energy solar neutrinos exit the Sun in the mass eigenstate
$\left| \nu_2 \right>$ (see Subsection~\ref{subsec:varying}), which is a propagation eigenstate
in vacuum, but not in matter;
hence they remain in the same state on their way from the Sun to the Earth,
but they oscillate when they travel through the Earth (see the review~\cite{pdg} for details).
Upward-going atmospheric neutrinos are also subject to Earth matter effects,
which modify their zenith angle distribution with respect to the case of vacuum
oscillations. This property can be exploited to determine the mass hierarchy
with neutrino telescopes (see Section~\ref{sec:future}).

\subsubsection{Medium with varying matter density                              %
\label{subsec:varying}}                                                                           %

When the density of the medium varies along the neutrino trajectory, $n_e(x) \neq const$,
the Hamiltonian governing the evolution of the system becomes time-dependent. As a result,
the matter eigenstates, energy levels and mixing angles in matter all depend on time,
and are called {\it instantaneous} quantities. 
The relations~(\ref{eq:H_diagonalization}) and~(\ref{eq:nu^m_i-nu_alpha}) are still valid,
but with instantaneous energy levels $E^m_i(t)$, mixing matrix $U_m(t)$ and matter
eigenstates $\left|\nu^m_i(t) \right>$. It follows that the evolution equations for the probability
amplitudes $\nu^m_i(t) = \left< \nu^m_i(t) | \nu(t) \right>$ are now coupled:
\be
  i \frac{d}{dt}\, \nu^m_i(t)\, =\, E^m_i(t) \nu^m_i(t)
    - i \sum_\gamma\, (U^*_m)_{\gamma i}(t) (\dot{U}_m)_{\gamma j}(t)\, \nu^m_j(t)\, ,
\eeq
where a dot on a quantity means derivative with respect to time, e.g. $\dot{U}_m(t) \equiv \frac{d}{dt}\, U_m(t)$.
In the two-flavour case, these equations reduce to
\be
  i \frac{d}{dt} \left(\!\! \begin{array}{c} \nu^m_1(t) \\ \nu^m_2(t) \end{array}\!\! \right)
    = \left(\!\! \begin{array}{cc} E^m_1(t) & -i \dot{\theta}_m(t) \\ i \dot{\theta}_m(t) & E^m_2(t) \end{array}\!\! \right)
    \left(\!\! \begin{array}{c} \nu^m_1(t) \\ \nu^m_2(t) \end{array}\!\! \right) .
\eeq
The terms proportional to $\dot{\theta}_m(t)$
in the evolution equations induce transitions between
the matter eigenstates $\left|\nu^m_1(t) \right>$ and $\left|\nu^m_2(t) \right>$, which are therefore
no longer propagation eigenstates: a neutrino has a certain probability to ``jump'' from one matter
eigenstate to another while travelling through a medium
of varying density~\cite{Smirnov:1986ij,Messiah:1986fc,Parke:1986jy}.
These transitions are more likely to occur when the neutrinos cross a resonance layer,
where the splitting between the energy levels $E^m_1$ and $E^m_2$ is minimal.
However, it turns out that in most physical environments (in particular in the Sun), the variation
of the matter density is slow enough that these off-diagonal terms can be neglected,
i.e. $|\dot{\theta}_m(t)| \ll |E^m_2 - E^m_1|$.
Physically, this amounts to say that the oscillation length at the resonance
is (much) smaller than the spatial width of the resonance~\cite{Mikheev:1986gs}.
In this case, the evolution of the neutrino system is said to be {\it adiabatic}\footnote{One can define
an adiabaticity criterion by introducing the parameter
$\gamma(t)\, \equiv\, \frac{|E^m_2(t) - E^m_1(t)|}{2 |\dot \theta_m(t)|}\, $;
the evolution is adiabatic when $\gamma \gg 1$~\cite{Smirnov:1986ij,Messiah:1986fc}.
It is {\it non adiabatic} when $\gamma < 1$, in which case transitions between different
matter eigenstates are possible.
If the neutrinos pass through a resonance layer, the adiabaticity parameter should be
evaluated at the resonance point:
$\gamma\, \equiv \left| \frac{E^m_2 - E^m_1}{2 \dot \theta_m} \right|_{\rm res}
    =\, \left| \frac{\Delta m^2 \sin^2 2 \theta}{2 E \cos 2 \theta} \right|
    \left| \frac{1}{n_{\rm res}} \left( \frac{d n}{d x} \right)_{\rm res} \right|^{-1}$.
Given the electron density profile
in the Sun and the values of the oscillation parameters $\Delta m^2_{21}$ and $\theta_{12}$,
this criterion is satisfied for all solar neutrino energies.}: a neutrino produced
in a given instantaneous matter eigenstate will follow the change in matter density during its
propagation and remain in the same matter eigenstate. However,
its flavour composition will evolve, since the instantaneous matter eigenstates change
as the matter density varies along the neutrino trajectory. This mechanism of neutrino flavour
transition is qualitatively different from oscillations in a medium with constant matter density,
and is called {\it adiabatic flavour conversion} or {\it MSW effect}~\cite{Mikheev:1986gs,Mikheev:1986wj}.

A particular situation, realized in dense astrophysical environments like supernovae or the Sun,
is the one of {\it level-crossing}. It arises when the electron density in the neutrino production
region (e.g. the center of the Sun) is much larger than the resonance density\footnote{The
terminology ``level-crossing'' is justified by the fact that the asymptotes (corresponding to
the limit $n \to \infty$) of the energy levels cross at the resonance.}, so that the neutrinos
cross the resonance layer during their propagation. When this is the case,
matter eigenstates approximately coincide with flavour eigenstates at production,
as can be seen by taking the limit $n_e / n_{\rm res} \to \infty$ in Eq.~(\ref{eq:sin_theta_m})
and~(\ref{eq:cos_theta_m}).
For instance, in the center of the Sun, one has $\left|\nu_e\right> \simeq \left|\nu^m_2 (r=0) \right>$
and $\left|\nu_\beta\right> \simeq - \left|\nu^m_1 (r=0) \right>$ (where $\left| \nu_\beta \right>$
describes a neutrino of either muon or tau flavour)
for neutrino energies
$E \gg 2\, \mbox{MeV}$, for which the condition $n_e(r=0) \gg n_{\rm res}$ is satisfied.
Hence high-energy solar neutrinos are produced as quasi pure eigenstates of propagation in matter,
and since their evolution is adiabatic, they remain in the instantaneous matter eigenstate
$\left|\nu^m_2 (r) \right>$ during their propagation. Eventually, they exit the Sun in the mass
eigenstate $\left|\nu_2 \right>$, since by continuity $\left|\nu^m_2 (r) \right> = \left|\nu_2 \right>$
at the radial coordinate $r$ where the electron density vanishes.
In other words, their flavour composition has changed as a result of the evolution
of the instantaneous matter eigenstates: produced as pure electron neutrinos,
they exit the Sun as mass eigenstates, i.e. as admixtures of all neutrino flavours.
Finally, mass eigenstates being eigenstates of propagation in vacuum, they reach
the Earth in the state $\left|\nu_2 \right>$, leading to a survival probability
$P_{ee} = | \langle \nu_e | \nu_2 \rangle |^2 \simeq \sin^2 \theta_{12}$~\cite{Mikheev:1986gs,Mikheev:1986wj}.
This formula describes reasonably well the behaviour of the high-energy part
of the $^8$B neutrino spectrum, which has been measured by the Super-Kamiokande
and SNO experiments. 
Low-energy solar neutrinos ($pp$ neutrinos), on the contrary, do not undergo
adiabatic flavour conversions, but their evolution is dominated by vacuum oscillations.
Indeed, they are characterized by a higher resonance density for which $n_e(r=0) \ll n_{\rm res}$,
so that matter effects can be neglected to a good approximation, leading to a survival
probability $P_{ee} \simeq 1 - \frac{1}{2}\, \sin^2 2 \theta_{12}$.

A more general formula for the electron neutrino survival probability on Earth,
valid for all solar neutrino energies, is given by~\cite{Blennow:2013rca}
\be
  P_{ee}\, =\, \sin^2 \theta_{12} + \cos^2 \theta^0_{12, m} \cos 2 \theta_{12}\, ,
\label{eq:Pee_general}
\eeq
in which $\theta^0_{12, m}$ is the $1$-$2$ mixing angle at the center of the Sun,
whose value is given by Eq.~(\ref{eq:cos_theta_m}) with $n_e = n_e (r=0)$. The two
limiting cases discussed above are easily recovered from Eq.~(\ref{eq:Pee_general}).
For low-energy solar neutrinos satisfying $n_{\rm res} \gg n_e(r=0)$, one has
$\cos \theta^0_{12, m} \simeq \cos \theta_{12}$ from Eq.~(\ref{eq:cos_theta_m}), implying 
$P_{ee} \simeq \sin^2 \theta_{12} + \cos^2 \theta_{12} \cos 2 \theta_{12} = 1 - \frac{1}{2}\, \sin^2 2 \theta_{12}$,
while for high-energy solar neutrinos satisfying $n_{\rm res} \ll n_e(r=0)$, one has instead
$\cos \theta^0_{12, m} \simeq 0$, from which $P_{ee} \simeq \sin^2 \theta_{12}$.
For intermediate solar neutrino energies, $0 < \cos \theta^0_{12, m} < \cos \theta_{12}$
and the formula~(\ref{eq:Pee_general}) interpolates between
$P_{ee} = 1 - \frac{1}{2}\, \sin^2 2 \theta_{12}$ and $P_{ee} = \sin^2 \theta_{12}$.
More precisely, $\cos \theta^0_{12, m}$ decreases monotonically when the neutrino energy
increases, and (using the experimental information that $\cos 2 \theta_{12} > 0$)
so does the survival probability. Thus, for the measured value of the 1-2 mixing angle,
adiabatic flavour conversions in the Sun are less efficient for intermediate-energy
solar neutrinos than for high-energy ones. This can be understood by noting that
the latter exit the Sun in the mass eigenstate $\left|\nu_2 \right>$, while the former
also have a $\left|\nu_1 \right>$ component, which has a larger probability
(namely $\cos^2 \theta_{12}$) of being detected as an electron neutrino.

\subsubsection{Three-flavour oscillations with matter effects       %
\label{subsec:matter_3f}}                                                              %

For long-baseline oscillation experiments in which matter effects cannot be neglected,
the vacuum formula~(\ref{eq:numu_nue_3f}) for the $\nu_\mu \to \nu_e$ oscillation probability
must be replaced by~\cite{Cervera:2000kp,Freund:2001pn}
\bea
  P (\nu_\mu \to \nu_e) &\! \simeq &\!
    \sin^2 \theta_{23}\, \frac{\sin^2 2 \theta_{13}}{(A-1)^2}\, \sin^2 \left[ (A-1) \Delta_{31} \right]
    +\, \alpha^2 \cos^2 \theta_{23}\, \frac{\sin^2 2 \theta_{12}}{A^2}\, \sin^2 (A \Delta_{31})  \nn \\
  &\! &\! +\ \alpha\, \frac{\cos \theta_{13} \sin 2 \theta_{12} \sin 2 \theta_{13} \sin 2 \theta_{23} \cos \delta_{CP}}{A (1-A)}\,
    \cos \Delta_{31} \sin (A \Delta_{31}) \sin \left[ (1-A) \Delta_{31} \right]  \nn \\
  &\! &\! -\ \alpha\, \frac{\cos \theta_{13} \sin 2 \theta_{12} \sin 2 \theta_{13} \sin 2 \theta_{23} \sin \delta_{\rm CP}}{A (1-A)}\,
    \sin \Delta_{31} \sin (A \Delta_{31}) \sin \left[ (1-A) \Delta_{31} \right]\, ,  \hskip 1cm
\label{eq:numu_nue_3f_matter}
\eea
where $\alpha \equiv \Delta m^2_{21} / \Delta m^2_{31}$, $\Delta_{31} \equiv \Delta m^2_{31} L / 4 E$
and the matter effects are encoded in the parameter
$A \equiv 2 V E / \Delta m^2_{31} = 2 \sqrt 2\, G_F n_e E/ \Delta m^2_{31}$.
The corresponding probability for $\bar \nu_\mu \to \bar \nu_e$ oscillations can be obtained
by switching the signs of the CP-violating phase $\delta_{\rm CP}$ and of the matter parameter $A$.
Like the vacuum formula~(\ref{eq:numu_nue_3f}), Eq.~(\ref{eq:numu_nue_3f_matter})
is an expansion to second order in the small quantities $\alpha$ and $\sin \theta_{13}$.
It assumes in addition a constant electron density $n_e$, which in practice is a good
approximation up to very long baselines, since neutrinos travel only through the mantle
for $L \lesssim 11000\, \mbox{km}$. However, it gives less good results for baselines
$L \gtrsim 11000\, \mbox{km}\, (E / 1\, \mbox{GeV}) (7.5 \times 10^{-5}\, \mbox{eV}^2 / \Delta m^2_{21})$,
for which $\Delta m^2_{21}$-driven oscillations cannot be linearized~\cite{Freund:2001pn}.
Furthermore, its validity is restricted to beam energies
$E  \gtrsim 0.34\, \mbox{GeV}\, (\Delta m^2_{21} / 7.5 \times 10^{-5}\, \mbox{eV}^2)
(2.8\, \mbox{g.cm}^{-3} / \rho)$, where $\rho$ is the Earth matter density in $\mbox{g.cm}^{-3}$,
because for smaller energies also the ``solar'' resonance (i.e. the matter effects
associated with the squared-mass difference $\Delta m^2_{21}$) should be taken
into account~\cite{Freund:2001pn}.

The terms in Eq.~(\ref{eq:numu_nue_3f_matter}) are in one-to-one correspondence
with the ones in the vacuum formula~(\ref{eq:numu_nue_3f}):
the first term corresponds to the dominant, $\Delta m^2_{31}$-driven oscillations;
the second one to the $\Delta m^2_{21}$-driven oscillations, as the presence of
the suppression factor $\alpha^2$ shows; the third and fourth ones
are the CP-even and the CP-odd terms, proportional to
$J \cot \delta_{\rm CP}$ and to $J$, respectively, where
$J = \cos \theta_{13} \sin 2 \theta_{12} \sin 2 \theta_{13} \sin 2 \theta_{23} \sin \delta_{\rm CP} / 8$
is the Jarlskog invariant.

The parameter $A$ can be written $A = n_e \cos 2 \theta_{13} / n_{\rm res}$, where
$n_{\rm res} = \Delta m^2_{31} \cos 2 \theta_{13} / (2 \sqrt{2}\, G_F E)$
is the ``atmospheric'' resonance density.
It therefore quantifies the importance of matter effects
on $\Delta m^2_{31}$-driven oscillations: the smaller $A$, the smaller
the matter effects (since $A \ll 1$ corresponds to $n_e \ll n_{\rm res}$).
One can check that, in the limit of negligible matter effects (i.e. $A \to 0$),
Eq.~(\ref{eq:numu_nue_3f_matter}) reduces to the vacuum formula~(\ref{eq:numu_nue_3f}).
As mentioned in Section~\ref{subsec:CPV}, the experimental challenge will be to disentangle
matter effects from CP violation, since both of them create an asymmetry between
$\nu_\mu \to \nu_e$ and $\bar \nu_\mu \to \bar \nu_e$ oscillations, depending on the
sign of $\Delta m^2_{31}$ for matter effects, and on the value of the ``Dirac'' phase
$\delta_{\rm CP}$ for CP violation.

\subsection{Sterile neutrinos         %
\label{subsec:steriles}}                  %

The number of light {\it active} neutrinos (i.e. of neutrinos with mass below $M_Z / 2$
and the same gauge interactions as the known left-handed neutrinos) is constrained
by the LEP measurement of the invisible decay width of the $Z$ boson~\cite{ALEPH:2005ab}:
\be
  N_\nu\, \equiv\, \frac{\Gamma^{\rm invisible}_Z}{\Gamma(Z \to \nu \bar \nu)_{\rm SM}}\,
    =\, 2.9840 \pm 0.0082\, ,
\eeq
where $\Gamma(Z \to \nu \bar \nu)_{\rm SM}$ is the partial decay width into a single
neutrino species computed in the Standard Model.
This rules out the possibility of a fourth light active neutrino beyond the three ones
already present in the Standard Model. It is not excluded, however, that additional neutrinos
without electroweak interactions may exist. Such neutrinos, called {\it sterile}, would interact
only through their mixing with the active neutrinos $\nu_{e L}$, $\nu_{\mu L}$, $\nu_{\tau L}$
and (provided that they are light enough) would affect their oscillations.
For this reason, sterile neutrinos with masses around 1 eV have been invoked
as a possible explanation of experimental anomalies that cannot be accounted for
in the framework of three-flavour oscillations (see Section~\ref{sec:anomalies}).
However, cosmological data, which constrains the number of effectively massless degrees
of freedom to be $N_{\rm eff} = 3.15 \pm 0.23\ (68\%\ \mbox{C. L.})$~\cite{Ade:2015xua}
and sets an upper bound on the sum of neutrino masses, is at odds with the hypothesis
of a light sterile neutrino.
In the following, we concentrate on the case of light (eV-scale) sterile neutrinos
that is relevant to short-distance oscillations
(for reviews on the subject, see e.g. Refs~\cite{Abazajian:2012ys,Gariazzo:2015rra}).

\subsubsection{Vacuum oscillations in the presence of sterile neutrinos  %
\label{subsec:vacuum_sterile}}                                                                  %

In the presence of $n$ light sterile neutrinos, the PMNS matrix is extended to a
$(3+n) \times (3+n)$ unitary matrix relating the left-handed mass eigenstates
$\nu_{1L}$, $\nu_{2L}$, \dots $\nu_{(3+n)L}$ to the left-handed flavour eigenstates
$\nu_{e L}$, $\nu_{\mu L}$, $\nu_{\tau L}$, $\nu_{s_1 L}$, \dots $\nu_{s_n L}$.
In the case of a single sterile neutrino ($n=1$), one has:
\be
  \left(\!\! \begin{array}{c} \nu_e(x) \\ \nu_\mu(x) \\ \nu_\tau(x) \\ \nu_s(x) \end{array}\!\! \right)_{\!\!L}
  =\, U\, \left(\!\! \begin{array}{c} \nu_1(x) \\ \nu_2(x) \\ \nu_3(x) \\ \nu_4(x)  \end{array}\!\! \right)_{\!\!L}
  =\, \left(\! \begin{array}{cccc} U_{e1} & U_{e2} & U_{e3} & U_{e4} \\
    U_{\mu1} & U_{\mu2} & U_{\mu3} & U_{\mu4} \\ U_{\tau 1} & U_{\tau2} & U_{\tau3} & U_{\tau4} \\
    U_{s1} & U_{s2} & U_{s3} & U_{s4}  \end{array}\! \right)\!
  \left(\!\! \begin{array}{c} \nu_1(x) \\ \nu_2(x) \\ \nu_3(x) \\ \nu_4(x) \end{array}\!\! \right)_{\!\!L} .
\label{eq:flavour_mass_relation_sterile}
\eeq
The $(3+n)$-flavour oscillation probability is given by the same formula~(\ref{eq:oscillation_probability})
as in the three-flavour case, but with $\alpha = e, \mu, \tau, s_1 \dots s_n$ and $i = 1, 2, 3, 4 \dots (3+n)$.
In practice, since sterile neutrinos are not detectable, we are only interested in oscillations
of active neutrinos, which involve the truncated $3 \times (3+n)$ matrix
$\{ U_{\alpha i} \}_{\alpha = e, \mu, \tau;\, i=1 \dots (3+n)}$ instead of the full $(3+n) \times (3+n)$
PMNS matrix. This truncated matrix can be parametrized by $3(n+1)$ mixing angles,
$2n+1$ ``Dirac phases'' and, in the Majorana case, $n+2$ additional ``Majorana phases''.
In the case of a single sterile neutrino ($n=1$), the 3 additional mixing angles are denoted
by $\theta_{14}$, $\theta_{24}$ and $\theta_{34}$; alternatively, one may parametrize
the active-sterile mixing by the PMNS matrix entries $U_{e4}$, $U_{\mu4}$ and $U_{\tau4}$.
Writing the $4 \times 4$ PMNS matrix as
$U = U_{34} U_{24} U_{14} U_{23} U_{13} U_{12} P$~\cite{deGouvea:2008nm},
where $U_{ij}$ is a unitary rotation in the $(i,j)$ plane and P a diagonal matrix containing
the Majorana phases ($P = \mathbf{1}$ in the Dirac case), one has, omitting CP-violating phases:
\be
  U_{e4} = s_{14}\, , \quad U_{\mu4} = c_{14} s_{24}\, , \quad U_{\tau4} = c_{14} c_{24} s_{34}\, ,
\eeq
in which $c_{ij} \equiv \cos \theta_{ij}$ and $s_{ij} \equiv \sin \theta_{ij}$.

For definiteness, we now focus on the case of a fourth neutrino with a mass $m_4 \gg m_{1,2,3}$
well separated from the other mass eigenstates, leading to
\be
  \Delta m^2_{\rm SBL} \equiv \Delta m^2_{41} \simeq \Delta m^2_{42} \simeq \Delta m^2_{43}\
  \gg\ |\Delta m^2_{31}|,\, |\Delta m^2_{32}|,\, \Delta m^2_{21}\, .
\eeq
This split spectrum is relevant to the anomalies reported by some short-baseline experiments,
which could be explained by oscillations governed by a squared-mass difference
$\Delta m^2_{\rm SBL} \sim 1\, \mbox{eV}^2$. The ``3+1'' spectrum
$m_4 \simeq \sqrt{\Delta m^2_{\rm SBL}} \gg m_{1,2,3}$ is favoured over
the ``1+3'' spectrum $m_{1,2,3} \simeq \sqrt{\Delta m^2_{\rm SBL}} \gg m_4$
by cosmological considerations, because the sum of neutrino masses would be
much larger in the latter case.
If one considers short-baseline oscillations for which
$\Delta m^2_{\rm atm} L / E \ll \Delta m^2_{SBL} L / E \lesssim 1$, it is
a reasonable approximation to set $\Delta m^2_{31} = \Delta m^2_{21} = 0$
in the (3+1)-flavour oscillation formula (which implies
$\Delta m^2_{43} = \Delta m^2_{42} = \Delta m^2_{41} \equiv \Delta m^2_{SBL}$). One arrives at
\be
  P (\nu_\alpha \to \nu_\alpha)\, =\, 1 - \sin^2 2 \theta^{\rm SBL}_{\alpha \alpha}\,
    \sin^2 \left( \frac{\Delta m^2_{41} L}{4E} \right) ,  \qquad  \sin^2 2 \theta^{\rm SBL}_{\alpha \alpha}\,
    \equiv\, 4 \left| U_{\alpha 4} \right|^2 \left( 1 - \left| U_{\alpha 4} \right|^2 \right) ,
\label{eq:SBL_disappearance}
\eeq
for disappearance probabilities, and
\be
  P (\nu_\alpha \to \nu_\beta)\, =\, \sin^2 2 \theta^{\rm SBL}_{\alpha \beta}\,
    \sin^2 \left( \frac{\Delta m^2_{41} L}{4E} \right) ,  \qquad
    \sin^2 2 \theta^{\rm SBL}_{\alpha \beta}\, \equiv\, 4 \left| U_{\alpha 4} U_{\beta 4} \right|^2 \qquad (\beta \neq \alpha)\, ,
\label{eq:SBL_appearance}
\eeq
for appearance probabilities.
Since these formulae describe effective two-flavour oscillations, they do not involve
any CP violation, i.e. the short-baseline appearance probability~(\ref{eq:SBL_appearance})
is the same for neutrinos and antineutrinos (disappearance probabilities are always CP conserving).
For short-baseline $\bar \nu_e$ disappearance at nuclear reactors (resp. $\nu_\mu / \bar \nu_\mu$ disappearance at accelerators),
one has:
\bea
  \sin^2 2 \theta^{\rm SBL}_{ee} &\!\! = &\!\! 4 \left| U_{e 4} \right|^2 \left( 1 - \left| U_{e 4} \right|^2 \right)
    =\, \sin^2 2 \theta_{14}\, ,  \label{eq:theta_ee_SBL}  \\
  \sin^2 2 \theta^{\rm SBL}_{\mu\mu} &\!\! = &\!\! 4 \left| U_{\mu 4} \right|^2 \left( 1 - \left| U_{\mu 4} \right|^2 \right)
    =\, \cos^2 \theta_{14} \sin^2 2 \theta_{24} + \sin^4 \theta_{24} \sin^2 2 \theta_{14}\, ,
  \label{eq:theta_mumu_SBL}
\eea
while for $\nu_e (\bar \nu_e)$ appearance in a $\nu_\mu (\bar \nu_\mu)$ beam:
\be
  \sin^2 2 \theta^{\rm SBL}_{\mu e}\, =\, 4 \left| U_{\mu 4} U_{e 4} \right|^2
    =\, \sin^2 \theta_{24} \sin^2 2 \theta_{14}\, .
\eeq
Since an order one mixing between the sterile neutrino and the active ones would
also affect long-baseline oscillations (including the oscillations of solar and atmospheric
neutrinos), the mixing angles $\theta_{14}$ and $\theta_{24}$ (or equivalently the
PMNS matrix entries $U_{e 4}$ and $U_{\mu 4}$) cannot be large. Approximating
$1 - |U_{e 4}|^2 \simeq 1$ and $1 - |U_{\mu 4}|^2 \simeq 1$ in Eqs.~(\ref{eq:theta_ee_SBL})
and~(\ref{eq:theta_mumu_SBL}), one obtains the relation
\be
  \sin^2 2 \theta^{\rm SBL}_{\mu e}\, \simeq\, \frac{1}{4} \sin^2 2 \theta^{\rm SBL}_{ee}
    \sin^2 2 \theta^{\rm SBL}_{\mu\mu}\, ,
\eeq
which, given experimental constraints on short-baseline $\nu_e$ and $\nu_\mu$
disappearance, provides an upper limit on short-baseline $\nu_\mu \to \nu_e$ oscillations.

Finally, let us briefly discuss the case of two additional sterile neutrinos.
We consider the ``3+2'' spectrum, in which the two heaviest mass eigenstates, which
are mostly sterile neutrinos, are split from the other three: $m_5 > m_4 \gg m_{1,2,3}$.
In this case, short-baseline oscillations with $\Delta m^2_{\rm atm} L / E \ll \Delta m^2_{41} L / E
< \Delta m^2_{51} L / E \lesssim 1$ depend on two independent squared-mass differences,
$\Delta m^2_{41}$ and $\Delta m^2_{51}$ (while $\Delta m^2_{54} = \Delta m^2_{51} - \Delta m^2_{41}$).
The short-baseline disappearance probabilities are given by~\cite{Karagiorgi:2006jf}
\bea
  P (\nu_\alpha \to \nu_\alpha) &\!\! = &\!\! 1
    - 4 \left| U_{\alpha 4} \right|^2 \left( 1 - \left| U_{\alpha 4} \right|^2 - \left| U_{\alpha 5} \right|^2 \right)
    \sin^2 \left( \frac{\Delta m^2_{41} L}{4E} \right)  \nn \\
  &\!\! &\!\! -\, 4 \left| U_{\alpha 5} \right|^2 \left( 1 - \left| U_{\alpha 4} \right|^2 - \left| U_{\alpha 5} \right|^2 \right)
    \sin^2 \left( \frac{\Delta m^2_{51} L}{4E} \right)
    - 4 \left| U_{\alpha 4} U_{\alpha 5} \right|^2 \sin^2 \left( \frac{\Delta m^2_{54} L}{4E} \right) ,  \hskip 1cm
\label{eq:SBL_disappearance_2}
\eea
while for appearance:
\bea
  P (\nu_\alpha \to \nu_\beta) &\!\! = &\!\!
    4 \left| U_{\alpha 4} U_{\beta 4} \right|^2 \sin^2 \left( \frac{\Delta m^2_{41} L}{4E} \right)
    + 4 \left| U_{\alpha 5} U_{\beta 5} \right|^2 \sin^2 \left( \frac{\Delta m^2_{51} L}{4E} \right)  \nn \\
  &\!\! &\!\! +\, 8 \left| U_{\alpha 4} U_{\beta 4} U_{\alpha 5} U_{\beta 5} \right|\,
    \sin \left( \frac{\Delta m^2_{41} L}{4E} \right) \sin \left( \frac{\Delta m^2_{51} L}{4E} \right)
    \cos \left( \frac{\Delta m^2_{54} L}{4E} - \eta \right) ,  \hskip 1cm
\label{eq:SBL_appearance_2}
\eea
where $\eta \equiv \arg \left( U_{\alpha 4} U^*_{\beta 4} U^*_{\alpha 5} U_{\beta 5} \right)$.
For $\bar \nu_\alpha \to \bar \nu_\beta$ oscillations, the $-$ sign in front of the CP-violating
phase $\eta$ in Eq.~(\ref{eq:SBL_appearance_2}) should be replaced by a $+$ sign.
Thus CP violation in short-distance appearance experiments is possible in the presence
of two sterile neutrinos, at variance with the 3+1 case.
The price to pay is a larger number of parameters, and an increased tension with cosmology.

\subsubsection{Matter effects in the presence of sterile neutrinos   %
\label{subsec:matter_sterile}}                                                           %

Sterile neutrinos have an impact on matter effects, because they do not interact
with the fermions constituting the medium, at variance with the electron, muon and tau
neutrinos. This can lead to an important enhancement of matter effects, e.g.
in oscillations of atmospheric neutrinos passing through the Earth~\cite{Liu:1997yb}.

The formalism presented in Subsection~\ref{subsec:matter} can be straightforwardly
extended to the case of $3+n$ flavours, where the $n$ additional states correspond
to sterile neutrinos. The matter Hamiltonian $H_{\beta \gamma}$ can be written
in a form similar to Eq.~(\ref{eq:H_components}):
\be
  H_{\beta \gamma}\, =\, \frac{1}{2E}\, \sum_i U_{\beta i} U^*_{\gamma i} m^2_i + V_\beta\, \delta_{\beta \gamma}\, ,
\label{eq:H_components_sterile}
\eeq
in which the indices $\beta, \gamma$ run over $e, \mu, \tau, s_1 \dots s_n$, the index $i$ runs
over $1, 2, 3, 4 \dots (3+n)$, and the potential $V_\beta$ receives the following charged current (CC)
and neutral current (NC) contributions:
\be
  V_{{\rm CC}, \beta}\, =\, \left\{\!\! \begin{array}{cl}
    \sqrt{2}\, G_F n_e(x) & \quad \beta = e \\ 0 & \quad \beta = \mu, \tau, s_1 \dots s_n \end{array} \right. ,  \qquad
  V_{{\rm NC}, \beta}\, =\, \left\{\!\! \begin{array}{cl}
    - \frac{G_F}{\sqrt{2}}\, n_n(x) & \quad \beta = e, \mu, \tau \\ 0 & \quad \beta = s_1 \dots s_n \end{array} \right.\! .
\label{eq:V_CC_NC_sterile}
\eeq
The main difference with the case of active neutrinos only is that $V_{\rm NC}$
is not flavour universal, hence its contribution cannot be subtracted from the matter Hamiltonian.
This has an impact, in particular, on the oscillations of atmospheric muon neutrinos in the Earth,
if they mix with a sterile neutrino~\cite{Nunokawa:2003ep}.
Finally, in problems where only one active flavour $\alpha$
and one sterile flavour $s$ are relevant, the matter Hamiltonian in the flavour eigenstate basis
$\{\, \left| \nu_\alpha \right>, \left| \nu_s \right> \}$ is given by Eq.~(\ref{eq:H_2f})
with the replacement
\be
  \begin{array}{ll} n_e\ \to\ n_e - \frac{1}{2}\, n_n & \quad \mbox{if}\ \, \alpha = e\, ,  \\
    n_e\ \to\ - \frac{1}{2}\, n_n & \quad \mbox{if}\ \, \alpha = \mu, \tau\, ,  \end{array}
\eeq
while the oscillation parameters in matter are given by Eqs.~(\ref{eq:Delta_Em})--(\ref{eq:cos_theta_m})
with the same replacement.

\subsection{Non-standard neutrino interactions    %
\label{subsec:NSI}}                                              %

Physics beyond the Standard Model can lead to new, subleading effects in neutrino oscillations.
Leaving aside the case of light sterile neutrinos discussed in the previous section,
these effects can be parametrized by four-fermion effective operators leading to so-called
{\it non-standard neutrino interactions} (NSI).
We shall give only a brief overview of NSI in this
section, and refer the reader to reviews such as Refs~\cite{Ohlsson:2012kf,Miranda:2015dra}.

One can distinguish between two kinds of NSI, of the charged-current type (CC-NSI)
and of the neutral-current type (NC-NSI):
\be
  -\, 2 \sqrt{2}\, G_F\, \epsilon^{q q'}_{\alpha \beta}
    \left( \bar \nu_{\alpha L} \gamma^\mu \ell_{\beta L} \right) \left( \bar q' \gamma_\mu P q \right)
    \qquad  \mbox{(CC-NSI, quarks)}
\label{eq:CC-NSI_q}
\eeq
\be
    -\, 2 \sqrt{2}\, G_F\, \epsilon^{\gamma \delta}_{\alpha \beta}
    \left( \bar \nu_{\alpha L} \gamma^\mu \ell_{\gamma L} \right) \left( \bar \ell_{\delta L} \gamma_\mu \nu_{\beta L} \right)
    \qquad  \mbox{(CC-NSI, leptons)}
\label{eq:CC-NSI_l}
\eeq
\be
  -\, 2 \sqrt{2}\, G_F\, \epsilon^f_{\alpha \beta}
    \left( \bar \nu_{\alpha L} \gamma^\mu \nu_{\beta L} \right) \left( \bar f \gamma_\mu P f \right)
    \qquad  \mbox{(NC-NSI) \phantom{lepton}}
\label{eq:NC-NSI}
\eeq
where $\alpha, \beta, \gamma, \delta = e, \mu, \tau$ are flavour indices, $q$ and $q'$ are quarks,
$f$ is a charged lepton or a quark, $P$ is either the chirality projector $P_L = \frac{1 - \gamma^5}{2}$
or $P_R = \frac{1 + \gamma^5}{2}$, and the $\epsilon_{\alpha \beta}$'s are dimensionless coefficients,
conventionally expressed in units of the Fermi constant $G_F$. Note that $\gamma \neq \delta$
in the purely leptonic CC-NSI, otherwise the operator would fall in the class of NC-NSI.
Since $\alpha$ can be different from $\beta$, non-standard interactions violate neutrino
flavour in general. In practice, only the operators with $(q, q') = (u, d)$, $(\gamma, \delta) = (e, \mu)$
and $f = u, d, e$ can affect neutrino oscillations.
The CC-NSI modify the neutrino production and detection processes, and can therefore
induce flavour change even at very short distances. For instance, the operators
$(\bar \nu_{e L} \gamma^\mu \mu_L) (\bar d \gamma_\mu P u)$ and
$(\bar \nu_{e L} \gamma^\mu \mu_L) (\bar e_L \gamma_\mu \nu_{e L})$
induce the decays $\pi^+ \to \mu^+ \nu_e$ and $\mu^- \to e^- \bar \nu_e \nu_e$,
respectively, both of which mimic $\nu_\mu \to \nu_e$ oscillations;
in the detector, $(\bar \nu_{\mu L} \gamma^\mu e_L) (\bar d \gamma_\mu P u)$
induces the reaction $\bar \nu_\mu + p \to n + e^+$, which mimics
$\bar \nu_\mu \to \bar \nu_e$ oscillations. The NC-NSI, on the other hand, affect
neutrino propagation in matter~\cite{Wolfenstein:1977ue}. Indeed, the operators
$(\bar \nu_{\alpha L} \gamma^\mu \nu_{\beta L}) (\bar f \gamma_\mu P f)$ modify
the forward coherent scatterings of neutrinos on the electrons and nucleons
of the medium, resulting in new, flavour-diagonal
and off-diagonal contributions to the neutrino potential in matter,
in addition to the standard contribution from $W$ and $Z$ boson exchange.
The NC-NSI also have an impact on neutrino detection processes that do not involve
the production of a charged lepton. For instance, the operators
$(\bar \nu_{\alpha L} \gamma^\mu \nu_{\beta L}) (\bar e \gamma_\mu P e)$
modify the neutrino-electron elastic scattering cross-section, thus affecting the solar neutrino
rate measured by experiments like Super-Kamiokande.

There are numerous constraints on non-standard neutrino interactions. The CC-NSI
contribute to a variety of electroweak processes; upper bounds on the coefficients
$\epsilon^{ud}_{\alpha \beta}$ and $\epsilon^{\mu e}_{\alpha \beta}$ are typically
of order $10^{-2}$ at the $90\%$ C.L. (see e.g. Ref.~\cite{Biggio:2009nt}).
The NC-NSI are constrained both by neutrino scattering experiments and by oscillation data
(see e.g. Ref.~\cite{Gonzalez-Garcia:2013usa}); the upper bounds on the coefficients
$\epsilon^u_{\alpha \beta}$, $\epsilon^d_{\alpha \beta}$ and $\epsilon^e_{\alpha \beta}$
are somewhat weaker than the ones on CC-NSI, of order $0.1$ ($10^{-2}$ for some of them).
These bounds leave some room for observable effects of non-standard
neutrino interactions at future long-baseline oscillation experiments\footnote{The presence
of NSI may also lead to degeneracies affecting the determination of the neutrino oscillation
parameters, see e.g. Ref.~\cite{Forero:2016cmb} for the degeneracy between NC-NSI
and $\delta_{\rm CP}$, and Ref.~\cite{Gonzalez-Garcia:2013usa} for a global analysis
of neutrino oscillation data in the presence of NC-NSI.}~\cite{Miranda:2015dra}.
Given however that the operators~(\ref{eq:CC-NSI_q})--(\ref{eq:NC-NSI}) must arise from some
gauge-invariant extension of the Standard Model, they are likely to be accompanied
by operators such as $(\bar \ell_{\alpha L} \gamma^\mu \ell_{\beta L}) (\bar f \gamma_\mu P f)$,
with coefficients related to the $\epsilon^f_{\alpha \beta}$.
Among these operators, the ones that violate lepton flavour (i.e. with $\alpha \neq \beta$) are severely
constrained by the non-observation of processes
like $\mu \to e \gamma$ or $\mu \to e$ conversion in nuclei,
yielding strong indirect upper bounds on the corresponding NSI. New physics models in which the connection
between non-standard neutrino interactions and charged-lepton flavour violation
can be avoided have been found (see e.g. Refs.~\cite{Bilenky:1993bt,Antusch:2008tz}),
but they generally fail to generate large NSI coefficients.
A possible way out is provided by models with light mediators (see e.g. Ref.~\cite{Farzan:2016wym}).

\section{The 1-2 sector}
\label{sec:solar}

\subsection{The Sun interior and neutrino fluxes}

The Sun is a star of the main sequence in the stable hydrogen fusion regime. It produces its energy by nuclear fusion reactions in its core, described by the overall equation:
\begin{equation}
4p \rightarrow {}{^4}{\rm He} + 2 e^+ + 2 \nu_e
\label{eq:ppreac}
\end{equation}
After positron annihilation with electrons, the energy balance is equal to $E_s =26.73$ MeV per helium nucleus produced, including the mean energy of the neutrinos, amounting to approximately $E_\nu =$ 0.6 MeV. The total flux $\phi$ of solar neutrinos on Earth can be estimated from the solar constant (or solar irradiance) F = 1.36 kW/m$^2$  as
$\phi = \frac{2F}{E_s - E_\nu} = 6.5 \times 10^{10} {\rm cm^{-2} s^{-1}}$.

\begin{figure}[htbp]
\begin{minipage}[c]{.46\linewidth}
   	      \includegraphics[width=0.9\linewidth]{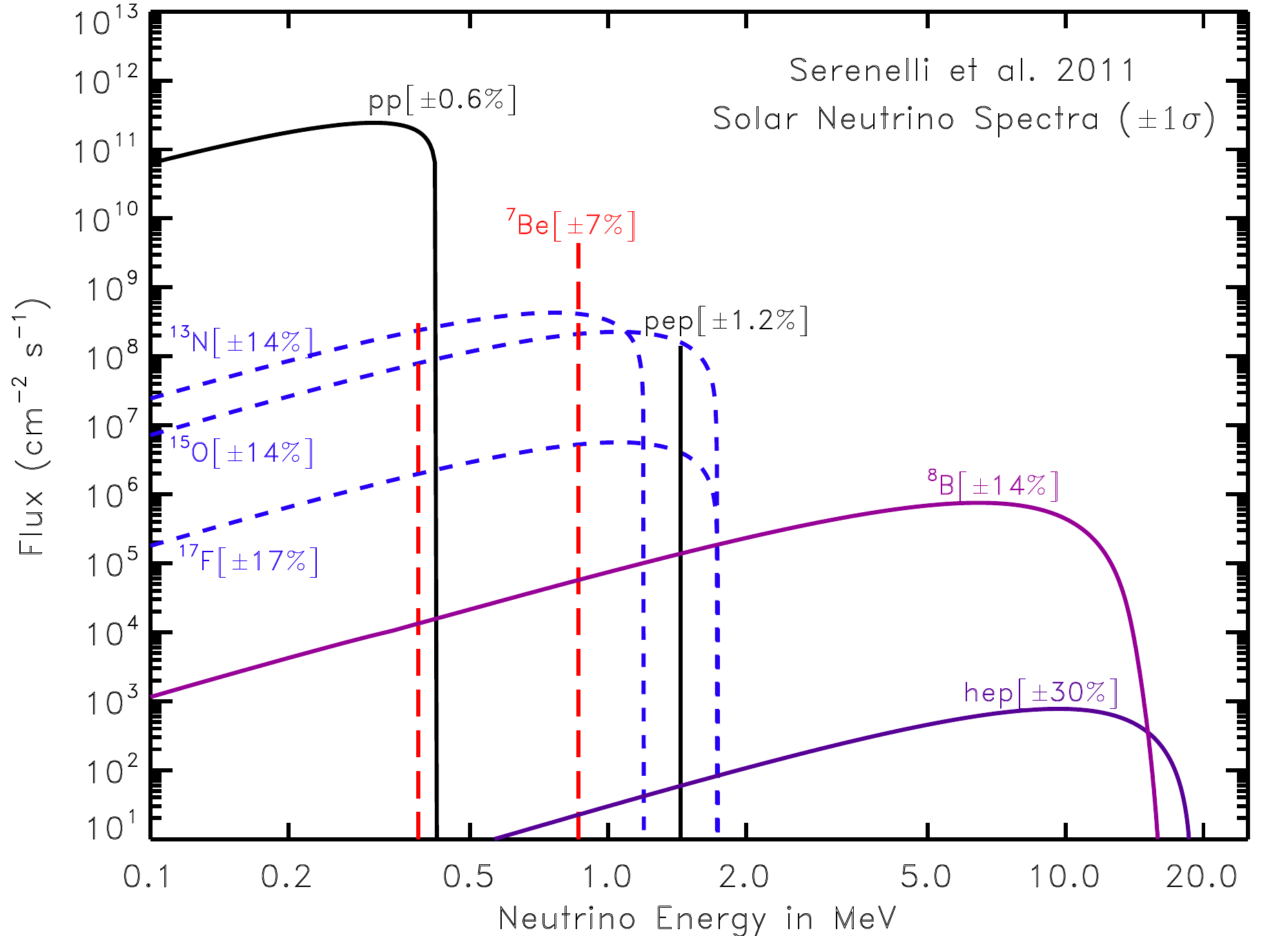}
   \end{minipage} \hfill
   \begin{minipage}{.46\linewidth}
      \includegraphics[width=0.9\linewidth]{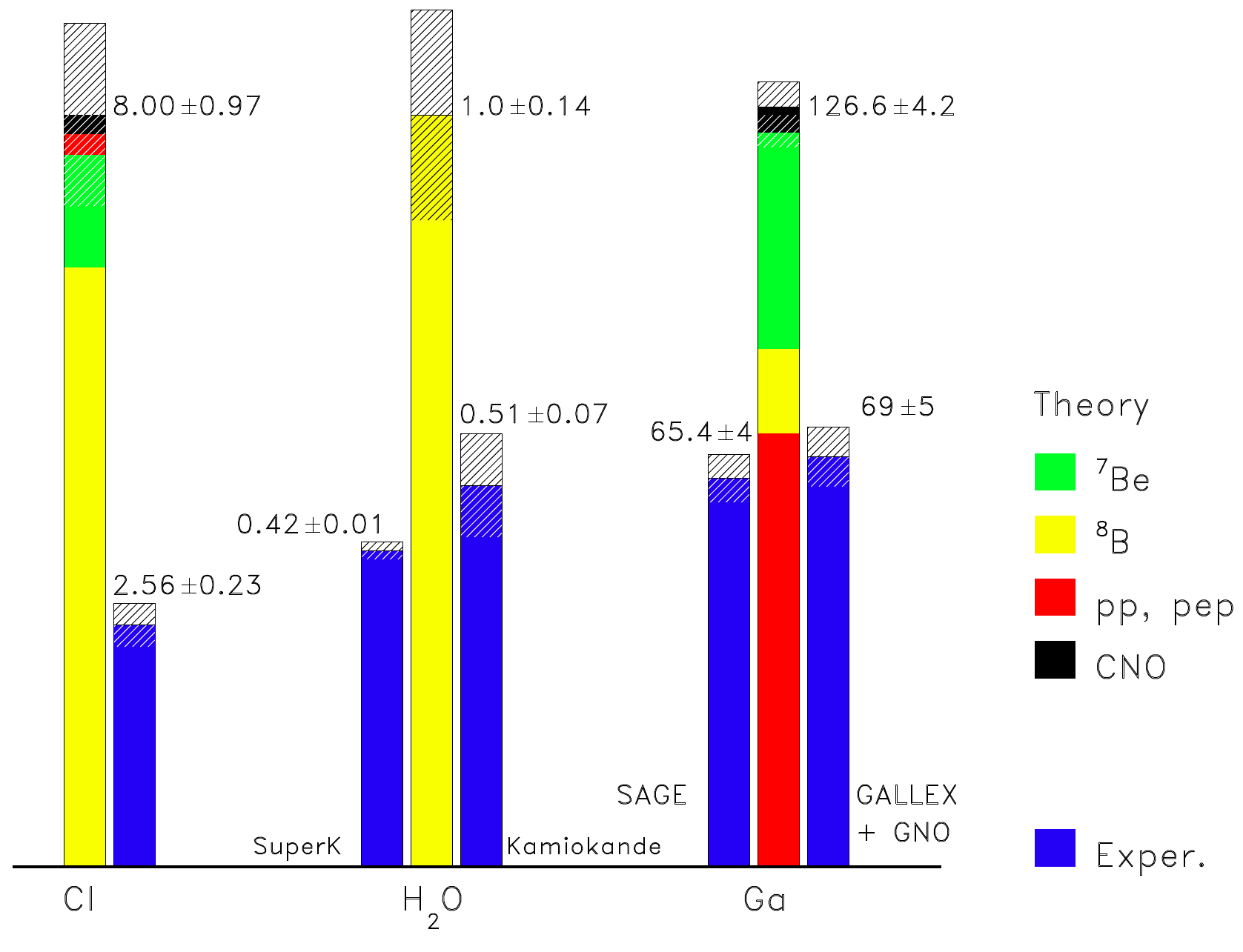}
   \end{minipage}
    \caption{
Left: solar neutrino spectra and SSM uncertainties \cite{serenelli}.
Right: measurements of the solar neutrino fluxes by various early experiments compared to the SSM prediction~\cite{serenelli}. Courtesy of Haxton {\it et al.}
Reproduced with permission from W.C. Haxton {\it et al.}, Ann. Rev. Astronom. Astrophys., 51, 21, 2013. Copyright by Annual Reviews.
}
\label{fig:sol-spectra}
\end{figure}

The resulting neutrino flux is due to several reaction chains contributing to the overall equation~(\ref{eq:ppreac}). The main reactions are shown in Table~\ref{tab:snuflux}, and the expected neutrino flux is shown in Fig.~\ref{fig:sol-spectra}. In addition to the $pp$ chain, the catalytic CNO cycle involving carbon, nitrogen and oxygen nuclei also takes place in the Sun. This cycle is dominant in more massive stars, while in the Sun it only contributes about 1\% of the total energy production~\cite{serenelli}. 
The neutrinos produced by a given reaction in the chain are usually referred to by their parent particles. For the following of this discussion, it is important to notice that neutrinos from different reactions span different energy ranges.
The vast majority of the neutrino flux is produced by the $pp$ neutrinos, which are very difficult to observe because of their end point energy of 0.42 MeV. 
The so-called $^{8}$B neutrinos are produced with energies up to 15 MeV and are easier to observe in experiments with high energy thresholds.
Other reactions produce monoenergetic neutrino lines, namely the two $^7$Be lines at 0.38 MeV ($10\%$)
and 0.86 MeV ($90\%$) from electron capture on $^7$Be nuclei, and the pep line at 1.44 MeV.
It must be noticed that these reactions have different dependences on the temperature and density and therefore the core regions where they take place differ. The $^8$B neutrinos are produced in the inner part of the core.

\begin{table}
\caption{Nuclear fusion reactions in the Sun and their neutrino fluxes (in units cm$^{-2}$ s$^{-1}$ times the factor in the fifth column) according to the SSM GS98-SFII and AGSS09-SFII~\cite{serenelli}, with relative uncertainties shown in the column labeled $\sigma$. The first five reactions are part of the $pp$ chain, the last three reactions are part of the CNO cycle.}
\centering
\begin{tabular}{|c|c|c|c|c|c|c|}
  \hline
  Reaction & $E_{MAX}$ (MeV) & GS98 & AGSS09& exp. & $\sigma$ (\%) &name \\ 
  \hline
$ p + p \rightarrow {^2}H + e ^+ + \nu_e$ & 0.42 & 5.98  & 6.03   & $10^{10}$ & 0.6 &pp \\
$ p + e^- + p \rightarrow {^2}H  + \nu_e$ & 1.44 & 1.44  &1.47& $10^{9}$  & 1.2 &pep \\
$ {^7}Be + e^- \rightarrow {^7}Li + \nu_e$ & 0.86 & 5.00 & 4.56& $10^{9}$  & 7 &$ {^7}$Be \\
$ {^8}B \rightarrow {^8}Be + e ^+ + \nu_e$ & $\simeq$ 15 & 5.58  &4.59  &  $10^{6}$  & 14 &$ {^8}$B\\
$  {^3}He + p \rightarrow {^4}He + e ^+ + \nu_e$ & 18.77 & 8.04  &8.31 &$10^{3}$  & 30 &hep \\
$ {^{13}}N \rightarrow {^{13}}C + e ^+ + \nu_e$ & 1.2 & 2.96 &2.17    &$10^{8}$  &14  &$ {^{13}}$N\\
$ {^{15}}O \rightarrow {^{15}}N + e ^+ + \nu_e$ & 1.73 & 2.23 &1.56    &$10^{8}$  & 15& $ {^{15}}$O\\
$ {^{17}}F \rightarrow {^{17}}O + e ^+ + \nu_e$ & 1.74 & 5.52  &3.40   &$10^{6}$  & 16 &$ {^{17}}$F\\
  \hline
\end{tabular}
\label{tab:snuflux}
\end{table}

A Standard Solar Model (SSM)~\cite{bahcall89} is a model describing the Sun's interior structure and the thermonuclear reactions therein taking into account hydrostatic equilibrium, energy production and transport through radiation and convection, opacity and chemical composition. The chemical composition is expressed as the initial abundance of hydrogen $X$, helium $Y$ and heavier elements $Z$, with $X+Y+Z = 1$.
The model also needs to comply with various boundary conditions, like the observed luminosity and radius of the Sun, as well as the chemical composition at its surface. From the point of view of neutrino physics, the main outcome of a SSM is
its prediction for the different components of the neutrino flux.


Several SSM have been developed since the pioneering calculations by J. Bahcall in the 1960s, with slightly different assumptions. For example, two representative SSM models~\cite{serenelli}, GS98-SFII and AGSS09-SFII, differ in their assumptions for the abundance of heavier elements, the so-called metallicity, which can be evaluated only indirectly from the solar atmosphere and from meteoritic abundances: $Z/X=0.023$ for GS98-SFII and $Z/X=0.018$ for AGSS09-SFII. These two models also predict slightly different fluxes of neutrinos, in particular the lower metallicity SSM AGSS09-SFII model predicts a 1\% cooler core, which results in a 10\% lower flux for $^8$B neutrinos and a smaller neutrino flux from the CNO cycle.  

These models have been confronted with results from helioseismology, the field that studies the sound waves propagating inside the Sun, giving informations on the density and pressure profiles.
These data show good agreement for GS98-SFII but some discrepancy for AGSS09-SFII~\cite{serenelli}. 
This problem is known as the solar abundance or metallicity problem.
New solar models have been proposed recently~\cite{new-ssm}.

\subsection{The early experiments}

The radiochemical detection of neutrinos using the reaction 
$\nu_e~^{37}Cl\rightarrow~^{37}Ar~e^-$ was first suggested by Bruno Pontecorvo in 1946~\cite{pontecorvo46}. This reaction has a threshold of 0.814 MeV and is therefore mainly sensitive to the $^8$B and $^7$Be solar neutrinos.

In 1965, after previous experiments with this technique on surface, Ray  Davis started  to construct a major experiment deep underground at Homestake (South Dakota), with 615~t of perchloroethylene (C$_2$Cl$_4$), which took data between 1968 and 2002~\cite{cleveland}.
The argon production rate induced by solar neutrino was 0.48 counts per day, over a background of 0.09 counts per day due to the interactions initiated by highly energetic muons with nuclei (so-called cosmogenics).
The $^{37}$Ar half-life of 35 days allows to expose the tank for a period of 60-70 days, followed by the efficient extraction of the produced argon by purging the liquid with helium.  
The isotope $^{37}$Ar decays by electron capture: 90 \% of these occur on the K-shell producing an average of 3.5 Auger electrons with 2.82 keV of total energy. Miniature gas proportional counters were developed to detect these decays with high efficiency and purity, achieving single-atom counting.
The efficiency of the whole extraction chain was calibrated injecting known quantities of other argon isotopes ($^{36}$Ar and $^{38}$Ar).  

The result of the Homestake chlorine experiment~\cite{cleveland} for the capture rate  is
\begin{equation}
(\sigma \phi)_{Cl} = 2.56 \pm 0.16 \pm 0.16 \: \mbox{SNU}
\end{equation}
where SNU stands for Solar Neutrino Units (1 SNU = 10$^{-36}$ captures/nucleus/second), while the prediction from the GS98-SFII solar model is $8.00 \pm 0.97$ SNU -- that is, the measured rate was about 30\% of the predicted rate.  
Starting from the first result in 1968, this major deficit triggered a thirty-year long debate, where many particle physicists were convinced that the SSM was not correct. Indeed, the production rate of $ {^8}$B neutrinos depends critically on the temperature of the Sun core as T$^{22}$~\cite{serenelli}, and helioseismological data were not yet available at that time to confirm the validity of the SSM. 

\begin{table}
\caption{Main characteristics of the different solar neutrino detectors.}
\centering
\begin{tabular}{|c|c|c|c|c|c|c|}
  \hline
  Detector & Ref & Composition & Active mass & Threshold (MeV)  \\ 
  \hline
Homestake & \text{\cite{cleveland}}& C$_2$Cl$_4$ & 615t &  0.814 \\
Kamiokande & \text{\cite{fukuda}}&H$_2$O & 3kt &  7  \\
SAGE & \text{\cite{abdurashitov}}&Ga & 57t &  0.233 \\
GALLEX/GNO & \text{\cite{hampel,altmann}}&Ga & 101t &  0.233  \\
Super-Kamiokande &  \text{\cite{abesk4}}&H$_2$O & 50kt &  3.5 \\
SNO & \text{\cite{aharmim}}&D$_2$O & 1kt &  3.5 \\
Borexino & \text{\cite{bellini}}& C$_9$H$_{12}$ & 278t &  0.2  \\
  \hline
\end{tabular}

\label{tab:snudet}
\end{table}

It took more than twenty years before other radiochemical experiments could probe solar neutrinos~(Table~\ref{tab:snudet}). The 
Soviet-American Gallium Experiment
(SAGE, 1989-2007)~\cite{abdurashitov} in the Baksan Laboratory (Russia) and the Gallium Experiment (Gallex, 1991-1997)~\cite{hampel} in the Gran Sasso Laboratory (Italy), which was followed by 
the Gallium Neutrino Observatory
(GNO, 1998-2003)~\cite{altmann}, used the reaction  $\nu_e  + {}^{71}{\rm Ga} \rightarrow {}^{71}{\rm Ge} + e^-$.
This reaction has a lower threshold of 0.233 MeV, and provides sensitivity to the $pp$ neutrinos, which constitute the majority of the flux. 

While the details of these experiments, related to the chemical properties of gallium, are different from Homestake, they follow the same overall scheme. They provided a confirmation of the solar neutrino deficit~\cite{abdurashitov,hampel,altmann,kaether}, although with a different reduction factor with respect the SSM prediction (approximately 50\%):
\begin{eqnarray}
(\sigma \phi)_{SAGE} & = & 65.4^{+3.1} _{-3.0} \; ^{+2.6} _{-2.8}\: \mbox{SNU} \\
(\sigma \phi)_{GALLEX} & = & 73.4  ^{+6.1}_{-6.0} \; ^{+3.7} _{-4.1}\: \mbox{SNU} \\
(\sigma \phi)_{GNO} & = & 62.9  ^{+5.5} _{-5.3} \; ^{+2.5} _{-2.5}\: \mbox{SNU} 
\end{eqnarray}
to be compared with a predicted flux of 126.6 $\pm$ 4.2 SNU (GS98) (Fig.~\ref{fig:sol-spectra}).

\subsection{Real time experiments: Kamiokande, Super-Kamiokande, SNO and Borexino}

In 1987-1995, the 2.14 kt water Cherenkov 
Kamioka Nucleon Decay Experiment (Kamiokande) in Japan, in its phases II and III, provided a different measurement of the solar neutrino flux. 

When a neutrino interaction takes place
in water, producing secondary charged particles with a speed 
exceeding that of light
in water, Cherenkov photons are radiated. These photons are emitted on a cone centered on the particle momentum, with an opening angle of 42 degrees for ultra-relativistic particles. They are subsequently detected by the photomultipliers instrumenting the walls of the detector. The image formed by the Cherenkov photons has the form of a ring for particles stopping inside the detector.

Neutrino elastic scattering on electrons, which is mainly sensitive to electron neutrinos, could be detected in this way with a threshold of 9 MeV, progressively reduced to 7 MeV.  
Kamiokande was the first experiment to detect solar neutrinos in real time. Since the scattered electron is emitted preferentially in the direction of the neutrino, their solar origin could be verified by correlating the electron momentum with the Sun direction. The combined result of Kamiokande~\cite{fukuda} is
\begin{equation}
\phi( ^8 B) = [2.80 \pm 0.19 ({\rm stat.}) \pm 0.33 ({\rm sys.})] \times 10^6 \: {\rm cm^{-2} s^{-1}}
\end{equation}
providing another indication of a large suppression of the flux predicted by the solar model.

Super-Kamiokande is a very large water Cherenkov detector located in the Mozumi mine (Japan), under an overburden of 1,000 m of rock, equivalent to 2,700 m of water. It is a stainless steel tank (41.4 m high, 39.3 m diameter) containing 50 kt of ultra-pure water. The detection volume is partitioned in an outer detector, composed of 1,885 8-inch Photomultiplier Tubes (PMT), and an inner detector with 11,146 20-inch PMTs. The fiducial volume is 22.5 kt. 

Super-Kamiokande, with a threshold of 7 MeV initially and 3.5 MeV today, started taking data in 1996 and confirmed the Kamiokande result with increased precision until its most recent update~\cite{abesk4}
\begin{equation}
\phi( ^8 B) = [2.308 \pm 0.020 ({\rm stat.}) ^{+0.040}_{-0.040} ({\rm sys.})] \times 10^6 \: {\rm cm^{-2} s^{-1}}.
\end{equation} 

Super-Kamiokande is still running at present and has recently observed for the first time a day-night effect for solar neutrinos~\cite{renshaw}. 
Let $r_D$ ($r_N$) be the day (night) event rate,
the day-night asymmetry $A_{DN}$ is defined as $A_{DN} = \frac{2 (r_D - r_N) }{r_D + r_N}$ and was measured to be
\begin{equation}
A_{DN} = (-3.2 \pm 1.1 ({\rm stat.}) \pm 0.5 ({\rm syst.}))\, \%,
\end{equation} 
deviating from zero by 2.7 $\sigma$.
This is similar to $K_S^0$ regeneration for a beam of $K_L^0$ passing through a material. It is the first indication of Earth matter effects on neutrino propagation.


The main limitations of radiochemical experiments is that they are only sensitive to charged current neutrino interactions, with a \nue producing an electron in the final state. Water Cherenkov experiments are sensitive to the elastic scattering of neutrinos off electrons, whose cross section is also mainly sensitive to the $\nu_e$ component of the flux. In this way, it is not possible to distinguish between a deficit in the \nue flux due to a wrong SSM prediction or a deficit due to neutrino oscillations.

The real breakthrough came from the Sudbury Neutrino Observatory (SNO, Canada), a Cherenkov detector located 2 km below ground and containing 1~kt of heavy water. 
With heavy water, three channels are available to probe the solar neutrino flux: 
\begin{itemize}
\item The Charged Current (CC) reaction for electron neutrinos $ \nu_e + d \rightarrow p + p + e^-$, where the produced electron carries off most of the neutrino energy, allowing to constrain the neutrino spectrum;
\item The Neutral Current (NC) reaction $ \nu_x + d \rightarrow \nu_x +  p + n $, which allows to count all neutrinos, independently of their flavour, above the threshold of 2.22 MeV; 
\item The Electron Scattering (ES) reaction $ \nu_x + e^- \rightarrow \nu_x + e^-$, also available in conventional water Cherenkov detectors, which is mainly sensitive to $ \nu_e$, as $\sigma_{ES}(\nu_e)\simeq 6 \; \sigma_{ES}(\nu_{\mu, \tau})$. 
\end{itemize}
SNO took data in three phases, which are distinguished by the technique used to detect the neutron produced in the neutral current dissociation of deuterium. Since the neutron is the only detectable particle in this reaction, this measurement is an experimental challenge and places severe requirements on the radiopurity of the whole apparatus. The successive neutron detection techniques improved the sensitivity to the NC reactions. In the first phase, neutron capture on deuterium was used. In the second phase, NaCl was dissolved in the water and neutrons were detected through the neutron capture on $^{35}$Cl followed by gamma emission. In the third phase, neutron detectors containing $^3$He were inserted in the detector.
The electron kinetic energy threshold was 5 MeV, making the experiment sensitive to the ${}^8$B neutrinos. 
The results of all phases of the detector were in agreement and yielded a measurement of the total flux of solar neutrinos~\cite{aharmim}, irrespective of their flavour,
\begin{equation}
\phi_{NC} (\nu \; active ) = [5.25 \pm 0.16 ({\rm stat.}) ^{+0.11}
_{-0.13} ({\rm syst.})] \times 10^6 \: {\rm cm^{-2} s^{-1}} 
\end{equation}
in agreement with the prediction of the solar models ($5.05 \times 10^6\, {\rm cm^{-2} s^{-1}}$), and the flux of $\nu_e$
\begin{equation}
\phi_{CC} (\nu_e ) =  (0.301 \pm 0.033) \; \phi_{NC} (\nu \; active ).
\end{equation}
As we will discuss in Section~\ref{subsec:solarinter}, this is clearly an indication of flavour conversion of solar neutrinos. 

Finally, another important experiment in the field of solar neutrino physics is the Borexino experiment, which is running since 2007. Borexino is a 278~t liquid scintillator detector installed in the Gran Sasso Laboratory (Italy). 
It is capable of measuring in real time low-energy neutrinos with a threshold of 200 keV, thanks to the high light yield of 10$^4$ photons per MeV, which is much higher than in a Cherenkov detector, and to the extreme levels of radiopurity reached inside the scintillator tank. 

Combining these two features, Borexino was able to measure the flux of neutrinos from different reactions. The original goal of the experiment was the measurement of the $^7$Be neutrino flux, which was found to be $(3.10 \pm 0.15) \times 10^9\: {\rm cm^{-2} s^{-1}}$, corresponding to 62\% of the GS98 SSM predicted flux~\cite{bellini}. It also provided the first determination of the pep flux: 
$(1.6 \pm 0.3) \times 10^8\: { \rm cm^{-2} s^{-1}}$, and a measurement of $^8$B neutrinos with a 3~MeV threshold. 
Finally, they recently published the first real time observation of $pp$ neutrinos,
as well as the most stringent upper limit on the CNO neutrino flux. This is interesting, since the flux of CNO neutrinos predicted by a given SSM is sensitive to the assumption made on metallicity.

\subsection{Confirmation with reactor neutrinos: KamLAND}

\begin{figure}[htbp]
\begin{minipage}[c]{.40\linewidth}
   	      \includegraphics[width=0.9\linewidth]{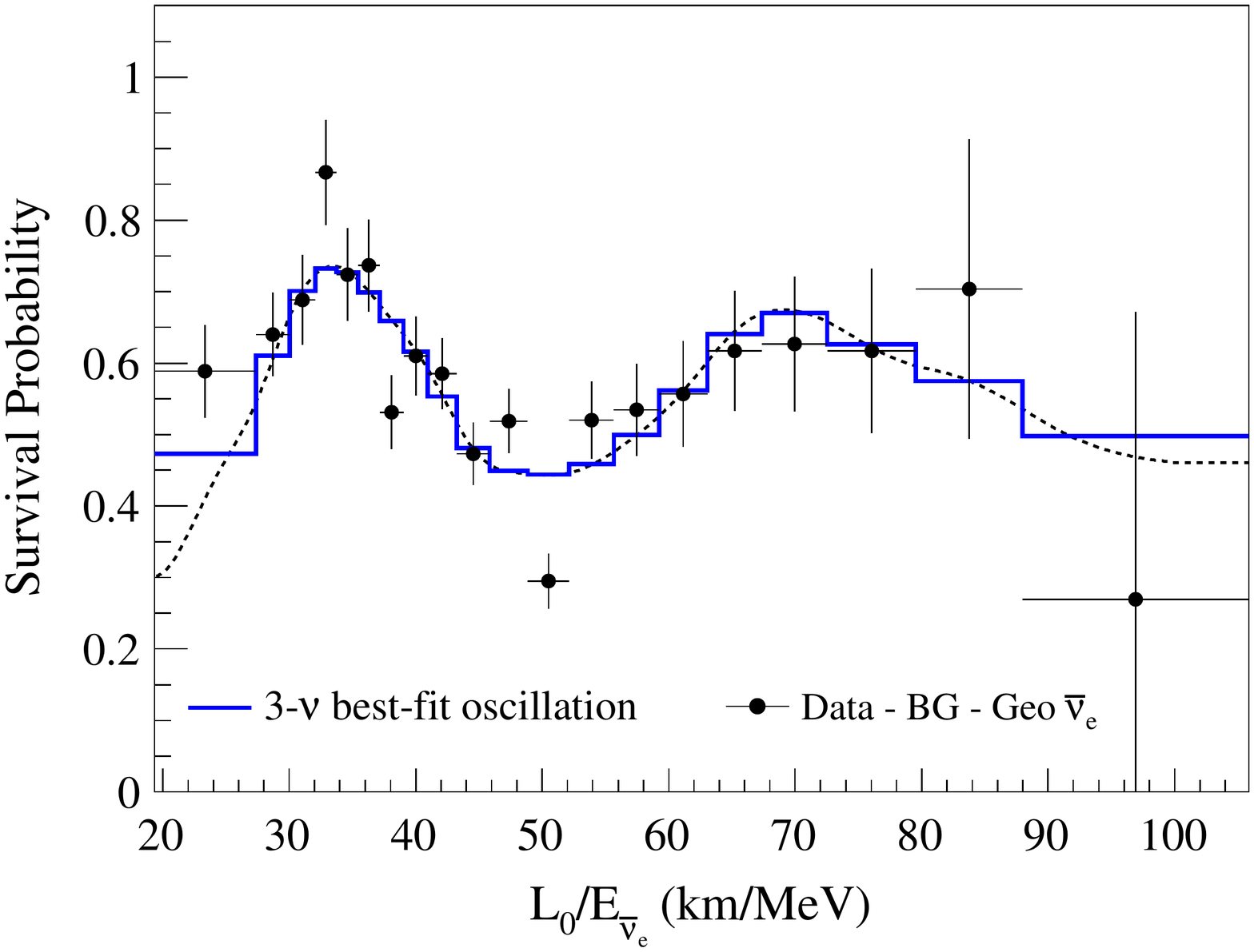}
   \end{minipage} \hfill
   \begin{minipage}{.52\linewidth}
      \includegraphics[width=1.0\linewidth]{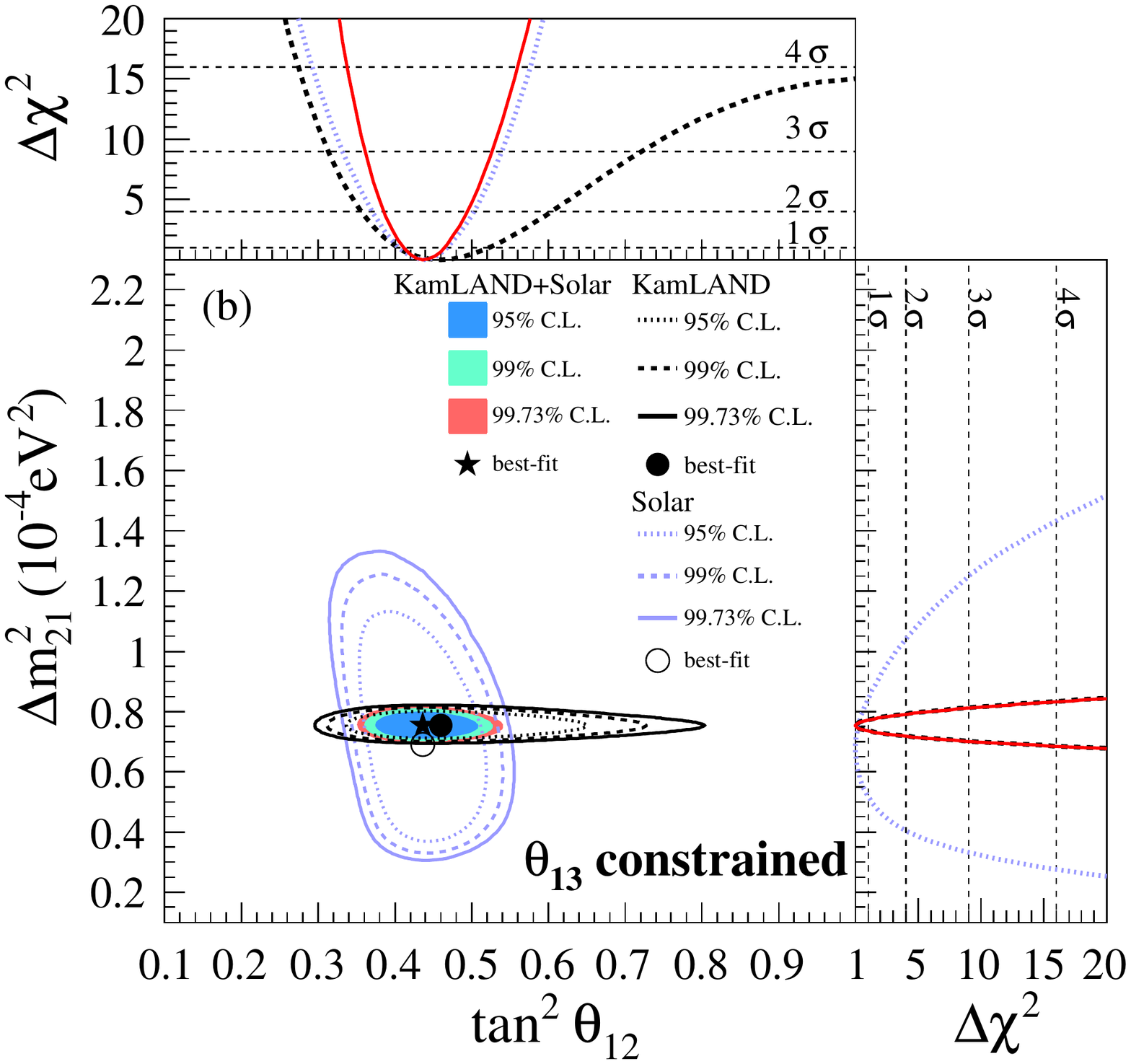}
   \end{minipage}
    \caption{KamLAND experiment. The left plot shows the ratio of the background- and geoneutrino-subtracted $\bar{\nu}_e$
spectrum to the expectation in the case of no-oscillation, as a function of
L/E. The curve and histogram show the expectation for the best fit oscillation hypothesis. The right plot shows the allowed region for neutrino oscillation parameters from
KamLAND and solar neutrino experiments~\cite{Gando:2013nba}. 
Courtesy of the KamLAND collaboration.
Reprinted figures with permission from A. Gando {\it et al.}, Phys. Rev. D, 88, 033001, 2013. Copyright 2013 by the American Physical Society.
} 
 \label{fig:sol-kam}
\end{figure}

KamLAND is a 1~kt liquid scintillator experiment located in the former site of the Kamiokande detector in Japan. As described in more detail in Section~\ref{subsec:reactorflux}, it is capable of detecting $\bar \nu_e$ emitted by nuclear reactors through the inverse beta decay reaction $\bar{\nu}_e \; p \rightarrow n \;  e^+$. The $\bar \nu_e$ interaction rate extends from the threshold (1.8 MeV) to about 7 MeV, and peaks around 3-4 MeV. 
KamLAND is surrounded by 55 Japanese nuclear reactors at an average distance of 150 km. In 2002, it reported the first evidence for the disappearance  of $\bar \nu_e$~\cite{kamland2002}. This measurement was very important for the interpretation of solar neutrino data, as it showed for the first time an oscillatory behaviour as a function of L/E, and it provided an independent and precise measurement of the oscillation parameters of interest.
In an updated report in 2013~\cite{Gando:2013nba}, they observed 2611 
events, to be compared with $3564 \pm 145$ expected from reactor neutrinos in the case of no oscillation, and  $364.1 \pm 30.5$ from background sources.
The KamLAND results are shown in Fig.~\ref{fig:sol-kam}. 

\subsection{Interpretation of the measurements and discussion}
\label{subsec:solarinter}

The first measurement of a deficit in the solar neutrino flux left several interpretations open. While Pontecorvo made the hypothesis of neutrino oscillations early on~\cite{Pontecorvo:1967fh}, modest changes to the Sun core temperature could also be invoked to explain the apparent deficit. 
The situation changed with the new measurements by the gallium experiments, as it was clear that the suppression factor depended on the neutrino energy. The SNO results were the first measurement of the total solar neutrino flux, and were also the first solid proof of flavour conversion as the explanation of the solar neutrino puzzle. 

Nevertheless, several different solutions were possible in the early days in terms of masses and mixing. The KamLAND result pinpointed one of them, the so-called Large Mixing Angle (LMA) solution which was already favoured by SNO data. Indeed, the results of KamLAND can be understood in terms of the simple two-flavour mixing formula~(\ref{eq:nuereactor}), $ P(\bar{\nu}_e \rightarrow \bar{\nu}_e ) = 1 - \sin^2 2 \theta_{12} \sin^2 (\frac{\Delta m^2_{21} L}{4 E}) $, neglecting matter effects and effects related to $\theta_{13}$. 

Since, as we will see later, the value of $\Delta m^2_{31}$ is much larger than $\Delta m^2_{21}$, it induces rapid oscillations that are averaged out given the large $L/E$ and the energy resolution of the detector. The $L/E$ pattern due to the $\Delta m^2_{21}$ term is prominent in the KamLAND data and beautifully confirms neutrino oscillations as the origin of the observed deficit (Fig.~\ref{fig:sol-kam}). The measurement of KamLAND, when combined with solar neutrino experiments, allows to determine $\Delta m^2_{21} = (7.53 \pm 0.18) \times 10^{-5}\, \mbox{eV}^2$ and $\tan^2 \theta_{12} = 0.44 \pm 0.03$. The agreement between solar neutrino experiments and KamLAND is generally good, as shown in Fig.~\ref{fig:sol-kam}.

The explanation of the deficit observed by the solar neutrino experiments involves flavour transitions of a different kind. As explained in Subsection~\ref{subsec:varying}, electron neutrinos are produced in the core of the Sun in a medium of high density. 
In the center of the Sun, the density $n(r=0)$ corresponds to the resonant density $ n_{res}= \frac{\Delta m^2_{21} \cos 2 \theta_{12}}{ 2 \sqrt{2}G_F E}$ for E = 1.9 MeV. For energies much higher than this value,
neutrinos are produced as $\nu^m_2$ matter eigenstates and, given that the evolution is adiabatic, exit the Sun as $\nu_2$. The probability to observe them as $\nu_e$ is then $| \langle \nu_e | \nu_2 \rangle |^2 = \sin^2 \theta_{12} \simeq 0.3$. 

\begin{figure}[htbp]
\centering
\includegraphics[width=0.6\linewidth]{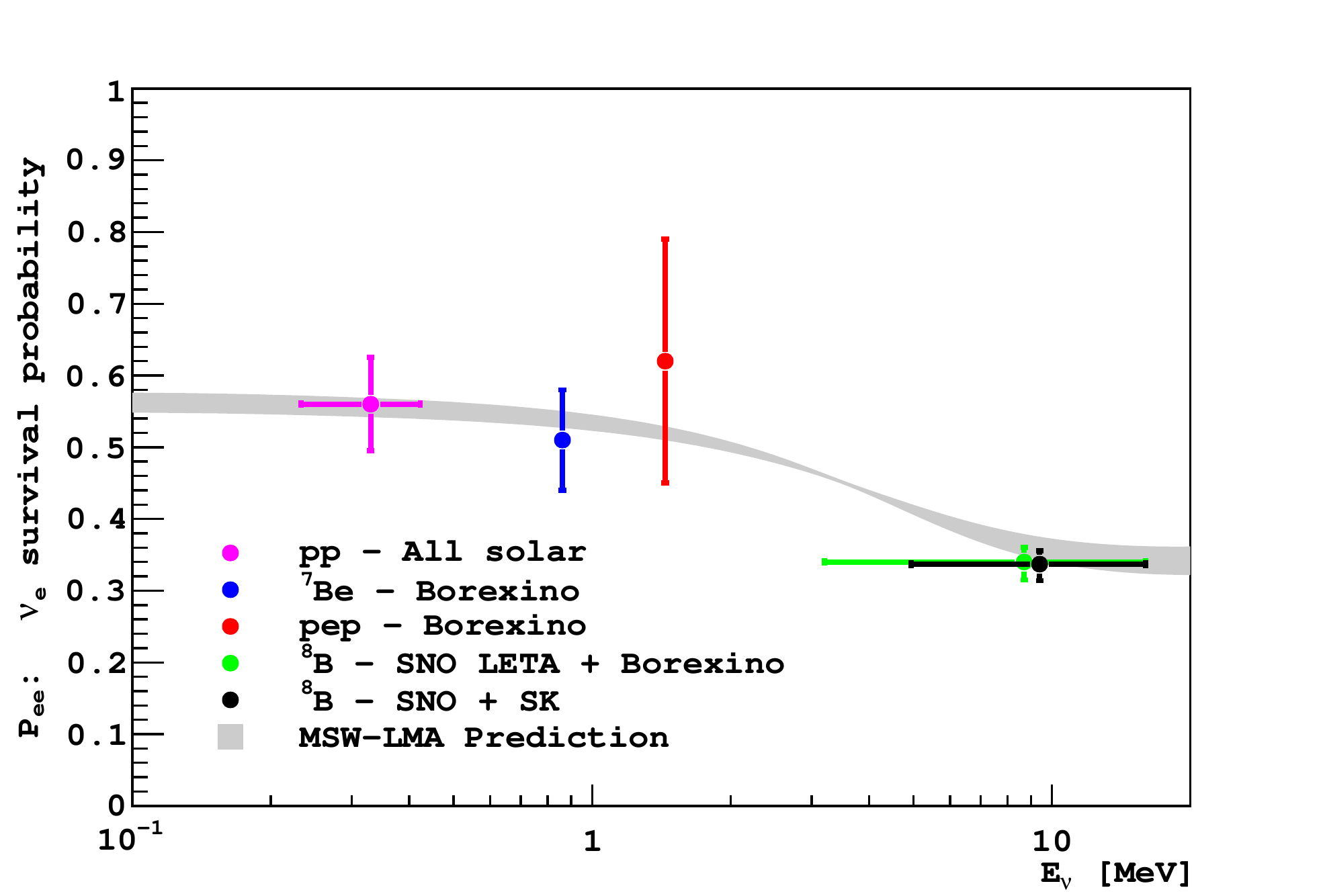}
  \caption{
  Solar neutrino survival probability as a function of the energy, including the Borexino results~\cite{derbin2016}. The gray band correspond to the $\pm1\sigma$ prediction of the MSW-LMA solution. Courtesy of the Borexino collaboration.
}
 \label{fig:sol-bor}
 \end{figure}

For neutrinos produced with energies much below this threshold, matter effect can be neglected and vacuum oscillations dominate. The suppression factor observed on Earth is then the result of the averaging of many oscillation periods:
$ P({\nu}_e \rightarrow {\nu}_e ) = 1 - \frac{1}{2} \sin^2 2 \theta_{12} \simeq 0.58$. 
This leads to a characteristic ``bath tub'' shape for the neutrino survival probability shown in Fig.~\ref{fig:sol-bor}
(see Subsection~\ref{subsec:varying} for a more detailed discussion). The adiabatic flavour conversion dominates for $^8$B neutrinos, while $pp$ are in the averaged vacuum oscillation region. 
This interpretation fits with all the solar neutrino data. 

The future of the field of solar neutrinos is twofold. The experiments will try to measure the CNO neutrinos as this is interesting both {\it per se} (the original motivation for studying solar neutrinos was in fact to understand how the Sun shines) and because it will shed some light on the solar abundance problem. This will be attempted by Borexino and SNO+~\cite{snoplus}, with liquid scintillator instead of the heavy water in the inner vessel. 

The upturn of the solar neutrino survival probability, that is, the transition from the MSW to the vacuum oscillation regime in the few MeV region, has not yet been observed. 
Experimentally, this is very difficult as it requires to lower the threshold and to have a good control of the backgrounds. The failure so far to observe this upturn, together with the value of the day-night asymmetry measured by Super-Kamiokande, is also related to the mild tension, recently reported by global fits of neutrino oscillation parameters, between the $\Delta m^2_{21}$ measured by KamLAND and the measurement using solar neutrino data~\cite{nufit}.
It has been argued that this tension could be alleviated by non-standard interactions modifying
the matter potential of solar neutrinos~\cite{Palazzo:2011vg,Gonzalez-Garcia:2013usa},
or by very light sterile neutrinos~\cite{deHolanda:2010am}.
Precision measurements of the solar neutrino flux are also a powerful probe of new physics scenarios.
 
On the other hand, long baseline reactor experiments of the KamLAND type can provide a very  clean measurement of the solar oscillation parameters. This is one of the physics goals of the 
Jiangmen Underground Neutrino Observatory
(JUNO) under construction in China, which will provide percent-level precision of the parameters $\Delta m^2_{21}$ and $\theta_{12}$.
The JUNO experiment will be described in Section~\ref{sec:future}.

\section{The 2-3 sector }

In this section, we describe the status of the so-called atmospheric neutrino oscillations, i.e. the 2-3 oscillations that were first discovered in the study of atmospheric neutrinos and were later precisely measured with the use of neutrino beams produced by particle accelerators.

\subsection{The atmospheric neutrino flux}

\begin{figure}[htbp]
\begin{minipage}[c]{.46\linewidth}
   	      \includegraphics[width=0.9\linewidth]{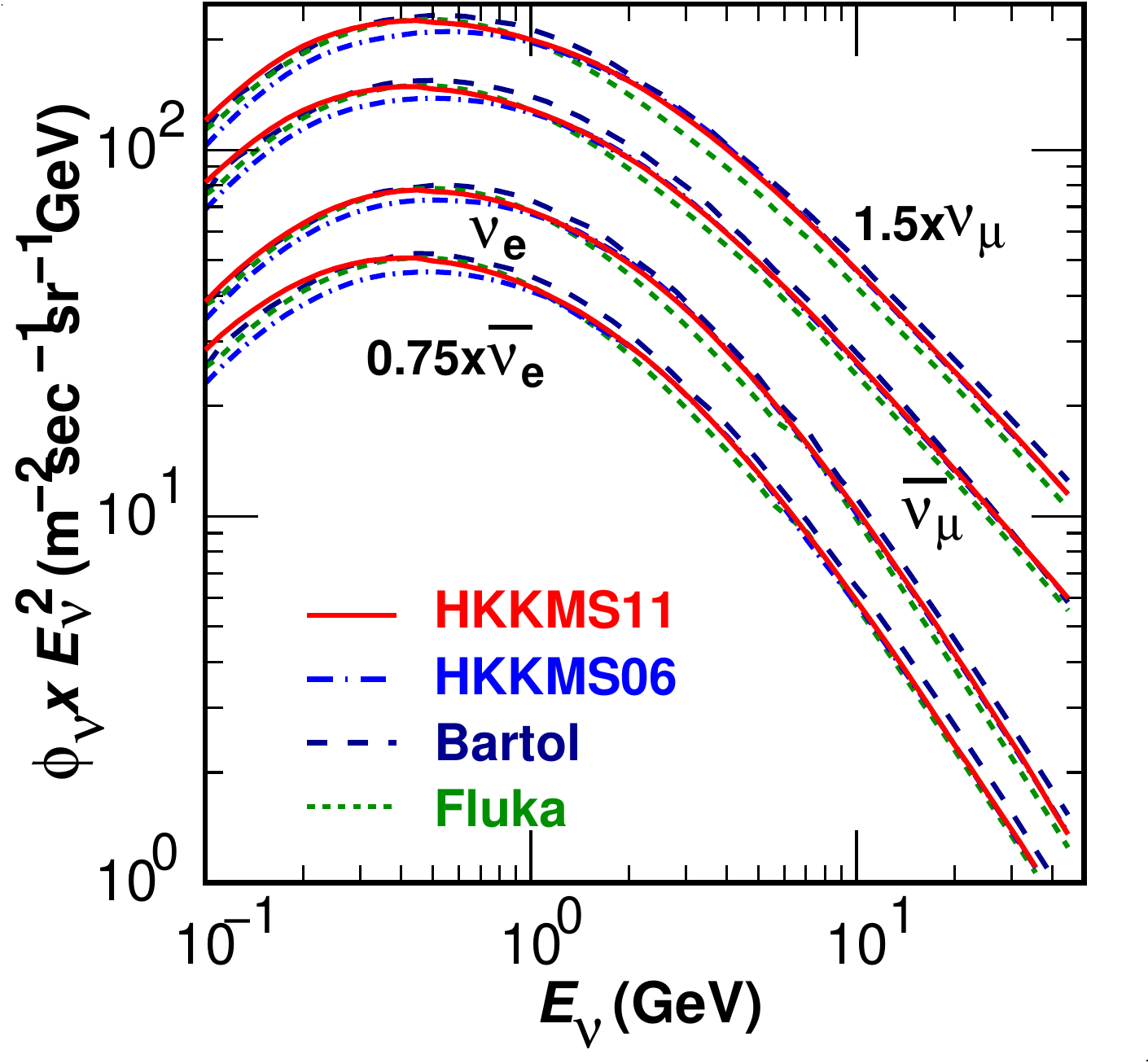}
   \end{minipage} \hfill
   \begin{minipage}{.46\linewidth}
      \includegraphics[width=0.9\linewidth]{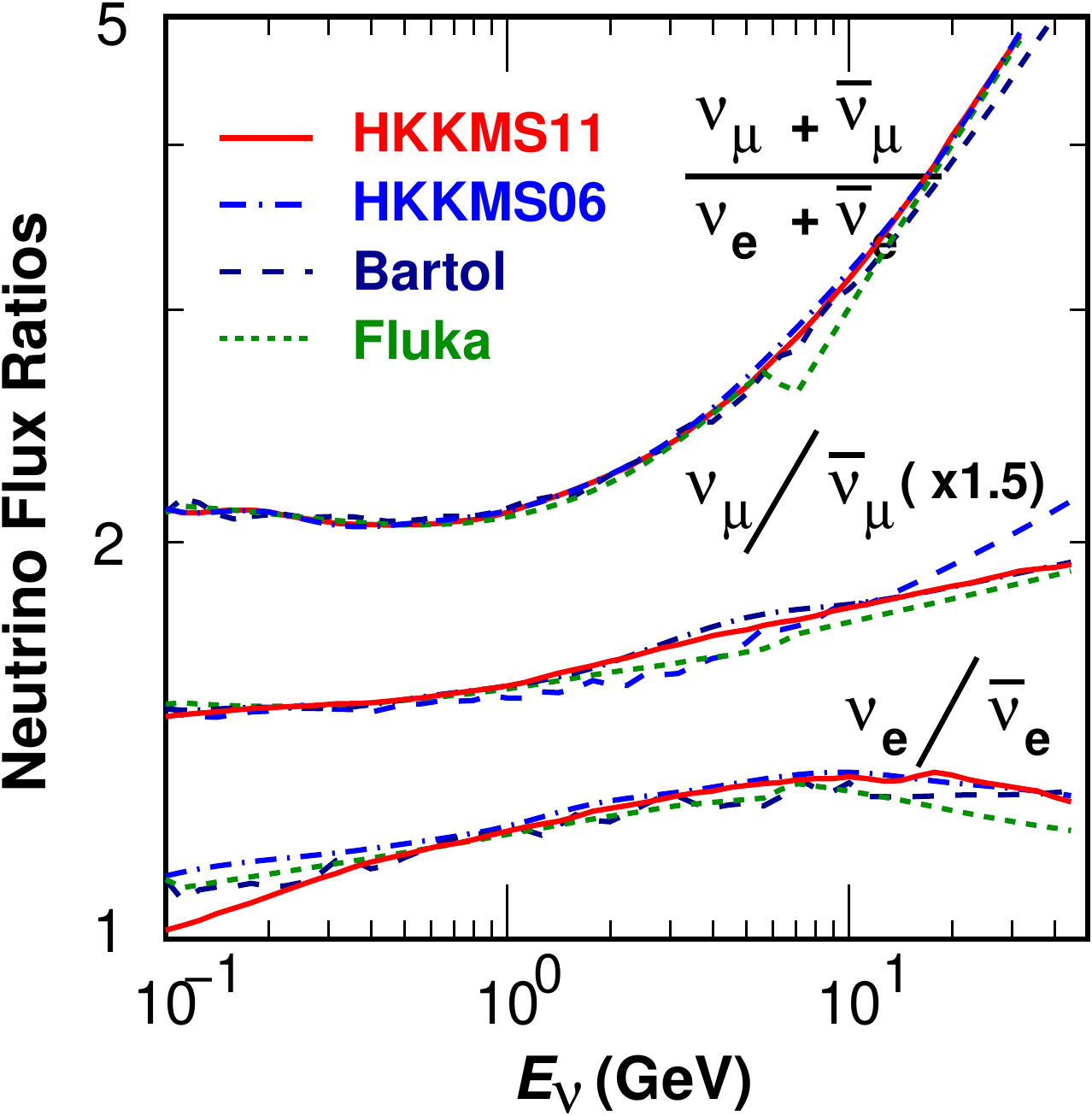}
   \end{minipage}
    \caption{The predicted flux of atmospheric neutrinos (left) as computed by several groups and the flux ratios (right)~\cite{PhysRevD.83.123001}. Courtesy of M. Honda et al. 
Reprinted figures with permission from M. Honda {\it et al.}, Phys. Rev. D, 83, 123001, 2011. Copyright 2011 by the American Physical Society.    
    }
 \label{fig:nuatmflux}
\end{figure}

The atmosphere is constantly bombarded by primary cosmic rays, composed mainly of protons, with a smaller component, $\simeq$ 5\%, of $\alpha$ particles, and an even smaller fraction of heavier nuclei. The interaction of these particles with atomic nuclei in the atmosphere produces hadronic showers composed mainly of pions and kaons. The decays of these mesons according to 
$\pi^+ \rightarrow \mu^+ \nu_\mu$ followed by 
$\mu^+ \rightarrow e^+ \nu_e \bar{\nu}_\mu $,
$K^+ \rightarrow \pi^+ \nu_\mu$ and $K_L \rightarrow \pi^+ e^- \nu_e$,
together with their charge conjugated processes, produces a flux of $\nu_\mu$, $\bar \nu_\mu$ and $\nu_e$, $\bar \nu_e$ with a steeply falling power-law spectrum (Fig. \ref{fig:nuatmflux}), approximately $E^{-2.7}$ in the region above 1 GeV.

The calculation of this neutrino flux (see Ref.~\cite{Gaisser:2002jj} for a review) relies on the knowledge of the primary cosmic ray flux and composition, of the Earth magnetic field, and of the hadro-production cross-section on the light nuclei of the Earth atmosphere. Detailed studies~\cite{PhysRevD.83.123001,Barr:2004br,Battistoni:2002ew,honda2007,honda2015} taking into account three-dimensional effects 
improve on previous efforts and reach a precision of 7-8\% for the flux in the 1-10 GeV range~\cite{honda2007}. It should be noticed that the ratio $N(\nu_\mu + \bar{\nu}_\mu)/N(\nu_e + \bar{\nu}_e)$ is predicted with a much better precision of 2\% ~\cite{honda2007}, as several systematic uncertainties cancel out in this ratio. In the limiting case where all the muons from pion decays decay in flight, this ratio is close to 2, as can be easily deduced from the decay processes mentioned above.
This case is approximately realized in the region of sub-GeV atmospheric neutrinos.


\subsection{Interaction of high-energy neutrinos with nuclei}

\begin{figure}[htbp]
\begin{minipage}[c]{.46\linewidth}
   	      \includegraphics[width=0.9\linewidth]{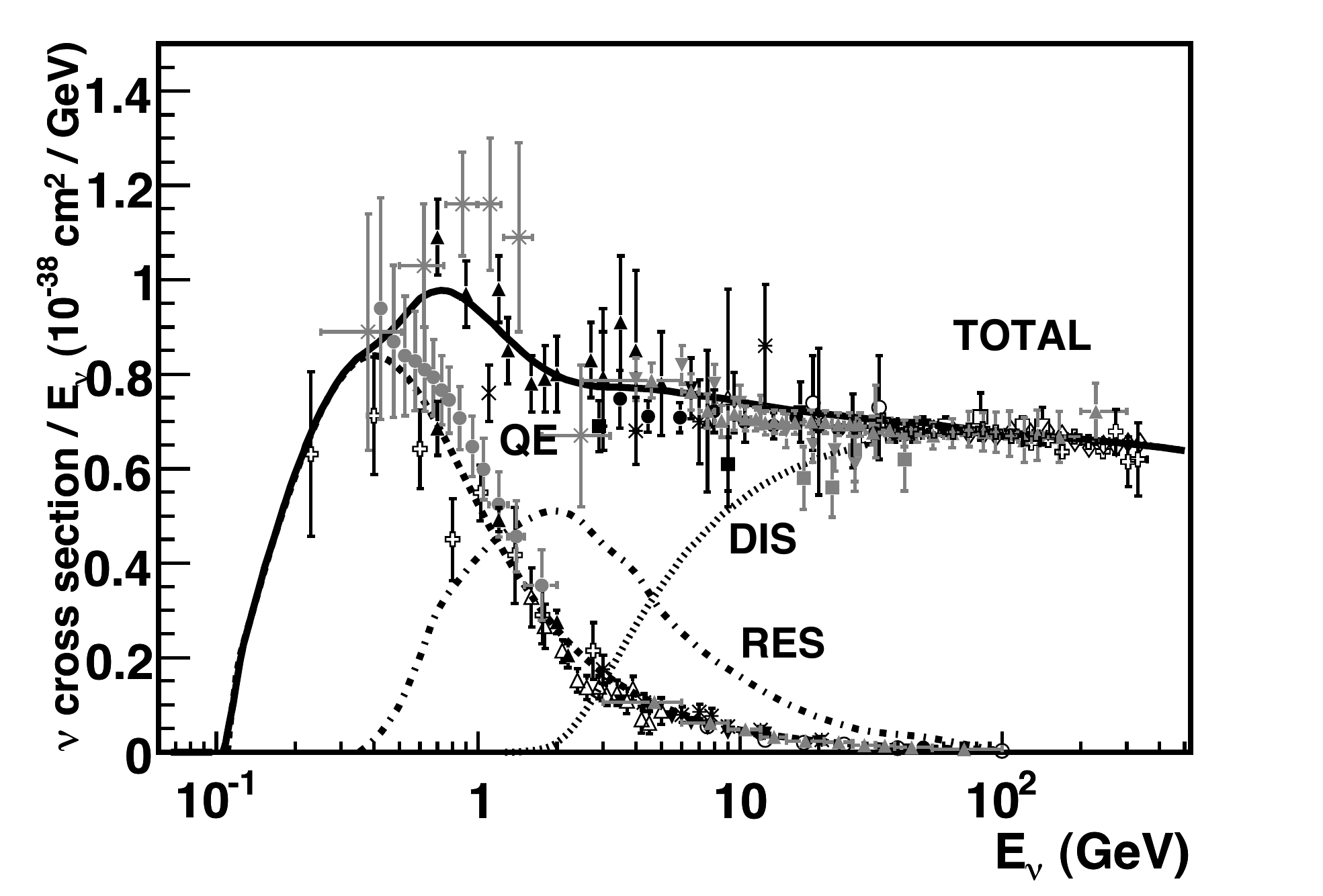}
   \end{minipage} \hfill
   \begin{minipage}{.46\linewidth}
            \includegraphics[width=0.9\linewidth]{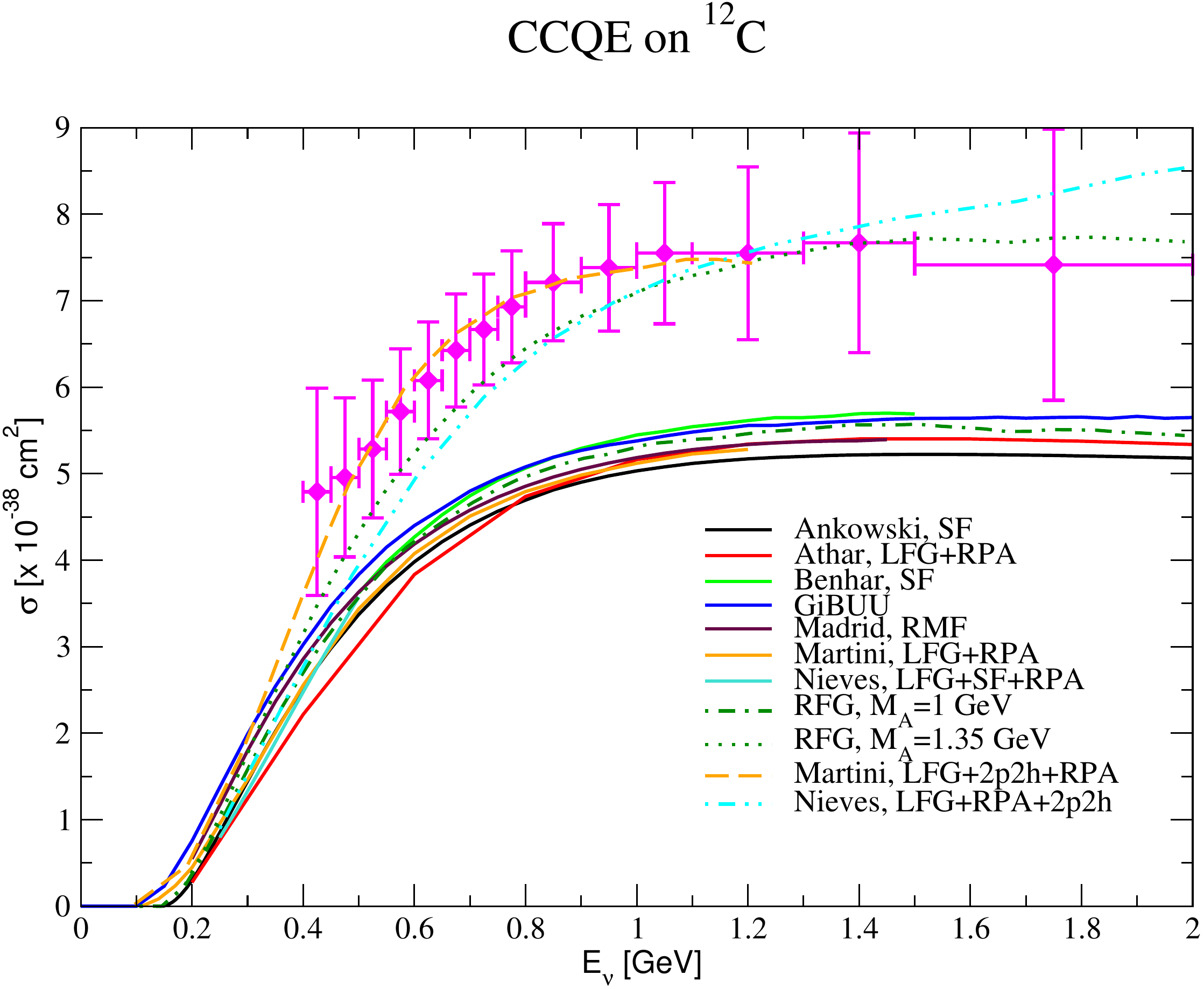}
   \end{minipage}
    \caption{ Left plot: Total CC neutrino cross-sections per nucleon  (for an isoscalar target) divided by neutrino energy and
plotted as a function of energy~\cite{formaggio}. Courtesy of J. A. Formaggio and G. P. Zeller. Reprinted figure with permission from J. A. Formaggio and G. P. Zeller, Rev. Mod. Phys., 84, 1307, 2012. Copyright 2012 by the American Physical Society. 
Right plot: CCQE cross-section on carbon~\cite{alvarez}.
The solid
lines denote various theoretical models.  The dash-dotted and
dotted lines are two models with $M_A=1$ and 1.35 GeV respectively. 
The dashed line takes into account the 2p-2h processes.  The data points
are from MiniBooNE. Courtesy of L. Alvarez-Ruso {\it et al.}.
}
 \label{fig:xsec}
\end{figure}

The interaction of neutrinos with matter takes place through charged current (CC) interactions (with the exchange of a $W$ boson and the production of a charged lepton in the final state) or neutral current (NC) interactions (with the exchange of a $Z^0$ boson). 
Above a few hundred MeV neutrino energy interactions with the nucleus have a typical $q^2$ corresponding to a scale of the order of one fermi, and the neutrino scatters off individual nucleons inside the nucleus. These interactions (Fig.~\ref{fig:xsec}) can be classified as:
\begin{itemize}
  \item (Quasi-) Elastic interactions, where the final state nucleon is ejected from the nucleus as a proton or neutron, like $\nu_\mu \: n \rightarrow \mu \: p$
  \item resonant single pion production like 
  $\nu_\mu \: n \rightarrow \mu \: N^*$ followed by $N^* \rightarrow n \pi$,
  where $N^*$ can be a resonance such as a $\Delta$,
  \item Deep Inelastic scattering at large energies and for large momentum transfers, where the neutrino interacts with a quark inside the nucleon.
  \end{itemize}  

The Charged Current Quasi Elastic (CCQE) process $\nu_l \: n \rightarrow l \: p$ (where $l=e,\mu$) plays a special role, because it is the dominant process below a neutrino energy of 1 GeV. Moreover its simple final state is very suitable to the reconstruction of the neutrino energy based on the measurement of the outgoing charged lepton. Neglecting Fermi momentum, two-body kinematics allows to write the reconstructed neutrino energy as 
\begin{equation}
E_\nu = \frac{m_p^2 - (m_n-E_b)^2 - m^2_l + 2 (m_n-E_b)E_l}{2 (m_n - E_b - E_l + p_l \cos \theta)}
\end{equation}
 where $E_b$ is the effective binding energy necessary to extract the nucleon from the nucleus; $m_p$, $m_n$ and $m_l$  are the proton, neutron and lepton mass; $E_l$, $p_l$ and $\theta$ are the measured lepton energy, momentum and angle with respect to the incoming neutrino direction. 
 In Cherenkov detectors, where the protons and most of the pions involved in these interactions are below the detection threshold, this process is particularly useful because the electrons and the muons can be easily detected, reconstructed and identified. 

In the simplest approach to the calculation of the cross-section, the impulse approximation (that is, the scattering from individual nucleons) is used and the nucleus is simply parametrized by the Fermi momentum and the binding energy $E_b$. The CCQE cross-section is then mainly governed by the axial vector form factor of the nucleon, where usually a dipole form is assumed,
$ G_A (q²) = \frac {g_A} {(1+q^2/M^2_A)^2}$.
Most parameters are determined from electron scattering for the vector form factors, and from nuclear $\beta$ decays for $g_A$, leaving the axial mass value $M_A$ as the main free parameter. This simple approach failed, as revealed by the fact that the value $M_A =  1.35 \pm 0.17$ GeV/c$^2$ measured by the Mini Booster Neutrino Experiment (MiniBoone)~\cite{miniboone-ccqe} for interactions on carbon was very different from the value 1.03 GeV/c$^2$ ~\cite{axial} obtained from the analysis of neutrino interactions on deuterium (Fig.~\ref{fig:xsec}).

A partial solution to this discrepancy came from considering additional processes like interactions of the neutrino with a correlated pair of nucleons inside the nucleus~\cite{Martini:2009uj}. These processes are known to exist in electron-nucleus scattering. From the experimental point of view, these processes, named two particle-two hole excitations (2p-2h), cannot be distinguished from CCQE interactions, as low momentum protons and neutrons are generally not detected. These processes can represent 15-25\% of the total CCQE cross-section~\cite{susa}, depending on the neutrino energy. Despite this progress, the normalization and shape of this additional contribution is still not well known and model dependent.

Other nuclear effects, like Final State Interactions (where the nucleon reinteracts within the same nucleus), or Secondary Interactions (where the nucleon interacts with other nuclei in the detector) introduce smearing and corrections which are under evaluation, especially for relatively large nuclei like carbon, oxygen, argon and iron generally used in these experiments. 
The anti-neutrino cross-sections are even less precisely established, as well as the $\nu_e$ cross-sections required by the present and future long-baseline oscillation experiments.  

All these cross-sections are the object of an intense theoretical and phenomenological activity~\cite{zeller,martini} given their relevance for oscillation analyses. The knowledge of these cross-sections is still model dependent and the precision does not exceed the 10-20\% level. Therefore, most of the accelerator experiments include near detectors to study the event rates before oscillations intervene. Clearly a more focused effort, combining phenomenological developments and well-controlled experimental data, is required for future high-precision oscillation experiments.

\subsection{The evidence for atmospheric neutrino disappearance}
\label{subsec:atmevidence}

The study of atmospheric neutrinos started in the 1960s: two experiments in very deep mines, in South Africa~\cite{Reines:1965qk} and in India~\cite{Achar:1965ova}, observed muons produced by atmospheric neutrino interactions. 
In the 1980s, several massive underground experiments, mainly motivated by the search for the proton decay predicted by Grand Unified Theories, started collecting data. These experiments needed to study in detail atmospheric neutrino interactions, as they constitute a background for proton decay searches.

In 1988, Kamiokande~\cite{kam88} reported a deficit in the number of $\nu_\mu$ single-ring candidates (85 events observed versus 144 expected), while the number of $\nu_e$ single-ring candidates agreed with the prediction (93 events observed versus 88.5 expected). 
This initiated the so-called atmospheric neutrino anomaly. A similar deficit was also observed by the Irvine-Michigan-Brookhaven (IMB) experiment~\cite{IMB} and later by the
Monopole, Astrophysics and Cosmic Ray Observatory
(MACRO)~\cite{macro} and Soudan-2~\cite{soudan}, while the Fr\'ejus~\cite{frejus} and the Nucleon Stability Experiment (NUSEX)~\cite{nusex} experiments observed no deficit. The situation evolved rapidly with the advent of Super-Kamiokande~\cite{sknim}, which started data-taking in 1996. 



Super-Kamiokande could rapidly accumulate a rather large data set of atmospheric neutrinos, measuring the direction of the produced lepton, its energy for fully contained events and their nature. For neutrino-nucleus interactions, the direction of the produced lepton is highly correlated with the direction of the neutrino above a lepton energy of 1 GeV~\cite{kajita2010}. 

Super-Kamiokande classifies events as Fully Contained (where the neutrino interaction takes place inside the inner detector and all the particles stop in the same volume), Partially Contained (some particles escape to the outer detector) and Upward Going muons, where the neutrino interacts in the rock below the detector. The events can be further classified according to the number of Cherenkov rings observed (if more than one ring is detected, the event is termed "multi-ring"), the observed total energy and the number of decay electrons. Electrons and muons can be reliably identified (the misidentification probability is 0.7\%~\cite{sk2005}) on the basis of the ring properties. Cherenkov rings produced by muons have sharp edges, while in the case of electrons the photons are produced by the numerous electrons and positrons in the electromagnetic shower, with angular dispersion due to scattering, resulting in a ring with diffuse edges. The momentum resolution for an isolated ring produced by a lepton of momentum $p$ is $0.6\% +2.6\%/\sqrt{p ({\rm GeV/c})}$~\cite{sk2005}. Sub-threshold muons and charged pions can be tagged by the presence of a delayed electron from muon decay, called Michel electrons.

\begin{figure}[htbp]
\centering
\includegraphics[width=0.95\linewidth]{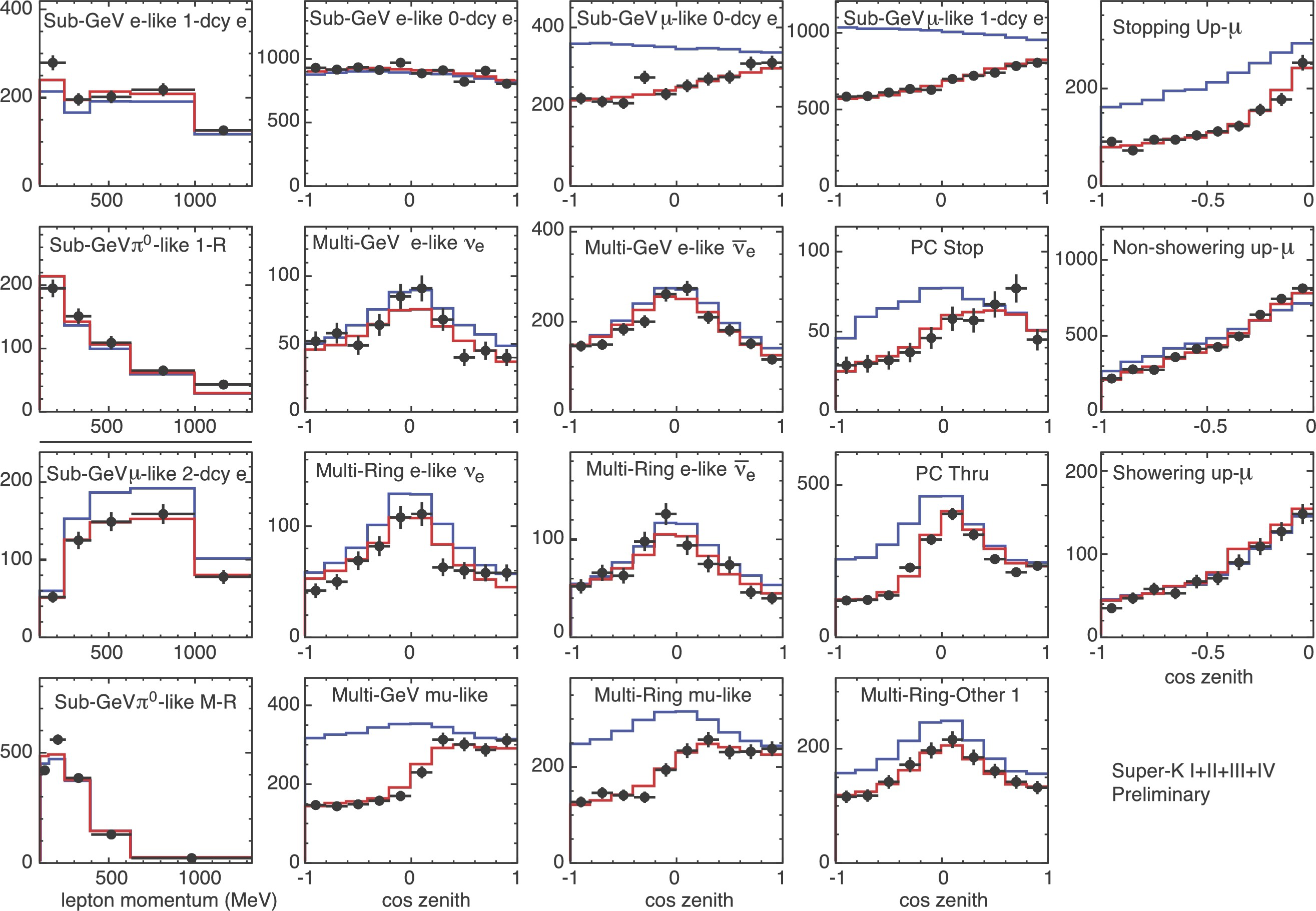}
  \caption{Zenith angle and
momentum distributions for atmospheric neutrino subsamples~\cite{skosc2016}
used in recent analyses by Super-Kamiokande, according to the ring type (e-like or $\mu$-like) and energy, the number of decay electrons, the number of rings, etc.
   The data are represented by points with
error bars. The blue histograms correspond to the expected distributions in the absence of oscillations,
while the red histograms give the
best-fit distributions in the oscillation hypothesis. Courtesy of the Super-Kamiokande collaboration.}
 \label{fig:sk-atm}
 \end{figure}

In 1998, the Super-Kamiokande collaboration presented their first analysis of atmospheric neutrinos~\cite{Fukuda:1998mi}, in particular the distributions of zenith angle for $ \nu_\mu$ and $\nu_e$ event selections (see Fig.~\ref{fig:sk-atm} for updated distributions) based on an exposure of 33 kton year. This was the first compelling evidence for neutrino oscillations as the explanation of the previously mentioned anomaly.  

Indeed, the neutrino path from the production to the detection varies from 15 km for down-going neutrinos (cosine of the zenith angle equal to 1) to more than 12,000 km for up-going neutrinos having traversed the whole Earth (cosine of the zenith angle equal to -1), thereby probing a large span of possible oscillation lengths. 

In a two neutrino scenario, the $\nu_\mu$ disappearance is governed by Eq.~(\ref{eq:numudisappApp}), namely
\begin{equation}
P(\nu_\mu \rightarrow \nu_\mu) \simeq 1 - \sin^2 2 \theta_{23} \sin^2 \left( \frac{\Delta m^2_{31} L}{4 E} \right) .
\label{eq:mudisapp}
\end{equation}
A glance at Fig.~\ref{fig:sk-atm}, especially at the distributions for Sub-GeV e-like, Multi-GeV e-like, Sub-GeV $\mu$-like and Multi-GeV $\mu$-like events, reveals several important overall features: 
\begin{itemize}
\item There is a strong disappearance of $\nu_\mu$, especially visible for up-going neutrinos. As the survival probability for very long baseline approaches $1- 1/2 \sin^2 2 \theta_{23}$, and the observed survival probability is close to 0.5, the mixing angle is therefore close to the maximal value $\pi/4$. 
\item The disappearance sets in for neutrinos close to horizontal zenith angle, and therefore the oscillation length should be of the order of 400 km for an energy around 1 GeV, corresponding to $\Delta m^2_{31} \simeq 10^{-3}$eV$^2$.  
\item There is no sizeable excess or deficit of $\nu_e$. Therefore the oscillations of $\nu_{\mu}$ should mainly involve either $\nu_{\mu} \rightarrow \nu_{\tau}$ or $\nu_{\mu} \rightarrow \nu_s$, where $\nu_s$ is an additional, sterile neutrino state.
\end{itemize}


Independently of any accurate prediction for the neutrino flux, the experimental observation of the distributions of Fig.~\ref{fig:sk-atm} is sufficient to make a strong case for neutrino disappearance. Indeed, above a few GeV, the neutrino flux is isotropic, as the primary cosmic rays are not deflected in a significant way by the geomagnetic field. The observation of a zenith-angle dependent deficit is therefore a sufficient argument to conclude that the fluxes of the different neutrino species are not conserved.  

While in 1998 other hypotheses like decay or decoherence were still open, the study of the L/E distribution by Super-Kamiokande~\cite{skle} and more recent data from long-baseline accelerator experiments  have ruled out all explanations apart from oscillations, because the alternative hypotheses imply a different L/E behaviour. 
    
\begin{figure}[htbp]
\centering
\includegraphics[width=0.6\linewidth]{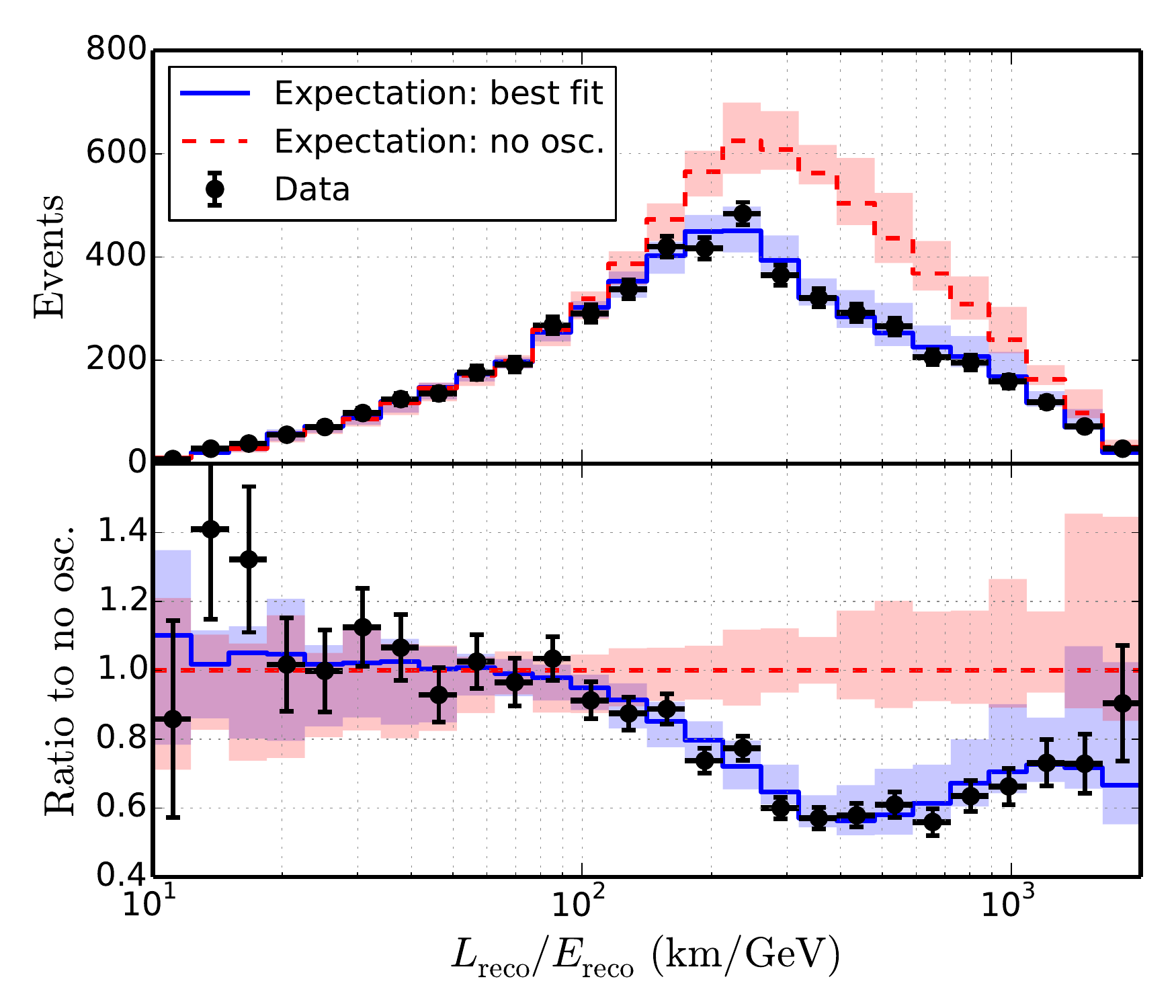}
  \caption{Distribution of atmospheric neutrino events measured by the IceCube experiment~\cite{Aartsen2016161} as a function of the
reconstructed L/E. Data are compared to the best fit and
expectation with no oscillations (top), and the ratio of data
and best fit to the expectation in the absence of oscillations is also shown
(bottom). Bands indicate estimated systematic uncertainties. Courtesy of the IceCube collaboration.}
 \label{fig:icecubeosc}
 \end{figure}

The IceCube experiment at the South Pole has recently completed the installation of DeepCore, a denser array of optical modules, aimed at significantly lowering the muon threshold. With data recorded between 2011 and 2014, corresponding to 5074 observed events, they have recently published an analysis~\cite{Aartsen2016161} of the disappearance of atmospheric $\nu_\mu$   in the range 10-100 GeV, requiring the zenithal angle to satisfy $\cos \theta < 0$ (see Fig.~\ref{fig:icecubeosc}). The sensitivity is similar to that of Super-Kamiokande 
for what concerns the muon neutrino disappearance, with the prospect of further improvements.

\subsection{Long-baseline neutrino beams }

Neutrino beams (see Ref.~\cite{Kopp2007101} for a review) based on particle accelerators have been used since the 1960s, when they provided the evidence for the existence of two types of neutrinos, $\nu_e$ and $\nu_\mu$. The general design is based on a high intensity proton beam impinging on a target and producing pions and kaons through interactions on the target nuclei. 

The long-baseline beams are based on the so-called wide-band beam concept. Here the secondary charged mesons are focused using a system of magnetic devices, called horns. The horns, usually with a cylindrical symmetry around the beam, are pulsed with a very intense current in coincidence with the arrival of the beam, and the direction of the current (the direction of the magnetic field) can be chosen in order to focus either positively or negatively charged particles. 
 
The mesons decay in a dedicated volume downstream of the target, via the following reactions: 
$\pi^+ \rightarrow \mu^+ \nu_\mu$ followed by 
$\mu^+ \rightarrow e^+ \nu_e \bar{\nu}_\mu $,
$K^+ \rightarrow \pi^+ \nu_\mu$ and $K_L \rightarrow \pi^+ e^- \nu_e$,
creating mainly a $\nu_\mu$ beam if positively charged pions are selected by the horns. Reversing the direction of the current in the horns will focus negatively charged particles and therefore produce a $\bar{\nu}_\mu$ beam, which allows the study of CP-violating effect, as we will discuss later. 
 
The flux is tuned in such a way that the phase $\Delta m^2_{32} L/ (4 E)$ reaches $\pi/2$ for the design baseline $L$ and the peak energy $E$, in order to probe the atmospheric oscillation sector with the $\nu_\mu$ beam. To do so, the proton energy, the target length and width, the focussing system and the decay volume length and width need to be accurately designed and optimized. The parameters of recent and future long baseline experiments are reported in Table~\ref{tab:lb}.

\begin{table}
\centering
\caption{Parameters of recent and future long-baseline experiments. Energy and power refer to the primary proton beam, L is the baseline, FD the mass of the far detector. POT (Protons On Target) represents the integrated dataset as of 2016 for the running experiments, and the total foreseen exposure for future projects.} \label{tab:lb}
\begin{tabular}{|c|c|c|c|c|c|c|}
  \hline
  Exp. & Ref & Energy (GeV) & Power (kW) & L (km) & FD mass (kt) & POT \\ 
  \hline
K2K & \text{\cite{Ahn:2006zza}} & 12 & 27 & 250 & 50 & 9.2 $\times$ 10$^{19}$\\
MINOS & \text{\cite{minos2014}}& 120 & 700 & 735 & 5.4 & 10.56 $\times$10$^{20}$\\
OPERA & \text{\cite{Agafonova:2015jxn}}& 450 & 300 & 730 & 1.8 & 1.8 $\times$10$^{20}$\\
T2K & \text{\cite{t2kprd}}& 30 & 750 & 295 & 50 & 2 $\times$10$^{21}$\\
\nova & \text{\cite{Adamson:2017qqn}}& 120 & 700& 810 & 14 & 6 $\times$ 10$^{20}$\\
HK & \text{\cite{hkdr}}& 30 & 1300 & 295 & 380 & 2.7 $\times$ 10$^{22}$\\
DUNE & \text{\cite{Acciarri:2015uup}}& 80 & 1070 & 1300 & 40 & 1.3 $\times$ 10$^{22}$\\
  \hline
\end{tabular}
\end{table}

\begin{figure}[htbp]
\centering
\includegraphics[width=0.6\linewidth]{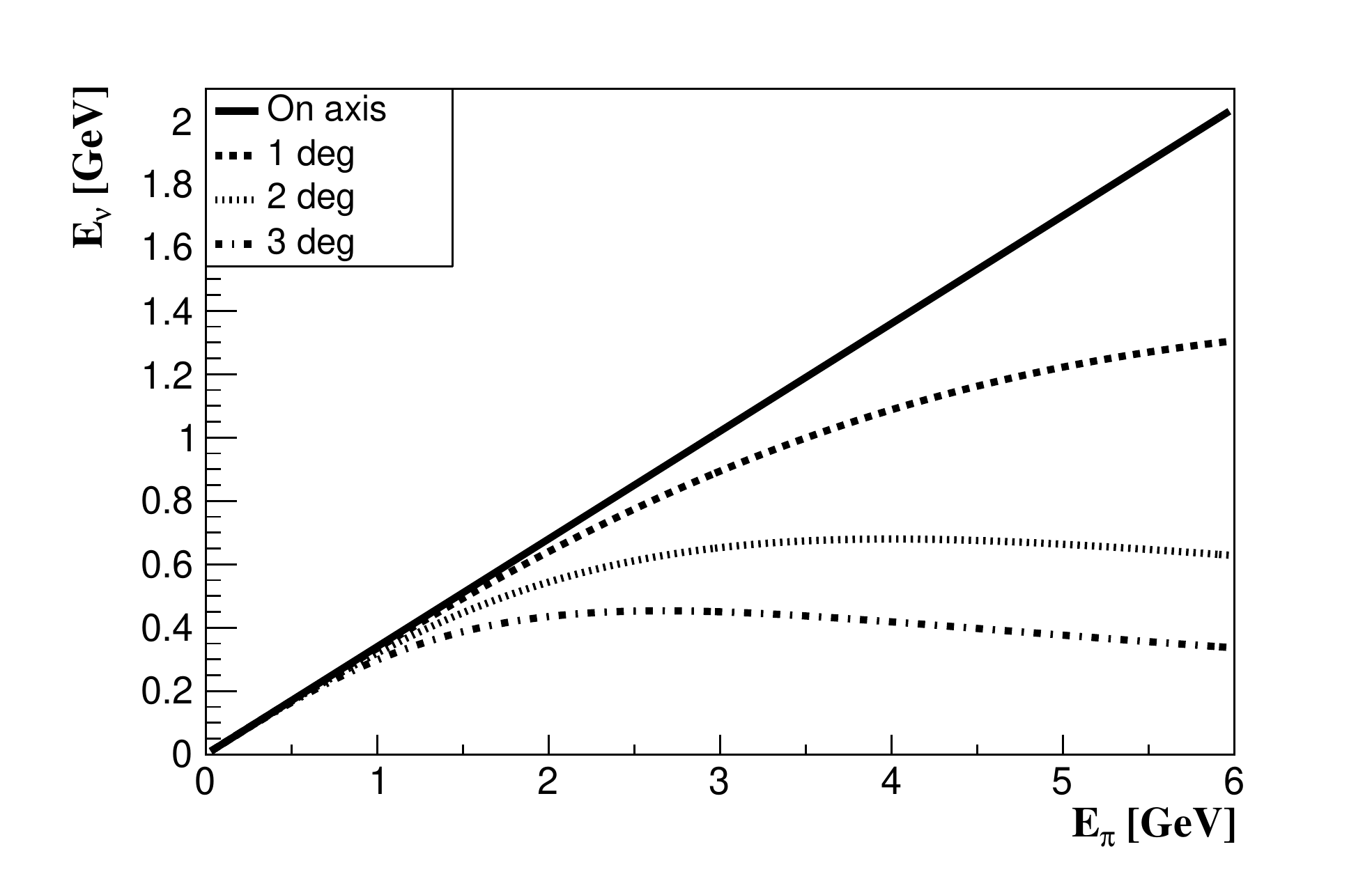}
  \caption{Neutrino energy as a function of the pion energy for on-axis decays and several off-axis angles. For a nonzero off-axis angle, the neutrino energy reaches a maximum. This feature is currently exploited in the T2K and \nova experiments. }
 \label{fig:offaxis}
 \end{figure}

An off-axis neutrino beam \cite{1995bnl} relies on the following idea: as the $\nu_\mu$ are mainly produced by the two-body decays of pions, there is a correlation between the pion energy $E_\pi$, the neutrino energy $E_\nu$ and the decay angle $\theta$ 
that can be understood as follows~\cite{nakaya}. For a perfectly focused beam of pions
$p_T = p^* \sin \theta^*$, $p_L = \gamma p^* (1 + \cos \theta^* ) \simeq E_\nu$,
where $p^*$ and $\theta^*$ are the decay momentum and angle in the rest frame of the pion.
Consider a fixed off-axis angle $\theta_{\rm lab} = p_T /p_L$.
Close to $\theta^* = \pi/2$, we have $\Delta p_T \simeq 0 $ for variations in $\theta^*$, and therefore $\Delta p_L = \frac{\Delta p_T}{\theta_{\rm lab}} \simeq 0$. This means that the variations of the neutrino energies as a function of the decay angle in the rest frame of the pion are suppressed, and that pions of various energies will contribute to a single neutrino energy determined by the off-axis angle (Fig.~\ref{fig:offaxis}).

\begin{figure}[htbp]
\centering
\includegraphics[width=0.4\linewidth]{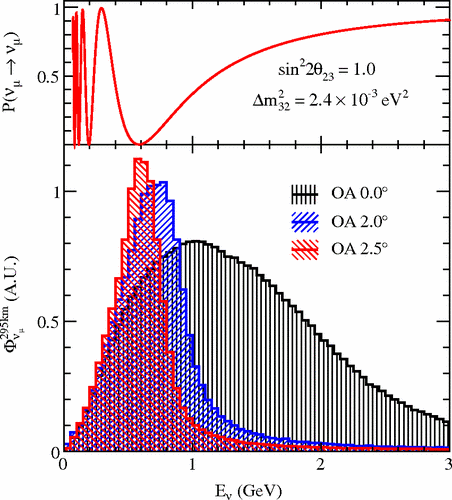}
  \caption{Neutrino flux in arbitrary units for the T2K setup with various off-axis angles~\cite{t2kflux}. The top part of the figure shows the muon neutrino survival probability at the far detector and illustrates how an off-axis beam can maximise the neutrino oscillation probability. Courtesy of the T2K collaboration.}
 \label{fig:offaxisflux}
 \end{figure}

Neutrinos emitted at a small angle with respect to the pion direction have therefore a distinct narrow spectrum peaking at a much lower energy with respect to the on-axis beam (Fig.~\ref{fig:offaxisflux}). This feature, which has been used by the T2K and \nova experiments, with off-axis angles of 43.6 and 14.6 mrad respectively, offers several advantages, because it avoids the large high-energy tail of the on-axis beam, thereby reducing some background reactions. 

As the neutrino beam is a tertiary beam, it is necessary to include in the experimental apparatus monitoring devices to ensure that it is stable in intensity and direction. To this effect, muon detectors sensitive to the muons produced by the pion decays are placed close to the end of the decay volume. Moreover, as the neutrino flux and cross-sections and the beam composition are not known with sufficient precision, a near detector is located close to the target station (typically within a few hundred meters). The near detector constrains the neutrino interaction rate, proportional to the product of the neutrino flux and the cross-section. Moreover, the near detector allows to measure the beam composition and to perform the study of several neutrino cross-sections.    

For recent long-baseline experiments, it is especially important that the beam is very pure in $\nu_\mu$. An irreducible $\nu_e$ component at the percent level is always present, due to the K$_{e3}$ semileptonic decays of the kaons and to the decays of the muons produced by the pions.

In the antineutrino mode, the fraction of neutrino interactions is relatively large. For instance, in the case of T2K, roughly 35\% of the interactions at SK in the antineutrino mode are produced by neutrinos~\cite{t2k-antinumu}. This is due to the the fact that more $\pi^+$ than $\pi^-$ are produced in proton-carbon interactions and that neutrino cross-sections are larger than antineutrino cross-sections. 
These impurities in the beam need to be measured, for instance with a magnetized near detector. 

\subsection{Detectors for long-baseline experiments}

The neutrino detectors in long-baseline experiments are quite similar to those used for atmospheric neutrinos, as the energy range of interest is similar. Moreover, in both cases, a very massive (tens of kt or more) underground detector is needed, and therefore the same detector can be used to study both sources. An example is Super-Kamiokande (see Section~\ref{subsec:atmevidence}), which is also used as the far detector for the long-baseline T2K experiment. 
Besides the detection of muons produced by $\nu_\mu$ interactions to probe the disappearance channel, the emphasis has been shifting recently to the study of $\nu_e$ appearance. This requires to discriminate between electron showers produced by CC reactions like $ \nu_e n \rightarrow e p$ and NC reactions like $\nu n \rightarrow \nu \pi^0 + X$.
Four main far detector technologies have been used or considered: 
\begin{itemize}
\item The water Cherenkov detection technique. It is used by Super-Kamiokande ~\cite{sknim} and has several advantages as it is well demonstrated and allows to instrument large detector masses. It is specially effective in the sub-GeV energy range where CCQE processes dominate the cross-section producing single Cherenkov rings, which are easier to reconstruct and measure. For fully contained events, the total lepton energy can be reconstructed and the neutrino energy inferred from the CCQE kinematics. The water Cherenkov technique is the only candidate to instrument very large detector masses approaching the megaton scale.
A similar technique is used by the IceCube~\cite{Aartsen2016161} and the Astronomy with a Neutrino Telescope and Abyss Environmental Research (ANTARES)~\cite{antares} experiments operating in an open natural medium, the Antarctic ice and the Mediterranean sea water, respectively.
\item The magnetized iron-scintillator calorimeters like the Main Injector Neutrino Oscillation Search (MINOS) experiment~\cite{PhysRevLett.110.251801}. They offer the advantage of measuring the charge of penetrating tracks, however they are limited by their ability to distinguish $\nu_e$ from NC $\pi^0$ production.
\item Totally active scintillator detectors like NOvA~\cite{Adamson:2017qqn}. Here the detection medium is a liquid scintillator contained in plastic cells read out by optical fibers. The totally active detector allows to reconstruct the event energy and the high granularity allows to distinguish $\nu_e$ from NC $\pi^0$ production.
\item Liquid Argon Time Projection Chambers. This is a new technology pionereed by the Imaging Cosmic And Rare Underground Signals (ICARUS) experiment~\cite{icarus}, which has been chosen for the future DUNE experiment~\cite{Acciarri:2015uup}. This detector offers superior performances in terms of information on the particles produced in the final state, with the possibility of particle identification and total energy reconstruction. An extensive program of R\&D is on-going at Fermilab and CERN to demonstrate the feasibility of this technology for large detectors of 10 kt.
\end{itemize}

A more specialized technology has been used by the 
Oscillation Project with Emulsion-tRacking Apparatus
(OPERA)~\cite{Agafonova:2015jxn} in order to observe $\nu_\tau$ appearance (see Section~\ref{subsec:opera}), with the main neutrino target consisting of nuclear emulsion detectors to reconstruct the short-lived $\tau$ leptons and their decay products.

\subsection{Flux and cross-section systematic uncertainties}
\label{sec:beamsyst}

Long-baseline neutrino oscillation experiments require a good control of the systematic uncertainties related to neutrino fluxes and neutrino cross-sections to perform precision measurements of oscillation parameters. 

A common strategy to reduce systematic uncertainties, used in most of the long-baseline experiments, is the use of a Near Detector complex, located few hundreds of meters from the beam target. In this way the neutrino spectrum and the flavour composition of the beam can be precisely measured before the oscillations, and this information is used to determine the expected spectra at the far detector.

Some long-baseline experiments, for example MINOS and \nova, use near detectors that have the same composition and technology as the far detector, so that cross-section uncertainties due to the different target nuclei or detector systematic uncertainties are minimized in the extrapolation. Other experiments, for example K2K and T2K, use a set of near detectors with different target nuclei in order to maximize the information on the cross-section models.

In this section, as an example, we will briefly describe how the T2K experiment uses the near detector data to reduce the systematic uncertainties in the oscillation analysis~\cite{t2kprd}. 
T2K is a long baseline experiment in Japan, using the 30 GeV proton beam generated by the Japan Proton Accelerator Research Complex (JPARC) to produce a neutrino beam aimed towards the Super-Kamiokande detector at a distance of 295 km.  

The T2K near detector, called ND280, at 280 m from the proton target, is magnetized, with several sub-detectors installed within the ex-UA1/NOMAD magnet that provides a 0.2 T magnetic field. For the oscillation analysis, the ND280 tracker system is used, comprising two Fine Grained Detectors (FGD) interleaved with three Time Projection Chambers (TPCs). The FGDs act as an active target for neutrino interactions (each with a mass of about 1~ton). Particles produced in the FGD enter the TPCs, where the charge and the momentum of the track is reconstructed and the particle identified based on the measurement of the energy loss in the gas. One important point to notice is that one of the two FGDs is a fully active detector, while in the second one scintillator layers are interleaved with inactive water layers, allowing to select neutrino interactions on carbon and on oxygen, the same target as the far detector, SK. 
    
At T2K energies, most of the \num charged current interactions are quasi-elastic events, producing only one visible track in the final state, the muon, since the proton has often a low momentum around 200 MeV/c and its track is too short to be reconstructed. The presence of the magnetic field allows to reconstruct the charge of the muons, distinguishing between negative muons produced by \num and positive muons produced by \numb interactions.
In some cases, neutrinos exchange enough energy with the target nuclei, producing also pions or protons that can enter the TPC. 
For \num selections, ND280 distinguishes three classes of events on the basis of the final state topology: only the muon is reconstructed (CC0$\pi$, with a 63 \% purity of CCQE events), the muon and one negative pion are reconstructed (CC1$\pi$), and events with more pions (CCother, essentially deep-inelastic events). For \numb selections, interactions with one track are separated from interactions with more than one track. The total sample selected in ND280 comprises about 25,000 events~\cite{t2kprd}.
  
The same selections are applied to interactions in both the FGDs, and all the samples are then fit to reduce systematic uncertainties on the flux and cross-section model. 
The main uncertainty on the flux model is due to the hadron production cross-section. Data from the NA61/SHINE experiment~\cite{shine} at CERN are used. In NA61/SHINE, a proton beam accelerated to 30 GeV/c strikes a target. The hadrons produced are measured with a system of TPCs and Time Of Flight detectors, and double differential cross-sections (in angle and momentum) are extracted. Thanks to NA61/SHINE data, the uncertainties on the neutrino fluxes, prior to the ND280 fits, are reduced to the 10\% level.

For the cross-section model, the priors are based, as much as possible, on external experiments such as the Mini Booster Neutrino Experiment
(MiniBooNE)~\cite{miniboone-ccqe} and
the Main Injector Experiment for $\nu$-A
 (MINERvA)~\cite{minerva}. An important aspect is that, for the first time, effects related to the multi-nucleon 2p-2h emissions are included in this analysis, based on the Nieves et al. model~\cite{Nieves:2011yp}.

To reduce flux and cross-section uncertainties, the ND280 samples are binned in the kinematic variables \ptheta according to the muon momentum and angle with respect to the beam direction. Data and Monte Carlo are fitted using a binned likelihood, where the prediction for each bin depends on the flux, cross-section and detector systematic parameters. 

The result of the fit is a set of point estimates and covariance for the systematic scaling factors for the unoscillated neutrino flux at SK. The impact of the ND280 fit on the total error budget in the T2K oscillation analysis is shown in Table~\ref{tab:t2ksyst}: a reduction of the systematic uncertainties from $12\%$ to $\sim5\%$ is obtained thanks to the near detector fit.

\begin{table}[htb]
\begin{center}
\caption{Sources of the systematic uncertainty on the predicted neutrino event rates at Super-Kamiokande in T2K oscillation analyses~\cite{t2k2016}.
The effect of the near detector (ND280) constraint on the flux and cross-section is particularly visible. The line (Other) reports the effect of Final State Interactions and Secondary Interactions.
}
\begin{tabular}{|l|c|c| } \hline 
Source of Uncertainty& $\nu_e$ & $\nu_{\mu}$ \\ 
 &$\delta N/N$&$\delta N/N$\\ \hline
Flux&3.7\%& 3.6\% \\ \hline
cross-section &5.1\%& 4.0\% \\ \hline
Flux+cross-section&& \\
(w/o ND280 Constraint)&11.3\% &10.8\% \\
(w/ ND280 Constraint)&4.2\% &2.9\% \\ \hline
Other & 3.5\%& 4.2\% \\ \hline
All & & \\
(w/o ND280 Constraint)& 12.7\%& 12.0\% \\ 
(w/ ND280 Constraint)& 6.8\%& 5.1\% \\ \hline 
\end{tabular}
\label{tab:t2ksyst}
\end{center}
\end{table}

In the case of \nova, the collaboration profits from the fact that the near and the far detectors are identical, and uses a calorimetric approach in which all the energy observed in the event at the near detector is reconstructed, and is then unfolded to obtain the expected true neutrino energy. The far to near ratio and the oscillation probabilities are then applied to obtain the expected true energy spectrum at the far detector. This method allows to reduce the total systematic uncertainty to a level similar to the one obtained by T2K.

\subsection{Results from long-baseline accelerator experiments}

KEK-to-Kamioka (K2K) was the first long-baseline neutrino beam, using Super-Kamiokande as its far detector at 250 km from the neutrino production source~\cite{t2k2016}. Operating between 1999 and 2004, it has measured the disappearance of $\nu_\mu$ governed by Eq.~(\ref{eq:mudisapp}): 112 events were observed, while 158.1$^{+9.2}_{-8.6}$ were expected without oscillation, a 4.3 $\sigma$ effect~\cite{Ahn:2006zza}. This measurement has confirmed neutrino oscillations as the explanation for the atmospheric neutrino disappearance. 

Further improved measurements of the $\nu_\mu \rightarrow \nu_\mu$ oscillations were reported by MINOS~\cite{minos2014}, a long-baseline experiment with a neutrino beam from Fermilab to the Soudan mine in Minnesota (baseline 735 km). Protons from the Main Injector with an energy of 120 GeV were used with a tunable setup allowing to vary the neutrino energy, mostly used in the position giving a 1-3 GeV neutrino beam.
The far detector was a 5.4 kton magnetized steel-scintillator calorimeter. 
In 2014 MINOS~\cite{minos2014} reported a combined fit using also a data set of atmospheric neutrinos (37.88 kton-year) yielding $|\Delta m^2_{32}| = [2.28-2.46]  \times 10^{-3}$ eV$^2$ (68\% C.L.) and $\sin^2 2 \theta_{23} =0.35-0.65 $ at 90\% C.L assuming the normal ordering, and $|\Delta m^2_{32}| = [2.32-2.53]  \times 10^{-3}$ eV$^2$ (68\% C.L.) and $\sin^2 2 \theta_{23} =0.34-0.67 $ at 90\% C.L for inverted ordering.
The final results from the MINOS experiment showed also agreement between the oscillation parameters for $\nu_\mu$ and $\bar \nu_\mu$~\cite{PhysRevLett.110.251801}.

The current generation of experiments, T2K and \nova, use a narrow band beam based on the off-axis design, which is particularly suited for the measurement of \num disappearance once the relevant squared-mass difference is known. For a given value of \dmsq and a given baseline, the oscillation probability depends on the neutrino energy, which can be tuned by changing the off-axis angle. In the case of T2K (L=295~km), the maximum of the oscillation is at an energy of 600 \mev, while in the case of \nova  (L=810~km) the maximum is at 1,800 \mev. 

While the most recent analyses published by T2K perform a global fit of appearance and disappearance channels in order to extract the maximal information on neutrino oscillation parameters, disappearance analyses only have also been done by T2K in order to test the PMNS framework. Since CP violation is possible only in appearance channels (see Section~\ref{subsec:vacuum}), the disappearance probability is expected to be the same for \num and \numb. 
The observation of a departure from this expectation would point to CPT violation.

\begin{figure}[htbp]
\centering
\includegraphics[width=0.6\linewidth]{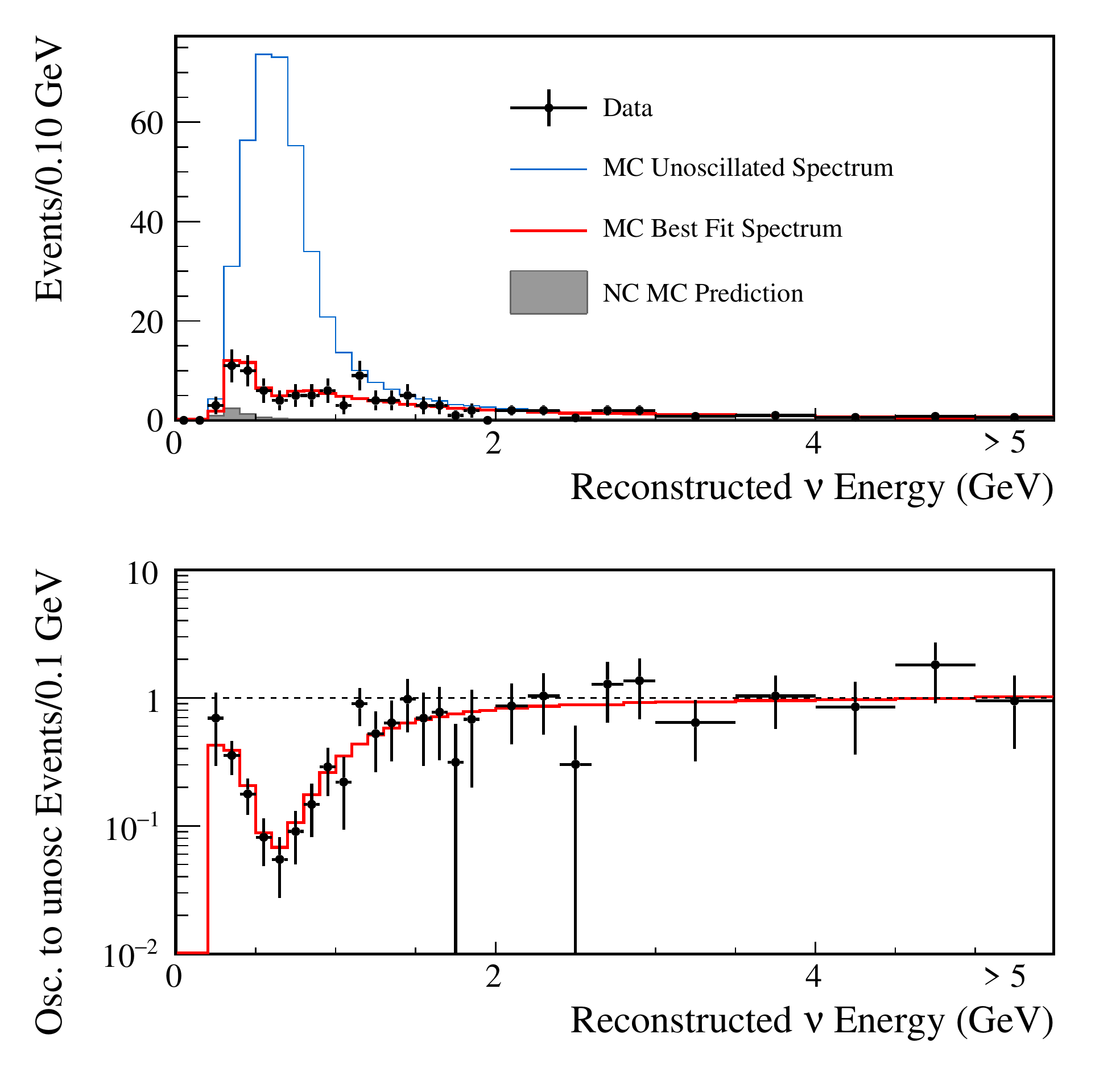}
  \caption{
Top: Reconstructed neutrino energy
spectrum for the T2K $\nu_\mu$ candidates, best-fit prediction, and unoscillated prediction~\cite{t2kprd}.
Bottom: Ratio of oscillated to unoscillated events as a function of
neutrino energy for the data and the best-fit spectrum.  Courtesy of the T2K collaboration.}
\label{fig:t2kdis} 
 \end{figure}

\begin{figure}[htbp]
\centering
\includegraphics[width=0.6\linewidth]{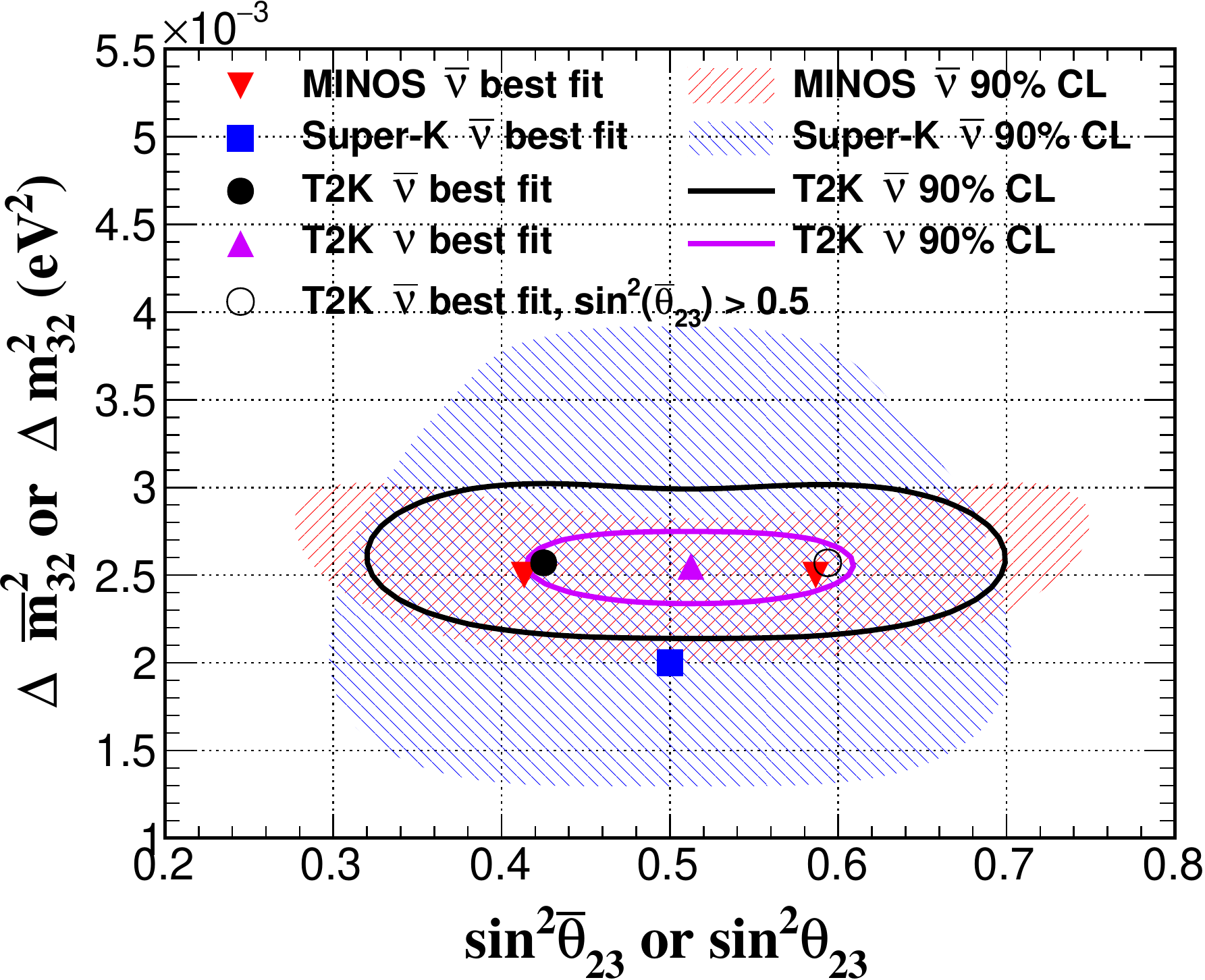}
  \caption{T2K  90\% CL region for $\Delta m_{32}^2 $ versus \stt for neutrinos and antineutrinos~\cite{t2knumubar2017}. Normal hierarchy is assumed. 90\% CL regions obtained by SK~\cite{SKnubar} and MINOS~\cite{minos2014} for antineutrinos are also shown.  
  Courtesy of the T2K collaboration.}
 \label{fig:atm-nunubar}
 \end{figure}

For the T2K experiment, fully contained \num events are selected in Super-Kamiokande in time with the beam, with one ring identified as produced by a muon and less than two decay electrons. The resulting sample is composed of 62\% of CCQE $\nu_\mu$ events, 32\% of non-CCQE $\nu_\mu$ events, the rest being essentially NC events~\cite{Abe:2017vif}.
Based on \nupot protons-on-target (POT) collected in the neutrino mode and \nubpot POT collected in the antineutrino mode, T2K has observed 135 (66) \num (\numb) candidates at SK with 522 (185) events expected in case of no-oscillations and 135.8 (64.2) events expected for maximum disappearance~\cite{t2k2016}. The reconstructed energy spectrum for  \num  events is shown in Fig.~\ref{fig:t2kdis}. Thanks to the use of the Near Detector data described in Section~\ref{sec:beamsyst}, the systematic uncertainties on these measurements are at the 5\% level, and a precise measurement of \thatm and \dmsq is obtained, as shown in Fig.~\ref{fig:atm-nunubar}. The measured oscillation parameters are in excellent agreement between neutrinos and antineutrinos, and both are compatible with a maximal \thatm mixing angle~\cite{t2k-antinumu}. 

%

\begin{figure}[htbp]
\centering
\includegraphics[width=0.5\linewidth]{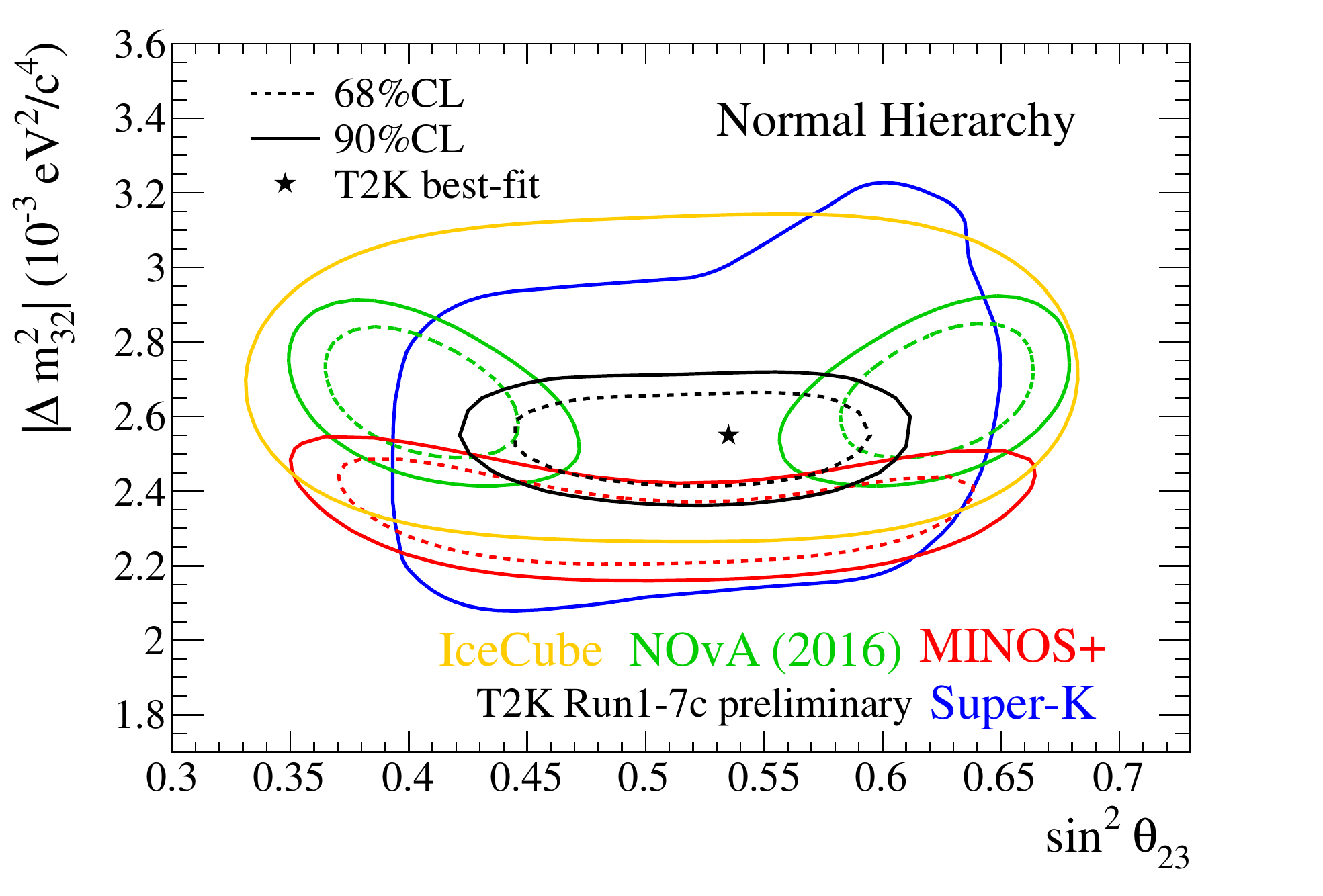}
  \caption{
Confidence regions in the plane  $\Delta m^2_{32}$ versus $\sin^2 \theta_{23}$, for T2K~\cite{t2k2016}, Super-K~\cite{Wendell:2015onk}, Minos+~\cite{PhysRevLett.110.251801}, \nova~\cite{Adamson:2017qqn} and IceCube DeepCore~\cite{Aartsen:2016psd}. Courtesy of the T2K collaboration.}
 \label{fig:atm-contour}
 \end{figure}

\nova is a long baseline experiment using the same neutrino beam as MINOS. Its main detector, located 810 km away in Ash River (USA), is 14.6 mrad off-axis : it consists of a 14 kt segmented liquid scintillator. 
\nova has reported a measurement of \num disappearance using \novapot POT and selecting \mmu-like candidates at the far detector~\cite{Adamson:2017qqn}. They observed 78 events in the far detector, while $473\pm30$ were expected without oscillations. This result leads to some tensions, especially with the T2K experiment, since maximal mixing is excluded by \nova at $2.5\sigma$ as shown in Fig.~\ref{fig:atm-contour}. 

\begin{figure}[htbp]
\centering
\includegraphics[width=0.6\linewidth]{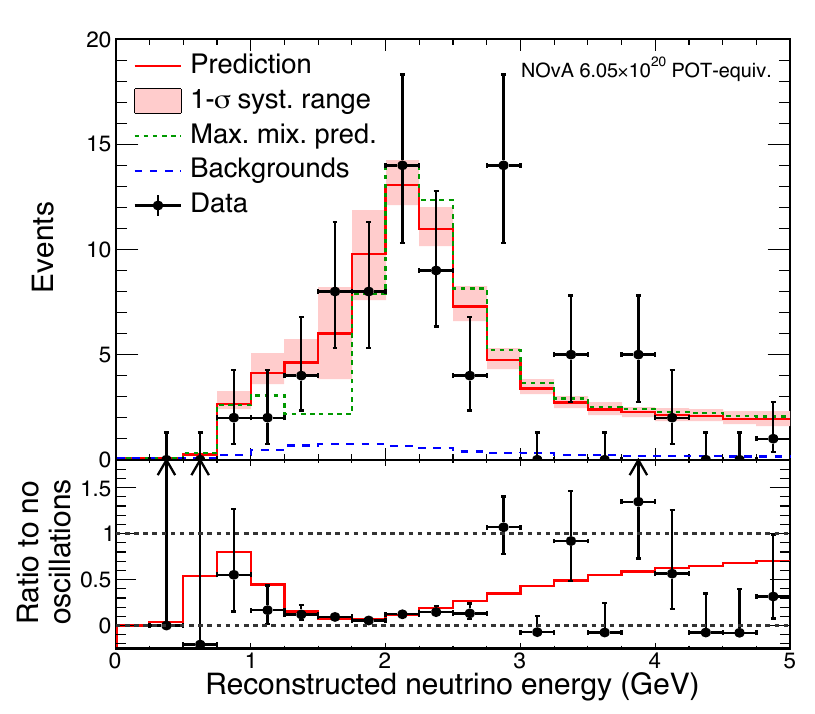}
  \caption{Reconstructed energy spectrum for \num candidates in \nova (black dots), compared with the best fit (solid red line) and the prediction assuming maximal mixing angle (dashed line)~\cite{Adamson:2017qqn}. Courtesy of the \nova collaboration.
Reprinted figure with permission from P. Adamson {\it et al.}, Phys. Rev. Lett. 118, 151802, 2017. Copyright 2017 by the American Physical Society.  
   }  \label{fig:atm-nova}
 \end{figure}

The two collaborations are working to understand this difference, which could simply be due to statistical fluctuations. A comparison between the reconstructed energy spectrum for the best-fit and the one obtained by assuming maximal mixing in \nova is shown in Fig.~\ref{fig:atm-nova}. 
Additional data from T2K and NOvA will help clarify the situation. 
 
 \subsection{Evidence for $\nu_\tau$ appearance}
\label{subsec:opera}

\begin{figure}[htbp]
\centering
\includegraphics[width=0.5\linewidth]{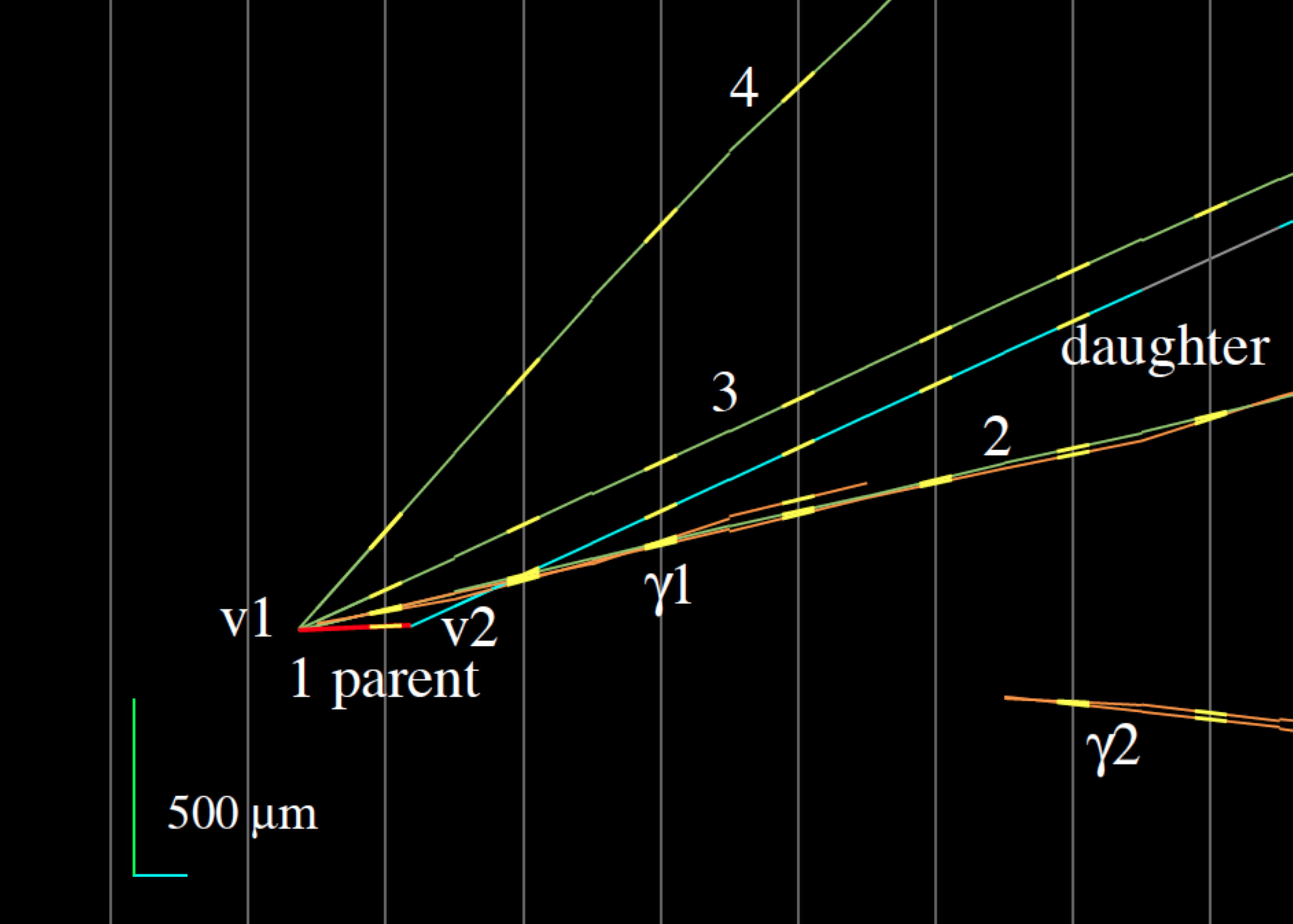}
  \caption{
Event display of the fourth $\nu_\tau$ candidate event from the OPERA 
experiment~\cite{DICRESCENZO2015186} in the
horizontal projection longitudinal to the neutrino direction.
The primary and secondary vertices are indicated as v$_1$ and
v$_2$, respectively. The kink between the parent and daughter track, a feature of $\tau$ lepton decays, is clearly visible. The yellow stubs represent the track segments as measured in the emulsion films.  Courtesy of the OPERA collaboration.
}
 \label{fig:opera}
 \end{figure}

The OPERA experiment~\cite{Agafonova:2015jxn} on the CERN to Gran Sasso neutrino beam, which took data between 2008 and 2012, was designed to test the $\nu_\mu \rightarrow \nu_\tau$ appearance hypothesis. The detector is based on the Emulsion Cloud Chamber technique, with 1,800 tons of nuclear emulsion detectors in the form of bricks, each brick being composed of a stack of nuclear emulsion films and lead plates. This target, capable of sub-micrometric track resolution, is devoted to the study of the neutrino interaction vertex and of the particles associated to it. The identification of the $\tau$ leptons relies mainly on their characteristic kink (Fig.~\ref{fig:opera}) due to the decay $\tau \rightarrow h \nu_\tau$, or $\tau \rightarrow l \nu_\tau \bar \nu_l$, where $h$ is a charged meson, and $l$ is an electron or a muon. Another signature is related to the decay $\tau \rightarrow 3 h \nu_\tau$ where the short $\tau$ track ends in a three-pronged vertex. The target detectors are complemented by scintillator trackers and muon spectrometers. 

OPERA has observed 5 $\nu_\tau$ candidate events~\cite{Agafonova:2015jxn} with a total background of 
$0.25 \pm 0.05$ events, mainly coming from decays of charmed particles. This corresponds to a 5.1 $\sigma$ observation of $\nu_\tau$ production in an oscillated $\nu_\mu$ beam. 

The Super-Kamiokande collaboration has also searched for $\nu_\tau$ appearance in multi-ring events due to atmospheric neutrinos to test the hypothesis of $\nu_\mu \rightarrow \nu_\tau$ oscillations~\cite{Abe:2012jj}. While the selected sample is affected by large backgrounds, there is an excess of tau-like events in the upward-going direction with a significance of 3.8 $\sigma$, offering a complementary confirmation of the OPERA result.  
 
\section{The 1-3 sector}

\subsection{Early limits and first indications}

The search for neutrino oscillation, and in particular for $\bar{\nu}_e$ disappearance, was  performed in the 1980s also using short-baseline experiments at nuclear reactors: Institut Laue-Langevin (ILL)~\cite{ill}, 
G\"{o}sgen~\cite{gosgen}, 
Bugey~\cite{bugey-1,bugey-2}, Rovno~\cite{rovno}, and Krasnoyarsk~\cite{krasnoyarsk} and later Savannah River~\cite{savannah}.
These experiments detected of the order of 10$^4$ events (10$^5$ events for the Bugey experiment) at a distance of 10-100 m from the reactor core. They did not find any evidence for oscillations and set limits down to $\Delta m^2$ of the order of 2 $\times$ 10$^{-2}$ eV$^2$.
In the 1990s, a second generation of experiments with Chooz~\cite{apollonio2003} and Palo Verde~\cite{paloverde} extended this search to a distance of the order of 1 km.
See Ref.~\cite{sblreview} for a review of these early reactor experiments.


After the discovery of atmospheric and solar neutrino oscillations, it became clear that the relevant oscillation length for the 1-3 sector of the PMNS matrix was governed by $\Delta m^2_{31}$, and that the  optimal baseline for a reactor neutrino experiment to see important spectral distortions due to the mixing angle $\theta_{13}$ was around 2 km.  Therefore, among the previous experiments, only Chooz and Palo Verde, with baselines close to 1 km, could set relevant limits in the context of three neutrino oscillations. The best limit on $\theta_{13}$, namely $\sin^2 2 \theta_{13}<0.17 $ at 90\% CL for $\Delta m^2_{31}=2.4 \times 10^{-3}$ eV$^2$, was held by the Chooz experiment~\cite{apollonio2003} for several years. 

Accelerator long baseline experiments also searched for $\theta_{13}$ in the appearance mode. At leading order, as we will describe in section~\ref{sec:questdelta}, the appearance probability reads $P(\nu_\mu \rightarrow \nu_e) = \sin^2 2 \theta_{13} \sin^2 \left( \frac {\Delta m^2_{31} L}{4E} \right)$. After the upper limits by K2K~\cite{k2knue} and MINOS~\cite{minosnue}, T2K was the first experiment to report, in 2011, an indication that $\theta_{13} \neq 0$ at $2.5\sigma$.
Prior to that, some hints of a nonzero value of $\theta_{13}$ were obtained from the combination
of atmospheric neutrino, solar neutrino and KamLAND data, with a statistical significance of about
$90\%$ C.L.~\cite{Fogli:2008jx} (see also the combinations of solar neutrino and KamLAND data
of Refs.~\cite{Gando:2010aa,aharmim}).
The $\theta_{13}$ mixing angle has subsequently been measured with excellent precision by the reactor experiments Daya Bay, with confirmations by the Reno and Double Chooz experiments.

\subsection{Reactor neutrinos: flux and detection}
\label{subsec:reactorflux}
Nuclear reactors are intense sources of electron antineutrinos. Indeed, the fission processes, starting from uranium enriched in the $^{235}U$ isotope, generate unstable neutron-rich nuclei that undergo $\beta^-$ decays. A typical example is $ {}^{235}{\rm U} + n \rightarrow {}^{140}{\rm Ba} + {}^{94}{\rm Kr} + 2n$, releasing an energy of 200 MeV. This is followed by six $\beta^-$ decays producing 6 electron antineutrinos. A 2800 MW reactor produces a flux of 5 $\times$ 10$^{20}$ $\bar{\nu}_e$ per second. 

\begin{figure}[htbp]
\centering
\includegraphics[width=0.6\linewidth]{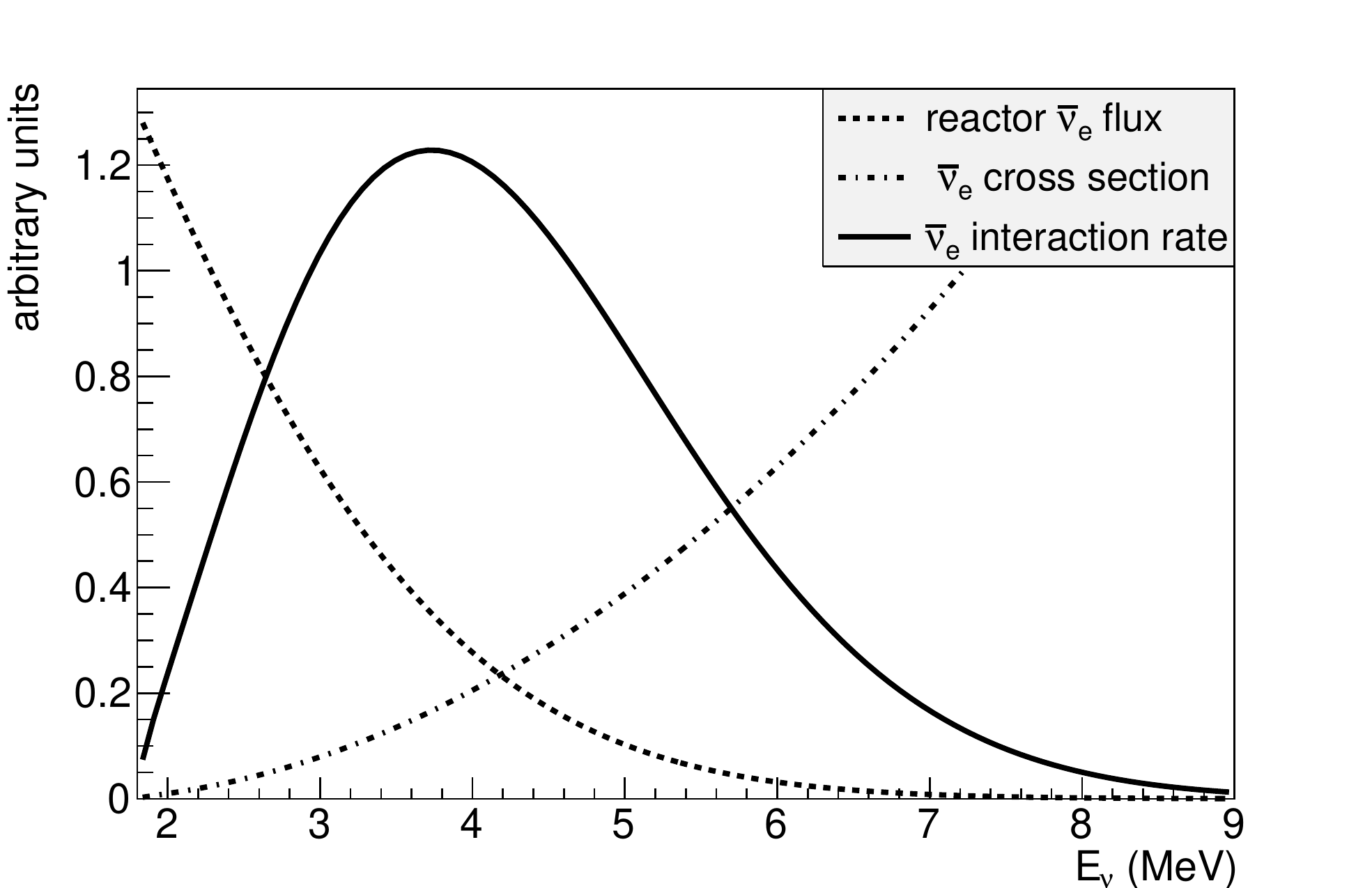}
  \caption{Typical reactor $\bar{\nu}_e$ flux, IBD cross-section and resulting interaction spectrum.}
 \label{fig:reactorflux}
 \end{figure}

The typical antineutrino energy is of the order of a few MeV. 
At these energies, an antineutrino interacting on a proton can initiate an Inverse Beta Decay (IBD) process $\bar{\nu}_e\, p \rightarrow n\, e^+$, which has a threshold of 1.8 MeV. The cross-section of this process can be calculated to high precision from the related neutron decay process~\cite{vogelbeacom, strumia}. The visible energy $E_{vis}$ produced by the positron, neglecting the small nuclear recoil, is related to the antineutrino energy $E_\nu$ by $E_{vis} =  E_{\nu} -(m_n - m_p) + m_e$, where $m_e$ is the electron mass. The convolution of the $\beta$ decay spectrum of antineutrinos (rapidly falling off with energy) with the IBD cross-section (rising with energy) results in a detected antineutrino spectrum peaking at approximately 4 MeV~(Fig.~\ref{fig:reactorflux}). 

In a scintillating medium the production of the positron can be easily detected  by ionization followed by annihilation. After thermalization, the neutron can be captured either on hydrogen or on nuclei providing larger neutron capture cross-sections, like cadmium or gadolinium. The capture is followed by the emission of several gamma rays. Depending on the capture nucleus and its concentration, the typical capture time can be of the order of a few tens of microseconds up to 200 $\mu $s.

The IBD and following neutron capture, taking place for instance in a liquid scintillator doped with gadolinium, is a very clean experimental method to detect antineutrinos. The positron signal followed by the signal related to the neutron capture in a delayed coincidence offers a clear and clean way to distinguish it from other background reactions. This experimental technique has been used since the discovery of the neutrino in the historic Savannah River experiment~\cite{reines56} and then by numerous experiments located close to nuclear reactors and searching for neutrino oscillations.

The precise calculation of the antineutrino flux from a reactor is a difficult task since it involves more than thousand beta decay branches involving unstable nuclei. For some of these little or no data is available. A series of experiments that took place in the 1980s at the ILL facility~\cite{ILL1,ILL2,ILL3} measured the electron spectrum emitted by the $\beta$ decays taking place in foils containing $^{235}$U, $^{239}$Pu, and $^{241}$Pu under a thermal neutron flux. The antineutrino spectrum can then be computed from the electron spectrum under some assumptions with the so-called inversion method, with a precision ranging from 2\% at 2 MeV to 10\% at 8 MeV. Ab-initio calculations were also performed recently (see Section~\ref{sec:anomalies}).   

\subsection{Results from reactor experiments}


In the PMNS framework, the $\bar \nu_e$ survival probability is given at leading order by Eq.~(\ref{eq:nuereactor}):
\begin{equation}
P (\bar \nu_e \rightarrow \bar \nu_e) \simeq 1 -\sin^2 2 \theta_{13} \sin^2 \left( \frac{\Delta m^2_{31} L}{4 E} \right) 
\end{equation}
for a baseline L and a neutrino energy E satisfying $\Delta m^2_{21} L/E \ll 1$.
The disappearance of $\bar \nu_e$ reaches a maximum at 2 km for $E_\nu=4$ MeV. A reactor neutrino experiment with a baseline of the order of 1-2 km offers therefore a good probe of the $\theta_{13}$ mixing angle.

In the last decade, a new generation of reactor experiments has been built to push the sensitivity even further: Double Chooz~\cite{dcproposal} in France, RENO~\cite{renoproposal} in South Korea and Daya Bay~\cite{DBdet} in China. The experimental features of these experiments are similar, and here we will describe the Daya Bay experiment.

The Daya Bay nuclear complex comprises six nuclear reactors with a total thermal power of 17.4 GW, grouped in two areas. Three underground experimental halls, two near sites close to the reactors and one far site, host eight identical antineutrino detectors (AD).
The flux-averaged baselines
for the three experimental halls are 520, 570 and 1590 m,
respectively.

Each AD consists of three concentric cylindrical volumes separated by acrylic vessels. The innermost volume contains 20 tons of gadolinium-loaded liquid scintillator, providing the target for the neutrino interactions. Most of the neutrons are captured inside this volume. The next volume is filled with liquid scintillator to detect the gammas and thereby reduce the energy leakage. The outermost volume is filled with mineral oil and provides a shield against the radioactivity produced by the PMTs and the walls of the tank, on which a total of 192 PMTs are mounted. The tank itself is immersed in ultra-pure water providing a shield against external radiation and cosmic rays that can be tagged by additional PMTs. 

After the selection of IBD events, more than 1 million candidate events have been detected. The largest background is due to accidental coincidences, followed by cosmogenic backgrounds and fast neutrons. In total the background represents 2\% (3\%) of the selected samples at the near (far) sites.

\begin{figure}[htbp]
\begin{minipage}[c]{.46\linewidth}
   	      \includegraphics[width=0.9\linewidth]{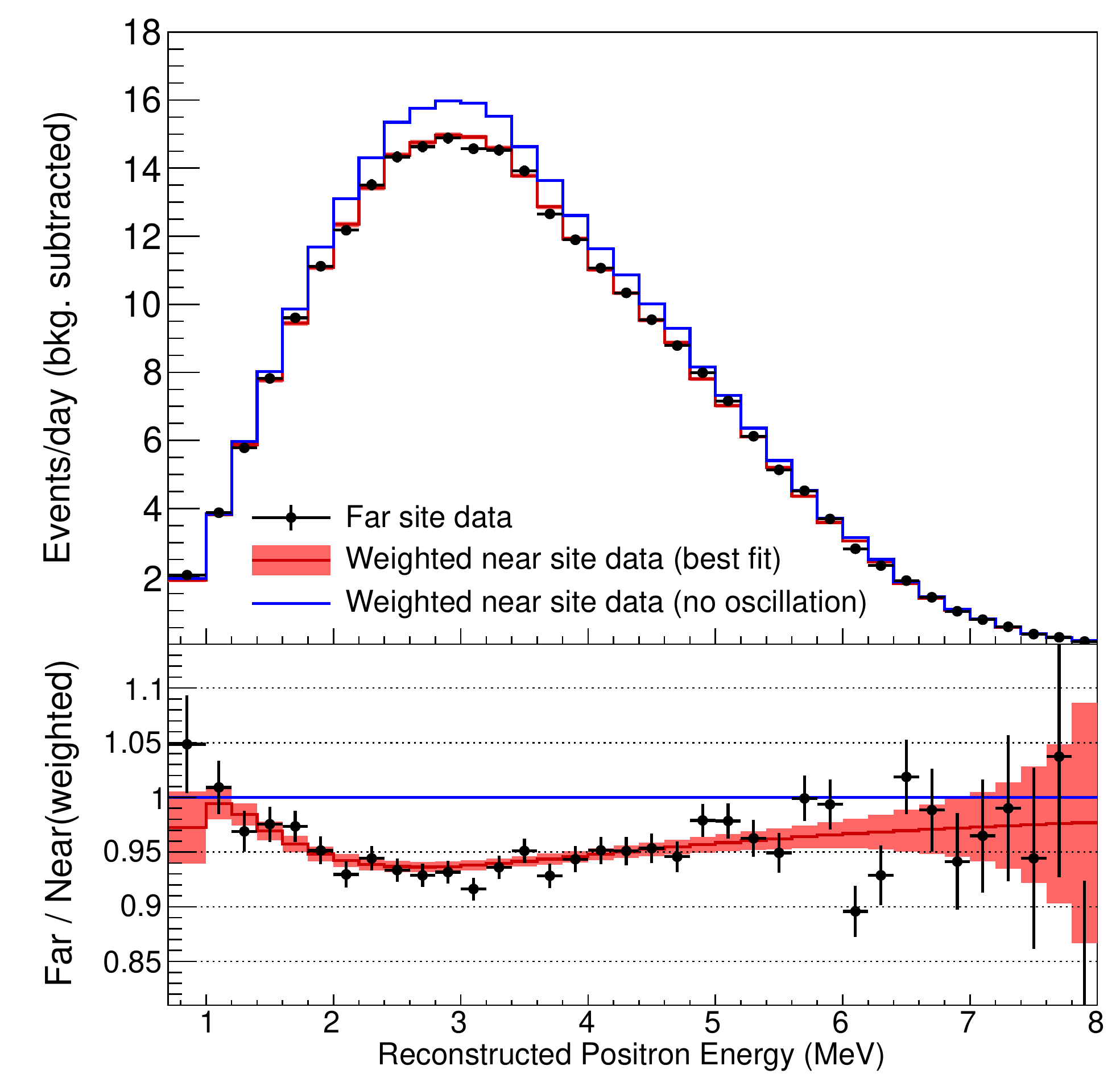}
   \end{minipage} \hfill
   \begin{minipage}{.46\linewidth}
      \includegraphics[width=0.9\linewidth]{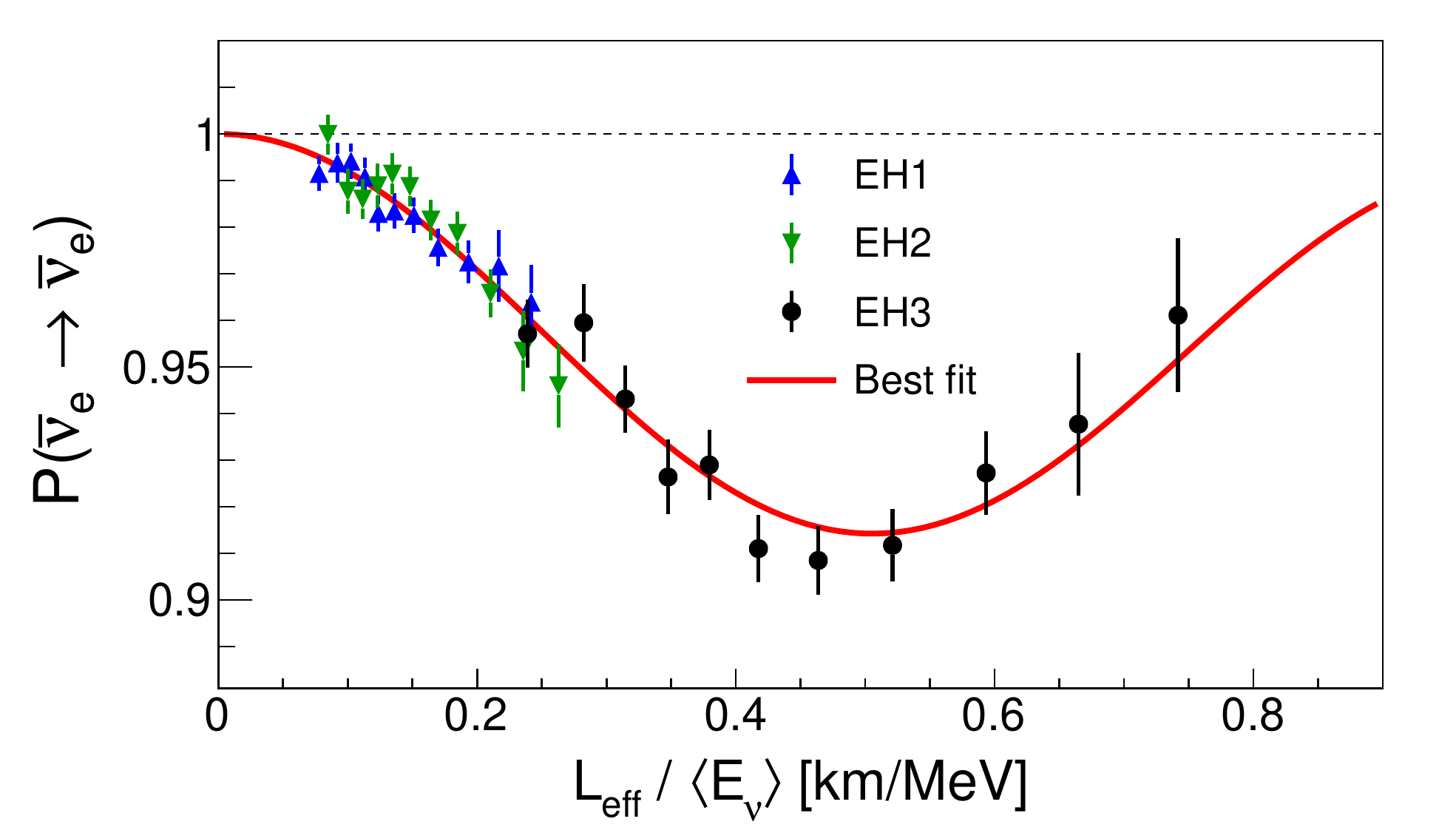}
   \end{minipage}
    \caption{Left plot: the reconstructed positron
energy spectrum observed by Daya Bay in the far site (black points), 
compared with the expectation derived from the near sites without (blue line) or
with (red line) oscillations. The lower plot shows the ratio of the spectra to the no-oscillation case and the shaded area 
includes the systematic and statistical uncertainties~\cite{An:2015rpe}. 
Right plot: Daya Bay electron antineutrino survival probability versus effective
propagation distance L divided by the average antineutrino energy
E~\cite{An:2015rpe}. Courtesy of the Daya Bay collaboration.
Reprinted figure with permission from F. P. An {\it et al.}, Phys. Rev. Lett., 115, 111802, 2015. Copyright 2015 by the American Physical Society.
}
 \label{fig:dayabay}
\end{figure}

Comparing the rate obtained in the near detectors to the rate in the far detectors (Fig.~\ref{fig:dayabay}), Daya Bay~\cite{DBth13} measures 
\begin{eqnarray}
\sin^2 2 \theta_{13} = 0.084 \pm 0.005 \\
\Delta m^2_{ee} = (2.42 \pm 0.11) \times 10^{-3} \: {\rm eV^2}
\end{eqnarray}
where $\Delta m^2_{ee}$ is the effective squared-mass difference governing short-baseline
$\bar \nu_e$ disappearance, which is given by
$\Delta m^2_{ee} = \cos^2 \theta_{12} \Delta m^2_{31} + \sin^2 \theta_{12} \Delta m^2_{32}$
to an excellent accuracy~\cite{parkemee} (see Section~\ref{subsec:3-flavour}).

The results from Double Chooz~\cite{doublechou},
$ \sin^2 2 \theta_{13}  = 0.088 \pm 0.033 $,
and RENO~\cite{RENO}, $ \sin^2 2 \theta_{13}  = 0.082 \pm 0.009 \pm 0.006 $, are in agreement with this result but with larger uncertainties. The mixing angle $\theta_{13}$ has thus recently gone from the last unknown angle in the PMNS matrix to the best known.

Precise measurements of the positron spectrum of IBD events have been compared with the expectation from the inversion method and found to disagree. The most notable feature is a shoulder around an energy of 5 MeV first observed by the RENO experiment~\cite{RENO,DBflux}. Background and detector effects have been disfavoured as a source of this discrepancy. 
As the measurement of $\theta_{13}$ is obtained by comparing the rate and shape measured at the far detector with those measured at the near detectors with minimal model dependence, it is practically not affected by this discrepancy. 

%

\section{Three-flavour effects}
\subsection{Electron neutrino appearance in long-baseline experiments}
\label{sec:questdelta}

As discussed in Section~\ref{subsec:CPV}, the measurement of nonzero values for all three mixing angles
$\theta_{12}$, $\theta_{23}$ and $\theta_{13}$ establishes a necessary condition for 
CP violation driven by the parameter $\delta_{CP}$.
This would be a genuine three-flavour effect. In the following, we will explain how these effects can be probed in a long-baseline neutrino experiment.

\begin{figure} [htbp!]
\begin{center}
\includegraphics[width=8cm]{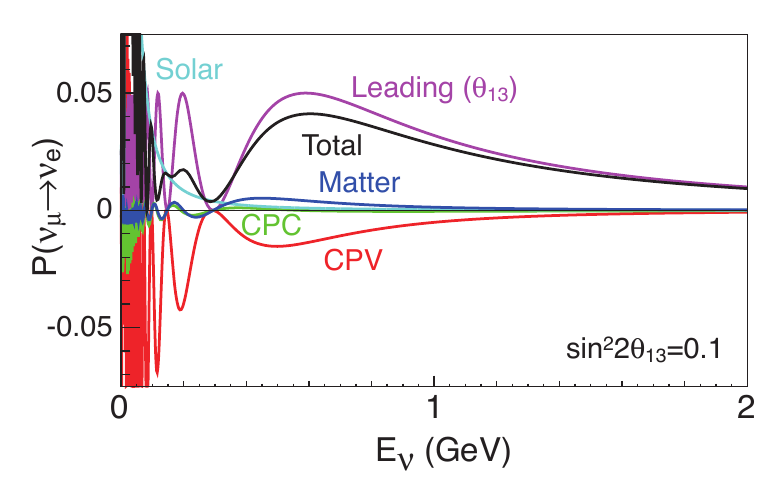}
\caption{\label{fig:t2kappprob} Oscillation probability $\nu_\mu \rightarrow \nu_e$ as a function of neutrino energy with L = 295 km, \stot = 0.1, \dcp= $\pi/2$ and normal mass ordering~\cite{hkpotential}. The contribution of each term in the oscillation probability~Eq.~(\ref{eq:theta13app}) is shown separately. Courtesy of the Hyper-Kamiokande collaboration.}
\end{center}
\end{figure}

The probability of electron neutrino appearance in a beam of muon neutrinos is given by Eq.~(\ref{eq:numu_nue_3f_matter}):
\begin{eqnarray}
\papp & = & \sin^2 \theta_{23} \frac{\sin^2 2 \theta_{13}}{(A-1)^2} \sin^2 [(A-1) \Delta_{31}] \nonumber \\
&+& \alpha^2 \cos^2 \theta_{23} \frac{\sin^2 2 \theta_{12}}{A^2} \sin^2 (A\Delta_{31}) \nonumber \\
&-& \alpha \frac{\sin 2 \theta_{12}\sin 2 \theta_{13}\sin 2 \theta_{23}\cos  \theta_{13}\sin  \delta_{CP}}{A ((1-A)} \sin \Delta_{31} \sin (A\Delta_{31})\sin[ (1-A) \Delta_{31}] \nonumber \\
&+& \alpha \frac{\sin 2 \theta_{12}\sin 2 \theta_{13}\sin 2 \theta_{23}\cos  \theta_{13}\cos  \delta_{CP}}{A ((1-A)} \cos \Delta_{31} \sin (A\Delta_{31})\sin[ (1-A) \Delta_{31}]
\label{eq:theta13app}
\end{eqnarray}
where $ \alpha = \Delta m^2_{21} / \Delta m^2_{31}$, $\Delta_{ji}= \Delta m^2_{ji} L / 4 E$ and $A= 2\sqrt 2 G_F n_e E/ \Delta m^2_{31}$.
The corresponding formula for \pappb can be obtained by reversing the signs of \dcp and $A$. The different terms contributing to \papp, which are described below, are plotted in Fig.~\ref{fig:t2kappprob} together with the total contribution from matter effects, assuming a baseline of 295 km.

The first term in Eq.~(\ref{eq:theta13app}) (labelled ``leading'' in Fig.~\ref{fig:t2kappprob}) is the leading order term and corresponds to the 1-3 sector oscillations driven by the squared-mass difference $\Delta m^2_{31}$. It should be noticed that it contains also a factor $\sin^2 \theta_{23}$ that is sensitive to the octant of $\theta_{23}$, while the leading term in the $\nu_\mu$ survival probability for atmospheric neutrinos and long-baseline experiments is proportional to $\sin^2 2 \theta_{23}$. In principle, a precise determination of the appearance probability would allow to determine the octant of $\theta_{23}$.

The second term (labelled ``solar'' in Fig.~\ref{fig:t2kappprob}) corresponds to the 1-2 sector oscillations and is suppressed because of the smallness of $ \alpha \simeq 0.03$. In other terms, for 
$\Delta m^2_{31} L / 4 E \simeq \pi/2$, which is the working point usually adopted for long-baseline experiments, the phase $\Delta m^2_{21} L / 4 E$ is lagging behind.

The third term (``CPV'') contains the CP-violating part and is proportional to the Jarlskog invariant $J$,
whose expression is given by Eq.~(\ref{eq:Jarlskog}).
It modifies the leading order term by as much as $\pm$ 30 \% for $\dcp = \mp \pi/2$, respectively, in the case of neutrinos.

The fourth term(``CPC'') is CP conserving, as it is proportional to $\cos \delta_{CP}$. Since its dependence on $\delta_{CP}$ is different from the one of the third term, it might be useful for resolving degeneracies. As this term is proportional to $\cos \Delta_{31}$, it is zero at the first vacuum oscillation maximum, $\Delta_{31}=\pi/2$. It must therefore be studied for L/E away from the first oscillation maximum. 

Matter effects enter through the $A$ term. It can be seen that $A$ is proportional to the ratio $E/E_{res}$, where $E_{res}$ is the energy for which the resonance condition is attained in the medium of constant density. Since the resonance condition can be satisfied either for neutrinos or for antineutrinos, depending on the sign of $\Delta m^2_{31}$ and therefore on the mass ordering, matter effects either enhance $ P (\nu_\mu \rightarrow \nu_e)$ or  $ P (\bar{\nu}_\mu \rightarrow \bar{\nu}_e)$. 
Values of $A$ for the various long-baseline experiments are shown in Table~\ref{tab:mateff}.
It can be seen that the matter effects are very small for T2K, sizeable for \nova and even more important for DUNE, as they enter non-linearly, notably through the factor $1/(A-1)^2$ in the first term of Eq.~(\ref{eq:theta13app}). Matter effects will either enhance or suppress the first term in Eq.~(\ref{eq:theta13app}). The separation between the two mass orderings will be more favourable for larger values of $\theta_{23}$ because of the factor $\sin^2 \theta_{23}$ appearing in the first term. 

\begin{table}
\caption{Matter effect parameter A for various baselines and energies for a constant matter density of 2.6 g/cm$^{-3}$.}
\centering
\begin{tabular}{|c|c|c|c|}
  \hline
  Experiment & L & E & A  \\ 
  \hline
  T2K & 295 & 0.6 & 0.046 \\
  \nova & 800 & 1.6 & 0.12\\
  DUNE & 1300  & 2.6  & 0.20 \\
  \hline
\end{tabular}

\label{tab:mateff}
\end{table}

On the other hand, CP-violating effects also enhance either the neutrino or the antineutrino appearance probability. Since the first term is proportional to
$\sin^2  \theta_{23}$, while the third, CP-violating term is proportional to 
$\sin 2 \theta_{23}$, the CP violation sensitivity is slightly better for lower values of $\theta_{23}$, for which the modulation induced by CP violation is larger.

In summary, a precise measurement of 
$ P (\nu_\mu \rightarrow \nu_e)$ and  $ P (\bar{\nu}_\mu \rightarrow \bar{\nu}_e)$
would allow to constrain the remaining unknowns in the PMNS matrix: the octant of $\theta_{23}$, $\delta_{CP}$ and the mass ordering, together with an independent determination of $\theta_{13}$. This explains why long-baseline neutrino experiments capable of carrying out this experimental program have received a lot of attention in the last years.

\begin{figure} [htbp!]
\begin{center}
\includegraphics[width=6cm]{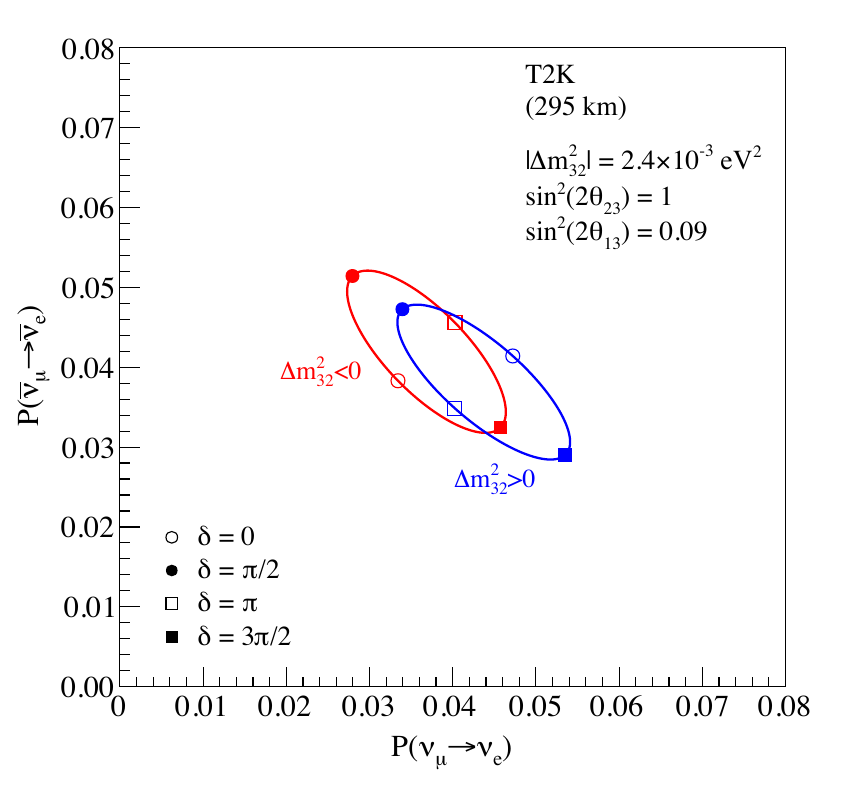}
\includegraphics[width=6cm]{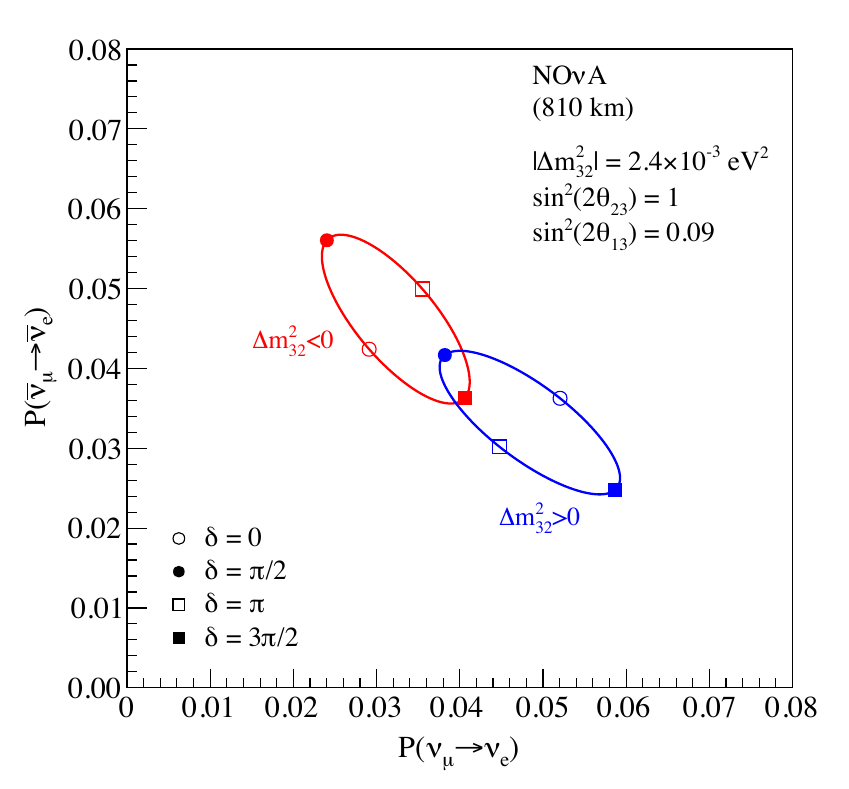}
\includegraphics[width=6cm]{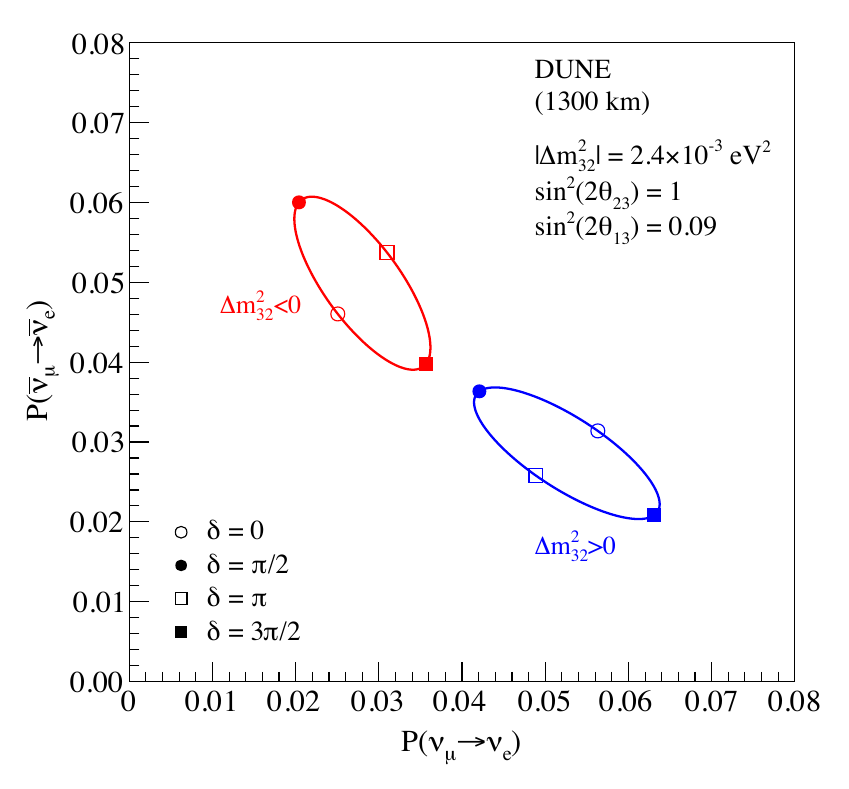}
\caption{\label{fig:novaellipse} Oscillation probabilities $ P (\nu_\mu \rightarrow \nu_e)$ and $ P (\bar{\nu}_\mu \rightarrow \bar{\nu}_e)$ for normal (blue) and inverted (red) mass ordering, assuming T2K (left), \nova (center) and DUNE (right) baselines and a representative L/E value of 0.4 km/MeV~\cite{Patterson:2015xja}. Courtesy of R. B. Patterson. 
Reproduced with permission of Ann. Rev. of Part. Sci., 65. Copyright by Annual Reviews.
}
\end{center}
\end{figure}

The interplay between the various terms in Eq.~(\ref{eq:theta13app}) can be best understood by plotting the probabilities $ P (\nu_\mu \rightarrow \nu_e)$ versus  $ P (\bar{\nu}_\mu \rightarrow \bar{\nu}_e)$ for fixed L and E, as shown in Fig.~\ref{fig:novaellipse}. The leading order term always lies on the diagonal. The CP-violating third term and the CP-conserving fourth term modulate the bi-probability in the shape of an ellipse as a function of \dcp.
Matter effects displace this ellipse either enhancing $ P (\nu_\mu \rightarrow \nu_e)$ or 
$ P (\bar{\nu}_\mu \rightarrow \bar{\nu}_e)$.

It has been pointed out that this experimental program is actually more difficult to achieve than it may appear because of multiple degeneracies~\cite{fogli-deg,burguet,minakata-deg,barger-deg}. 
A measurement at a single energy and baseline of $ P (\nu_\mu \rightarrow \nu_e)$ and  $ P (\bar{\nu}_\mu \rightarrow \bar{\nu}_e)$, even with infinite precision, would lead to multiple ambiguities in the determination of the underlying parameters. This is due to three independent two-fold parameter degeneracies: $(\delta_{CP},\theta_{13})$, the sign of $\Delta m^2_{31}$ (that is the mass ordering) and $(\theta_{23}, \pi/2-\theta_{23})$~\cite{barger-deg}. The first degeneracy is solved by precision measurements of $\theta_{13}$ at reactors. The second degeneracy is lifted by comparing measurements at different baselines, like T2K and \nova for the running experiments, or DUNE and Hyper-Kamionande in the next decade (see for instance Refs.~\cite{yasuda-deg,ghosh-deg}). Moreover, both DUNE and Hyper-Kamiokande are considering the possibility of multiple L/E within the same experiment. DUNE will achieve this with a wide-band beam, capable of some sensitivity at the second oscillation maximum. Hyper-Kamiokande is considering a second detector in Korea on the same beam line, but at a very different L around 1100~km. In addition, Hyper-Kamiokande will benefit from a large sample of atmospheric neutrinos. In any case, the precise determination of $\theta_{23}$ will become increasingly difficult if the true value lies very close to $\pi/4$~\cite{parkedeg}.

\subsection{Experimental results from T2K and \nova}

\begin{figure} [htbp!]
\begin{center}
\includegraphics[width=9cm]{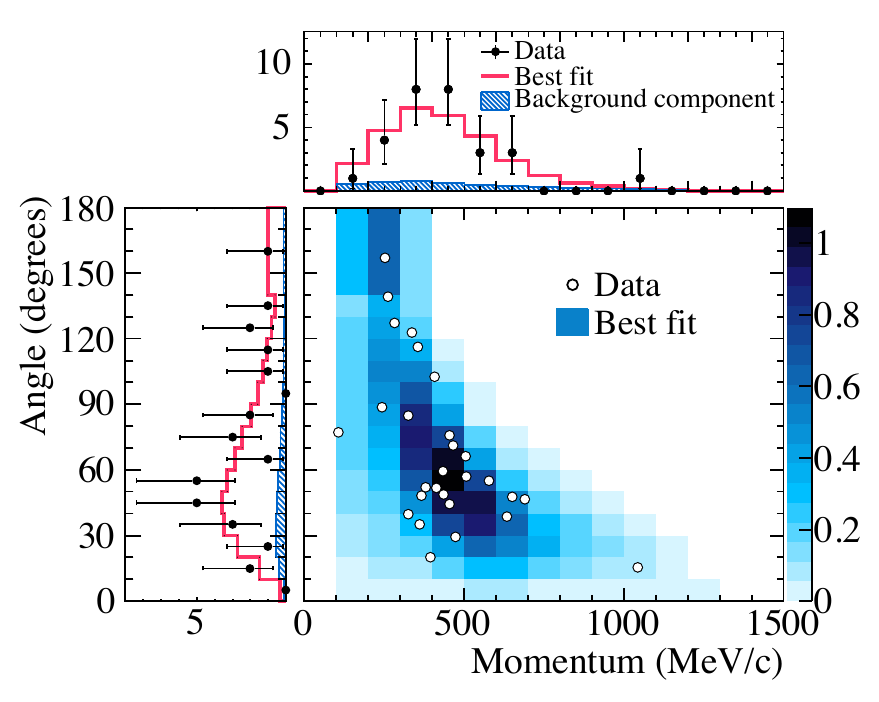}
\caption{\label{fig:t2kapp} Electron \ptheta distribution for the \nue candidate events observed in the T2K far detector~\cite{Abe:2013hdq}. Courtesy of the T2K collaboration.}
\end{center}
\end{figure}

After having obtained the first precious indication~\cite{t2k2011} of a non-zero value of \thint in 2011, the \nue appearance phenomenon has been observed for the first time by T2K in 2013~\cite{Abe:2013hdq}. A total of 28 electron neutrino candidates were detected in Super-Kamiokande, while $4.92\pm0.55$ background events were expected for $\thint=0$. The selection of these events in Super-Kamiokande is performed requiring fully contained, one-ring electron-like events. The background due to NC interactions with $\pi^0$ production is reduced by a factor 9 with a special algorithm~\cite{t2kprd} testing the hypothesis of two electromagnetic showers, corresponding to $\pi^0 \rightarrow \gamma \gamma$. The purity of the final sample is 80\% for $\nu_\mu \rightarrow \nu_e$ oscillated events, while the remaining background is mainly due to the irreducible beam $\nu_e$ background (15\%) and to NC events (4.4~\%). 
The distribution of lepton momentum and angle (with respect to the beam direction) for these events is shown in Fig.~\ref{fig:t2kapp}. The significance of this measurement corresponds to $7.3~\sigma$. 


In the coming years, before the new projects described in Section~\ref{sec:future} will come online, the quest for the unknown parameters in the PMNS scheme (\dcp, mass ordering, \thatm octant) will be led by the two long-baseline accelerator experiments that are currently taking data, namely T2K and \nova.

T2K has recently published its first oscillation analysis obtained combining data taken with neutrino and antineutrino beams with approximately the same amount of POT~\cite{t2k2016}. With this data set, 32 e-like candidates are observed at SK in neutrino mode and 4 in antineutrino mode. The expected number of events depends on the value of \dcp and on the mass ordering, as shown in Table~\ref{tab:evtnue}. They vary between 19.6 and 28.7 (17.1 and 25.4) for normal (inverted) ordering in neutrino mode and between 6.0 and 7.7 (6.5 and 7.4) in antineutrino mode. The value $\dcp=-\pi/2$ maximizes the \nue appearance probability and minimizes the \nueb appearance probability while the opposite happens for $\dcp=\pi/2$. 

\begin{table}[htbp]
    \centering
    \caption{Number of events for the $\nu_e$ and $\overline{\nu}_e$ selection in $\nu$ and $\bar{\nu}$ modes expected for various values of $\delta_{CP}$ and both mass orderings,
    compared with the observed numbers in the T2K experiment~\cite{t2k2016}.}
    \label{tab:evtnue}
    \begin{tabular}{|c|c|c|c|c|c|}
        \hline
        Normal & $\delta_{CP}= -\pi/2$ & $\delta_{CP}= 0 $ & $\delta_{CP}= \pi/2$ &  $\delta_{CP}= \pi$  & Observed\\
        \hline 
        $\nu$ mode &   28.7 & 24.2& 19.6& 24.1& 32 \\
        $\overline{\nu}$ mode &  6.0 &6.9& 7.7 &6.8 &4 \\     
        \hline
        \hline
        Inverted & $\delta_{CP}= -\pi/2$ & $\delta_{CP}= 0 $ & $\delta_{CP}= \pi/2$ &  $\delta_{CP}= \pi$  & Observed\\
        \hline 
        $\nu$ mode 			& 25.4 	& 21.3	& 17.1	& 21.3	& 32 \\
        $\overline{\nu}$ mode 	& 6.5 	& 7.4		& 8.4		& 7.4		&4 \\    
\hline
    \end{tabular}
\end{table}

\begin{figure}[htbp]
\begin{minipage}[c]{.46\linewidth}
   	      \includegraphics[width=0.9\linewidth]{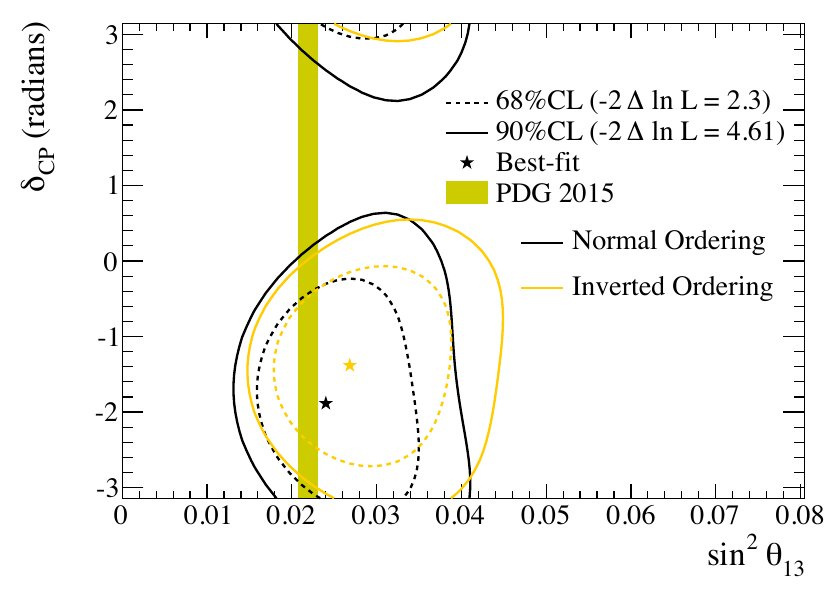}
   \end{minipage} \hfill
   \begin{minipage}{.46\linewidth}
      \includegraphics[width=0.9\linewidth]{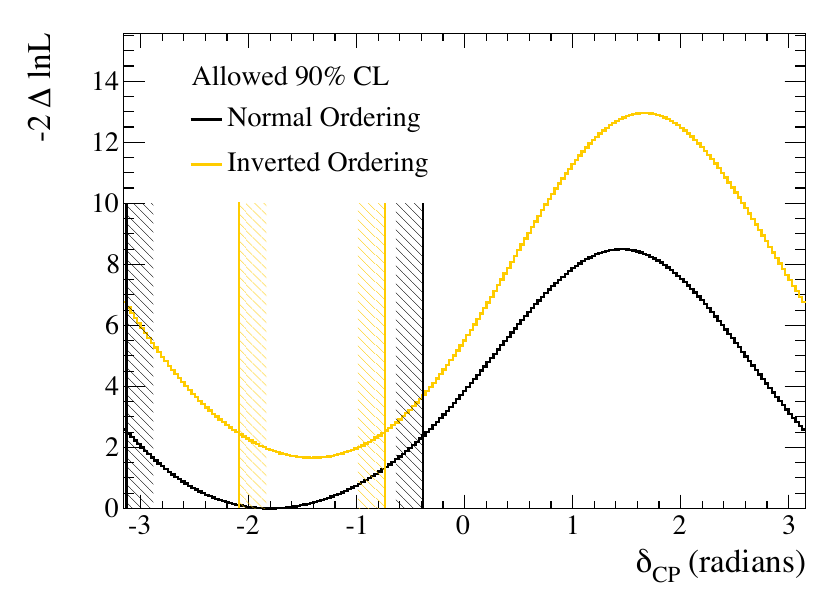}
   \end{minipage}
    \caption{Left plot: T2K 68\% and 90\%  confidence regions in the \dcp --$\sin^2 \theta_{13}$ plane~\cite{t2k2016} are shown by the dashed
(continuous) lines, computed independently for the normal
(black) and inverted (yellow) mass ordering. The best-fit point
is shown by a star for each mass ordering hypothesis. The
68\% confidence region from reactor experiments on $\sin^2 \theta_{13}$ is
shown by the yellow vertical band. 
Right plot: T2K constraint expressed as $-2\Delta ln L$ as a function of \dcp for the normal (black)
and inverted (yellow) mass ordering~\cite{t2k2016}. The vertical lines show the
corresponding allowed 90\% confidence intervals. The reactor measurement of   $\sin^2 \theta_{13}$ is used as prior. Courtesy of the T2K collaboration.
}
 \label{fig:t2kjoint}
\end{figure}

Since T2K observes a mild excess of \nue candidates with respect to the most favourable value and a deficit of \nueb candidates, a value of \dcp close to $-\pi/2$ and the normal ordering are preferred by the data even without the inclusion of the reactor constraint for the measurement of \thint. This is shown in Fig.~\ref{fig:t2kjoint}. For the first time, long-baseline experiments have a sensitivity to \dcp without considering independent informations from reactors, and a good agreement between the value of \thint measured by the reactors and the one measured by T2K is observed.

%

When the average value of \thint measured by the reactor experiments is used, additional information on the value of \dcp can be obtained. The T2K collaboration has obtained confidence intervals for \dcp (see Fig.~\ref{fig:t2kjoint}) and values conserving CP (\dcp= 0 or $\pi$) are excluded at 90\% C.L.

\begin{figure} [htbp!]
\begin{center}
\includegraphics[width=8cm]{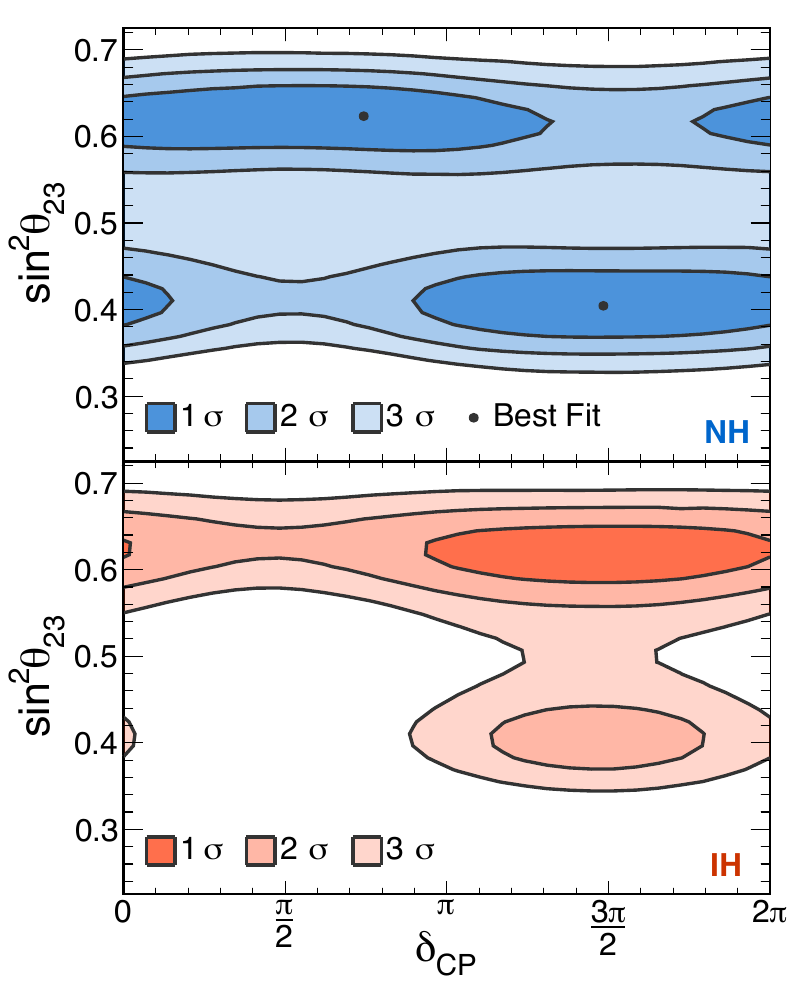}
\caption{\label{fig:novadcp}  Confidence regions of \dcp versus \stt consistent with the observed spectrum of \nue and \num candidates in \nova. The top panel corresponds to the normal mass hierarchy (NH) and the bottom panel to the inverted hierarchy (IH)~\cite{Adamson:2017gxd}. Courtesy of the \nova collaboration.
Reprinted figure with permission from P. Adamson {\it et al.}, Phys. Rev. Lett., 118, 231801, 2017. Copyright 2017 by the American Physical Society.}
\end{center}
\end{figure}

\nova has also released new results in 2016 for \nue appearance in the neutrino mode~\cite{Adamson:2017gxd}. The collaboration observed 33 \nue candidate events at the far detector with $8.2\pm0.8$ background events expected. The prediction for the signal events vary from 28.2 (normal ordering, $\dcp=-\pi/2$) to 11.2 (inverted ordering, $\dcp=\pi/2$). A joint analysis of appearance and disappearance channels has been performed and the results in the \stt-\dcp plane are shown in Fig.~\ref{fig:novadcp}. Given the non-maximal $\theta_{23}$ value found by \nova, there are two degenerate best fit points, both in
the normal ordering, (\stt = 0.404, \dcp = 1.48 $\pi$) and
(\stt = 0.623, \dcp = 0.74 $\pi$). This is a practical illustration of the degeneracy problem discussed previously.
The inverted mass ordering in
the lower octant is disfavoured at greater than 93\% C.L.
for all values of \dcp, and excluded at greater than 3 $\sigma$
significance outside the range 0.97 $\pi$~$<$~\dcp~$<$~1.94$\pi$.
These data are in agreement with T2K results but clearly both experiments are still dominated by the statistical uncertainty. The updates of these results will be of great interest for the determination of the remaining unknown neutrino oscillation parameters.


\subsection{Three-flavour effects with atmospheric neutrinos}

Three-flavour oscillation effects can also be studied with atmospheric neutrinos. The Super-Kamiokande collaboration has separated
its atmospheric neutrino data, registered since 1996 and corresponding to 5326 days, into three topologically distinct categories: fully contained, partially contained, and upward-going muons. Fully contained events are divided into subsamples based on their number of reconstructed Cherenkov rings, their
energy, the particle ID (PID) of their most energetic ring and the number of observed Michel electrons.  
After all selections there are 19 analysis samples (Fig.~\ref{fig:sk-atm}).
Some separation is possible between $\nu_e$ and $\bar{\nu}_e$ events and this enhances the sensitivity of the analysis.
The fit of these samples shows some preference for the normal mass ordering ($\chi^2_{IO}-\chi^2_{NO}=4.3$) and $\dcp = 3 \pi / 2$~\cite{li2016}.

\section{Global fits of the PMNS model}
\label{sec:summary}

The large set of experimental results on neutrino oscillations, with the exception of the few anomalies that will be presented in Section~\ref{sec:anomalies}, supports the global picture of three active neutrino mixing parametrized by the PMNS mixing matrix and two independent squared-mass differences.

\begin{figure}[htbp]
\begin{center}
   	      \includegraphics[width=0.8\linewidth]{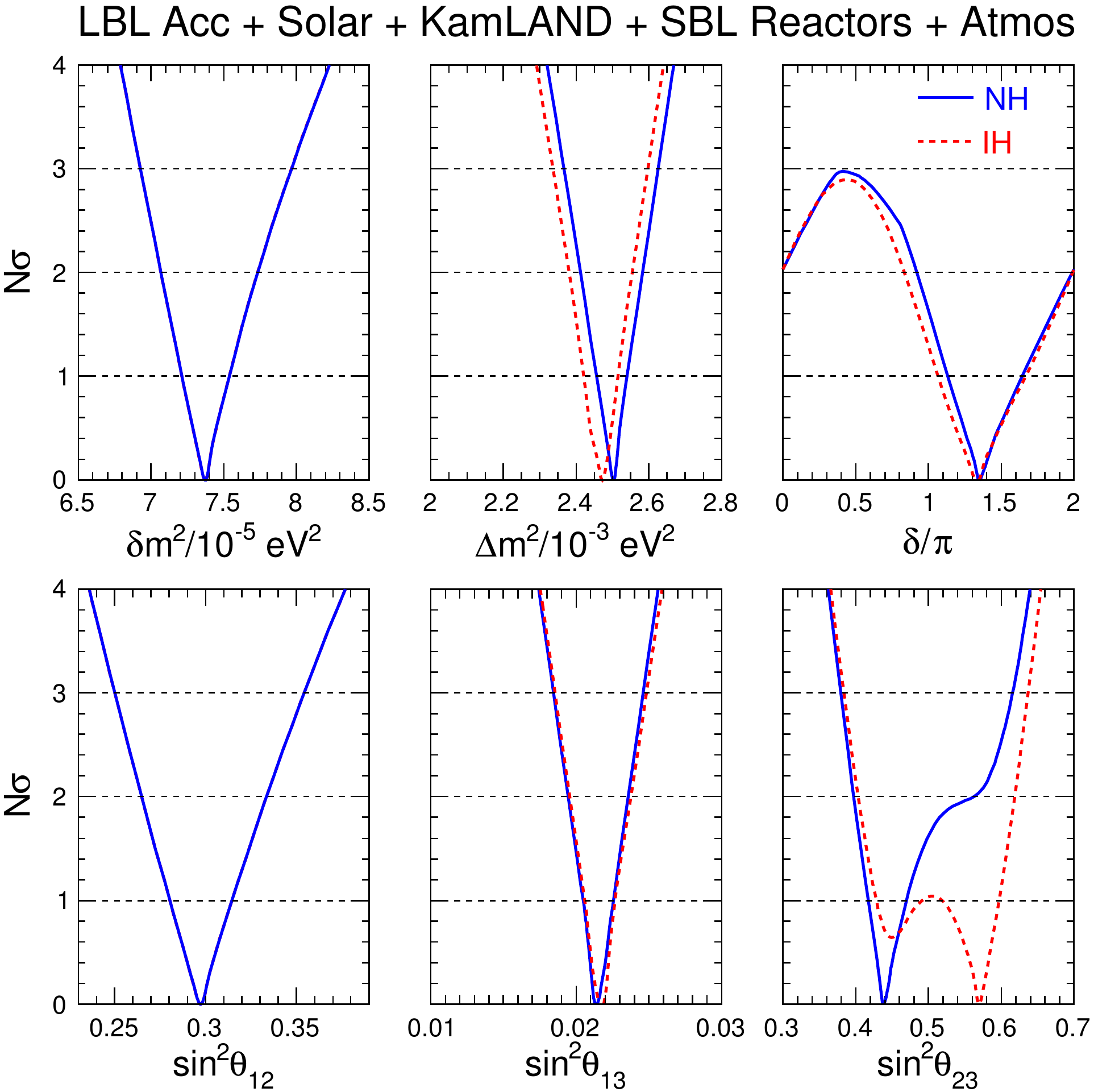}
    \caption{Bounds on the mass and mixing parameters in terms of
the number of standard deviations $\sigma$ from the best fit, for either normal ordering (solid lines) or inverted ordering (dashed lines)~\cite{Capozzi:2016rtj}. The bounds on $\Delta m^2_{21}$ and \thsol are
independent of the ordering. The horizontal dotted lines mark the 1-, 2- and 3$\sigma$ levels for each parameter. The parameters $\delta m^2$ and $\Delta m^2$ are defined as 
$\delta m^2 = m^2_2 - m^2_1$ ($ = \Delta m^2_{21}$) and $\Delta m^2 = m^2_3 - (m^2_2 + m^2_1)/2$
($ = (\Delta m^2_{31} + \Delta m^2_{32}) / 2$).
This fit does not include the latest 2016 T2K results for $\delta_{CP}$.
Courtesy of F. Capozzi et al.}
 \label{fig:globalfit}
 \end{center}
\end{figure}

Global fits of the neutrino oscillation data have been performed by three groups~\cite{nufit,Capozzi:2016rtj,valle-fit}, with results in good agreement.
The results of the most recent fits are reported in Table~\ref{tab:globalfit}~\cite{nufit} and in Fig~\ref{fig:globalfit}~\cite{Capozzi:2016rtj} and in the following of this section.

\begin{table}[htbp]
\centering
\begin{tabular}{|c|c|c|}
  \hline
  Parameter & Normal Ordering & Inverted Ordering  \\ 
  \hline
$\theta_{12}$ (deg)& $33.56\:^{+0.77}_{-0.75}$ &  $33.56\:^{+0.77}_{-0.75}$\\  
  $\theta_{23}$ (deg)& $41.6\:^{+1.5}_{-1.2}$ &  $50.0\:^{+1.1}_{-1.4}$\\  
  $\theta_{13}$ (deg)& $8.46\:^{+0.15}_{-0.15}$ & $8.49\:^{+0.15}_{-0.15}$ \\  
  $\delta_{CP}$ (deg)&  $261\:^{+51}_{-59}$& $277\:^{+40}_{-46}$ \\  
  $\Delta m^2_{21}$ ($10^{-5}$eV$^2$)& $7.50\:^{+0.19}_{-0.17}$ & $7.50\:^{+0.19}_{-0.17}$ \\  
  $\Delta m^2_{3l}$ ($10^{-3}$eV$^2$)&  $2.524\:^{+0.039}_{-0.040}$&  -$2.514\:^{+0.038}_{-0.041}$\\  
  \hline
\end{tabular}
\caption{
PMNS parameters determined by a recent global fit to the world neutrino data \cite{nufit} in the hypothesis of normal ordering (second column) and inverted ordering (third column). The parameter $\Delta m^2_{3l}$ is equal to $\Delta m^2_{31}$ for NO and to $\Delta m^2_{32}$ for IO. }
\label{tab:globalfit}
\end{table}

Presently, there is no strong preference for the normal or inverted ordering of the neutrino mass eigenstates. The measurement of the angles $\theta_{12}$ and 
$\theta_{13}$ and of the squared-mass differences has already reached the few percent precision level. This is not the case for the angle $\theta_{23}$, whose 3$\sigma$ range spans the interval [38,52] degrees (see also Fig.~\ref{fig:globalfit}) because of a mirror solution in the higher octant, and because the data do not exclude the possibility of a maximal $\theta_{23}$. 
The precise determination of $\theta_{23}$, of the CP-violating phase $\delta_{CP}$ and of the mass ordering remains a task for future experiments.

Another long-standing feature of the data is the minor (2$\sigma$) tension between the determination of $\Delta m^2_{21}$ by KamLAND on one side, and by the Super-Kamiokande, SNO and Borexino solar neutrino results on the other side.
This tension is related to the non-observation of the turn up on the lower part of the energy spectrum as predicted by the MSW effect. The observation of the day-night effect for solar neutrinos by Super-Kamiokande is also contributing to this tension.

The $3\sigma$ ranges for the magnitude of the entries of the PMNS matrix, as determined from the global fit
of Ref.~\cite{nufit}, are
\begin{equation}
|U| = \begin{pmatrix}
0.800 - 0.844 & 0.515 - 0.581 & 0.139 - 0.155 \\
0.229 - 0.516 & 0.438 - 0.699 & 0.614 - 0.790 \\
0.249 - 0.528 & 0.462 - 0.715 & 0.595 - 0.776 \\
\end{pmatrix}
\end{equation}
The Jarlskog invariant cannot be precisely determined today, however its maximum value, with $1\sigma$ uncertainty,
is~\cite{nufit}
\begin{equation}
J_{CP}^{max}
= 0.0329 \pm 0.0007 
\end{equation}
to be compared with the Jarlskog invariant of the CKM matrix
$J_{CP}^{CKM} = (3.04 \: {}^{+0.21} _{-0.20} ) \times 10^{-5}$~\cite{pdg}.

\section{Experimental anomalies beyond the PMNS model}
\label{sec:anomalies}

The vast majority of experimental measurement related to neutrino oscillations can be described in the PMNS framework as explained in the previous sections. There are however some unresolved anomalies: 
\begin{itemize}
\item The Liquid Scintillator Neutrino Detector (LSND)~\cite{lsnd}, collecting data between 1993 and 1998, reported an excess of $87.9 \pm 22.4 \pm 6.0$ events consistent with $\bar{\nu}_e p \rightarrow e^+ n$ scattering while studying $\bar{\nu}_\mu$ (endpoint energy 52 MeV) from $\mu^+$ decaying at rest, with a distance from the source to the liquid scintillator detector (167 tons) of 30~m. This might be interpreted as evidence for $\bar{\nu}_\mu \rightarrow  \bar{\nu}_e$ neutrino oscillations with $\Delta m^2$ in the 0.2-10 eV$^2$ range. 
Instead the KArlsruhe Rutherford Medium Energy Neutrino (KARMEN) experiment~\cite{karmen} at the spallation neutron source ISIS used $\bar{\nu}_\mu$  from $\mu^+$ decay at rest in the search for neutrino oscillations in the appearance mode $\bar{\nu}_\mu \rightarrow  \bar{\nu}_e$, and did not observe an indication for oscillation. 
\item The MiniBooNE experiment~\cite{miniboone1,miniboone2} was built at the Fermi National Accelerator Laboratory (FNAL) to confirm the LSND claim with a muon neutrino beam , whose peak energy is 600 MeV for neutrinos and 400 MeV for antineutrinos. It consists of a tank containing 806 tons of mineral oil equipped with 1,520 PMTs at 540~m from the beam target. It reported unexplained excesses in the low-energy region of electron neutrinos (162.0 $\pm$ 47.8 events) and antineutrinos (78.4 $\pm$ 28.5 events) at the 3.4 and 2.8$\sigma$ levels. The energy distribution of the excess in the neutrino channel is only marginally compatible with a simple two-neutrino effective model. 
\item The GALLEX and SAGE Gallium solar neutrino experiments have been calibrated using intense radioactive sources ($^{51}$Cr and $^{37}$Ar) placed inside the detectors~\cite{gallex-ga-1, gallex-ga-2, gallex-3, sage-ga-1,sage-ga-2, sage-ga-3,abdurashitov}. The measured event rate is lower than expected, with the ratio of observed over expected rate $0.86\pm 0.05$ deviating from 1 at the $2.8\sigma$ level~\cite{gallium}. This constitutes the Gallium anomaly. 
\item A recent re-evaluation~\cite{mueller} of the reactor neutrino flux, combining an {\it ab initio} calculation of the spectrum related to $^{238}$U and a revised $\beta$ inversion method for the other actinides, resulted in an upward shift of the normalization by 3\% on average. A comparison of the expected with the observed antineutrino rate in short-baseline reactor experiments (L $< 100$~m and down to 9~m for the ILL experiment~\cite{ill} and 15 m for the Bugey-4 experiment~\cite{mueller}) results in a ratio of $ 0.943 \pm 0.023$~\cite{bugey-2}. This constitutes the reactor anomaly. 
\end{itemize}

All these anomalies can be interpreted as hinting to the existence of oscillations of active neutrinos towards a sterile state with a $\Delta m^2$ around 1 eV$^2$. 
It must be stressed that, while all measurements related to the various sectors of the PMNS matrix have been confirmed by several experiments with very different techniques (for instance, the $ \theta_{23}$-induced oscillations have been observed by experiments using atmospheric neutrinos and by long-baseline experiments), each of these anomalies is rather isolated. Moreover, the statistical significance is far from being compelling and does not exceed 3$\sigma$. 
Repeated and numerous attempts have been made to interpret these anomalies in models where the three known active neutrinos mix with a fourth sterile state (3+1) or with two new sterile states (3+2).
 In the 3+1 model, in the limit where the $\Delta m^2_{21}$- and $\Delta m^2_{31}$-driven oscillations can be neglected, the $\nu_e$ ($\bar \nu_e$) survival probability is given by Eq.~(\ref{eq:SBL_disappearance}), namely
\begin{equation}
P(\nu_e \rightarrow \nu_e)
= 1 - \sin^2 2\theta_{ee} \sin^2 \left( \frac{\Delta m^2_{41}L}{4E} \right)
\end{equation}
where the effective mixing angle $\theta_{ee}$ is defined as $\sin^2 2 \theta_{ee} = 4 |U_{e4}|^2 (1 - |U_{e4}|^2)$.
On the other hand, the $\nu_\mu \to \nu_e$ appearance probability is given by Eq.~(\ref{eq:SBL_appearance}):
\begin{equation}
P(\nu_\mu \rightarrow \nu_e)
= \sin^2 2 \theta_{\mu e} \sin^2 \left( \frac{\Delta m^2_{41}L}{4E} \right)
\end{equation}
where the effective mixing angle $\sin^2 2 \theta_{\mu e} = 4 |U_{e4} U_{\mu 4}|^2$ depends on the two
$4 \times 4$ PMNS matrix entries $U_{e4}$ and $U_{\mu 4}$. Therefore, to explain the results of the LSND
and MiniBooNE experiments, both $U_{e4}$ and $U_{\mu 4}$ must be different from zero.
The formulae are more complex for the 3+2 model, which contains a larger number of free parameters.
The new feature of this scheme with respect to the 3+1 model is that it allows for CP violation in short-baseline
appearance experiments (see subsection~\ref{subsec:steriles}).

Global fits to the world dataset including these anomalies and all other experimental results on neutrino oscillation measurements and searches have been attempted \cite{giuntirev,kopp}. The result is that there is a strong tension between appearance experiments like LSND and MiniBooNE and the negative results from disappearance experiments like CERN-Dortmund-Heidelberg-Saclay (CDHS)~\cite{dydak}. Oscillations in the $\bar{\nu}_\mu \rightarrow  \bar{\nu}_e$ channel inevitably entail disappearance in the 
$\bar{\nu}_\mu \rightarrow  \bar{\nu}_\mu$ channel with the same L/E range. This disappearance has not been observed. The compatibility of appearance and disappearance data is at the level of $10^{-4}$ for the 3+1 model~\cite{kopp}.
More complex schemes like the 3+2 model, which was considered initially to reduce the tension between
the neutrino and antineutrino MiniBooNE data (this tension has almost completely disappeared
from the final data, reducing the motivation for this model),
do not improve the compatibility between appearance and disappearance data.
The authors of Ref.~\cite{giuntirev} propose to exclude the MiniBooNE low-energy bins to improve the goodness of fit.

\begin{figure}[htbp]
\begin{minipage}[c]{.46\linewidth}
   	      \includegraphics[width=0.9\linewidth]{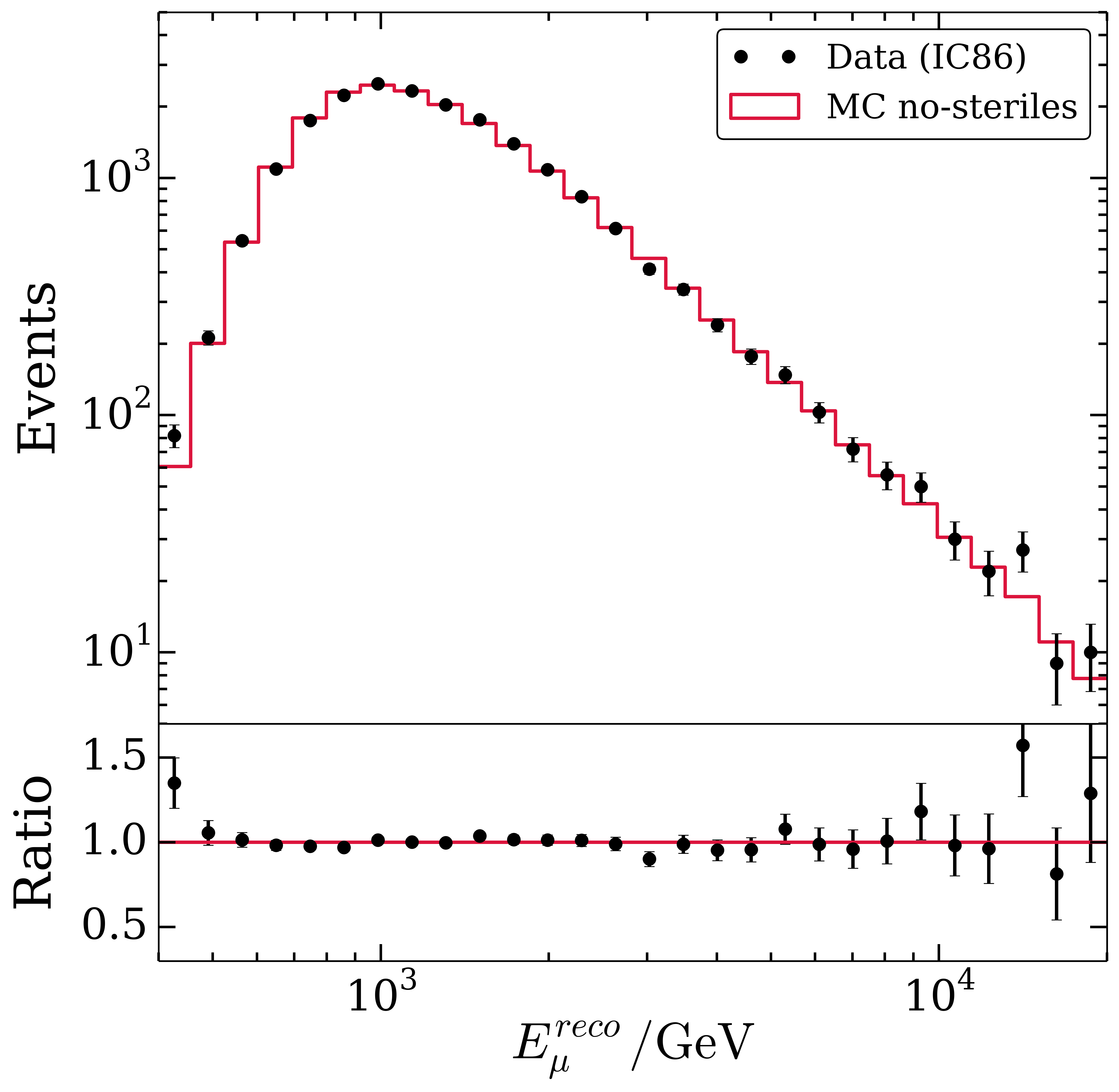}
   \end{minipage} \hfill
   \begin{minipage}{.46\linewidth}
      \includegraphics[width=0.9\linewidth]{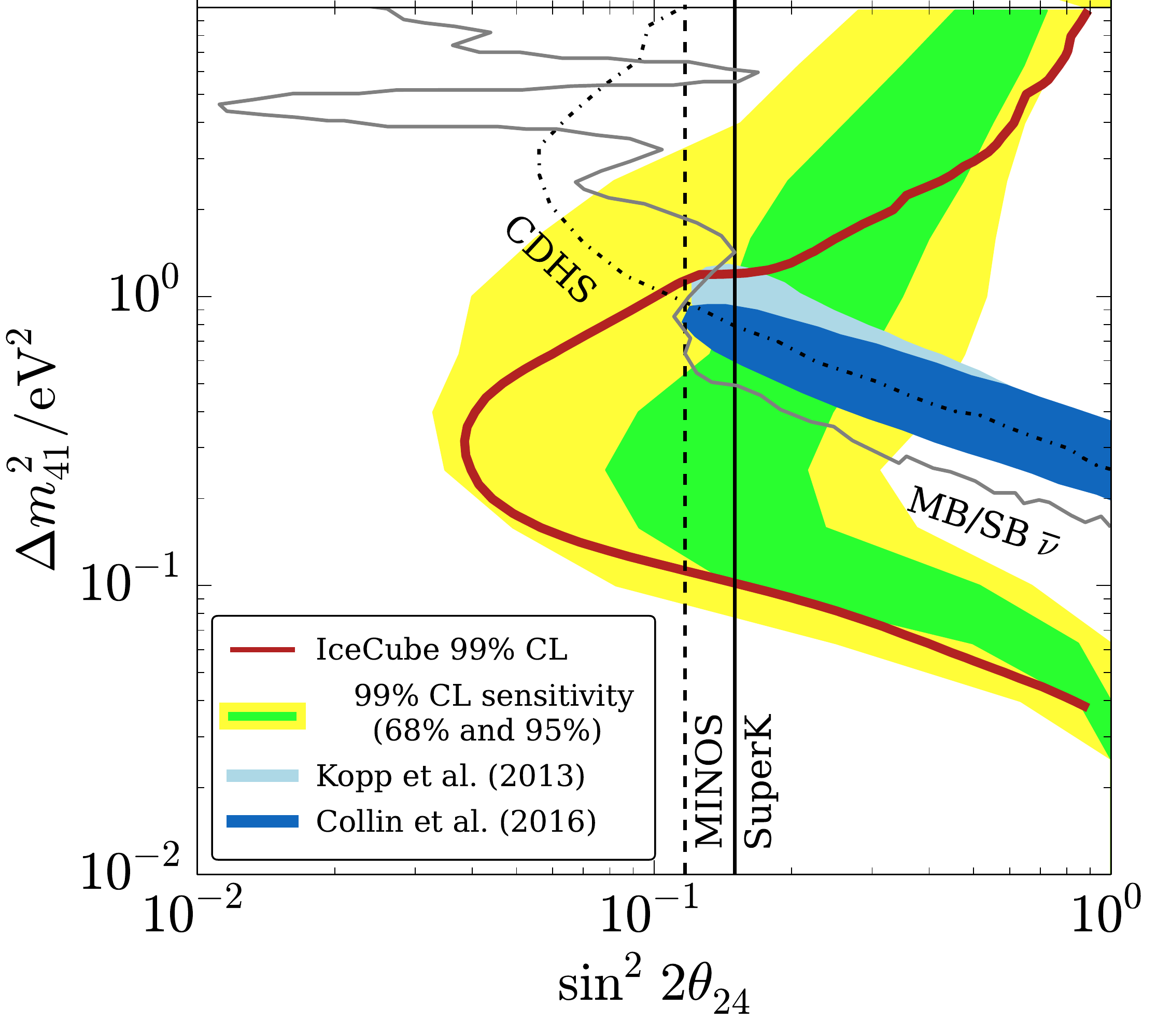}
   \end{minipage}
    \caption{Left plot: the reconstructed energy distribution of atmospheric $\nu_\mu$ events observed by the IceCube experiment compared to the prediction without sterile neutrinos~\cite{TheIceCube:2016oqi}. 
Right plot: the 99\% CL contour (red solid line) in the $\Delta m^2_{41}$ versus $\sin^2 2 \theta_{24}$ plane obtained by the IceCube experiment~\cite{TheIceCube:2016oqi} is shown with bands containing
68\% (green) and 95\% (yellow) of the 99\% CL contours in simulated pseudo-experiments. The contours and
bands are overlaid on 90\% CL exclusions from previous experiments~\cite{Abe:2014gda, Adamson:2011ku, Cheng:2012yy, dydak} and the MiniBooNE / LSND 90\% CL allowed
region with different assumptions on  $|U_{e4}|^2$~\cite{Kopp:2013vaa,Conrad:2012qt}. Courtesy of the IceCube collaboration.
Reprinted figure with permission from M. G. Aartsen {\it et al.}, Phys. Rev. Lett., 117, 071801, 2016. Copyright 2016 by the American Physical Society.
}
 \label{fig:icecube-sterile}
\end{figure}

Very recent data actually bring more experimental evidence against the LSND and MiniBooNE anomalies.
The IceCube has studied the atmospheric muon neutrino spectrum in the energy range 320 GeV to 20 TeV~\cite{icecubeste}. Muon neutrino mixing with a sterile state with $\Delta m^2_{41}$ in the 1 eV$^2$ range would significantly deplete the antineutrino spectrum in the 3+1 model, due to MSW resonance effects related to neutrino propagation in the Earth. The observed deficit can reach 11\% for the bin with the largest effect in the distribution of reconstructed energy and zenith angle. Similar effects would deplete the neutrino spectrum in the 1+3 model. No depletion is observed in a sample of 20,145 well reconstructed muons registered with an 86 string configuration in 2011 and 2012. This allows to set strong limits in the 
$\Delta m^2_{41}$ versus $\sin^2 2 \theta_{24}$ plane. The allowed region from the appearance experiments LSND and MiniBooNE is excluded at the 99\% CL (Fig.~\ref{fig:icecube-sterile}).  

An interesting limit has been set by a combined analysis of the MINOS, Daya Bay and Bugey-3 data~\cite{dayabaycomb}. MINOS is sensitive to a sterile neutrino as active (mainly $\nu_\mu$ in this case)-sterile mixing would modify the shape and normalization of charged current and neutral current events in the near and far detectors.
It can therefore set limits on $|U_{\mu 4}|^2$. Bugey-3 and Daya Bay data can be used to constrain electron antineutrino disappearance and place limits on $|U_{e 4}|^2$. The combination of these three experiments excludes the phase space allowed by the LSND and MiniBooNE experiments for $\Delta m^2_{41}$ below 0.8 eV$^2$ at 95\% CL . 

The reactor and Gallium anomalies are not in such a conflict with other measurements, since they do not require $U_{\mu 4}$ to be different from zero. In this sector, recent developments have focussed on the reliability of the systematic uncertainty related to the understanding of the reactor antineutrino flux. The observation of the 5 MeV shoulder~\cite{RENO,DBflux} has indeed brought under close scrutiny the method to compute the reactor antineutrino flux. The conclusions of recent studies \cite{hayes,vogel} are that several uncertainties might have been underestimated. For instance, 30\% of the flux comes from forbidden transitions. The effect of various shape corrections applied to forbidden transitions has been investigated and it leads to a 4-5\% uncertainty on the reactor antineutrino flux. Taking into account this uncertainty would considerably release the tension at the origin of the reactor anomaly.

A new generation of reactor neutrino experiments (STEREO~\cite{Manzanillas:2017rta}, SoLiD~\cite{Ryder:2015sma},
Neutrino-4~\cite{Serebrov:2017wml}, DANSS~\cite{Alekseev:2016llm}, NEOS~\cite{Ko:2016owz},
PROSPECT~\cite{Ashenfelter:2015uxt}, see Ref.~\cite{othervsbl} 
for a review and references therein) has been built to investigate in detail the possible mixing of $\nu_e$ with a sterile state with $\Delta m^2_{41}$ in the eV$^2$ range.
Indeed, $\bar{\nu}_e \rightarrow \bar \nu_s$ oscillations would induce a deformation of the spectrum if the detector is placed close enough (L~$\simeq$~10~m) to the reactor core. Several of these experiments have started data taking in 2016 or 2017, and new results on this hot topic are expected soon.
A different technique~\cite{cribier} is used by the 
Short distance neutrino Oscillations with
BoreXino
(SOX) experiment~\cite{sox} that will deploy an intense $^{144}$Ce antineutrino source very close to the Borexino detector with data taking foreseen in 2018.

The three Liquid Argon detectors Short-Baseline Neutrino Program (SBN)~\cite{sbnfnal} is under construction at FNAL. It comprises the 170 t MicroBooNE Liquid Argon detector~\cite{microboone}, built on the Booster Neutrino beam (the same beam used by MiniBooNE) at L = 470 m from the target. It will be completed by the refurbished 760 t ICARUS detector~\cite{icarus} placed at 600~m, and by the 220 t LAr1-ND near detector~\cite{lar1nd} at 110~m from the production target, for the investigation of the reported anomalies both in the appearance and in the disappearance channels. This set of detectors will be able to probe the LSND appearance excess with a sensitivity of 5$\sigma$. The installation will be completed in 2018.

\section{Next steps in the investigation of the PMNS paradigm}
\label{sec:future}

After the precise measurement of \thint, we know that all three mixing angles of the PMNS matrix are different from zero, opening the possibility for CP violation in the lepton sector. The discovery of this new source of CP violation, corresponding to the observation of a value of sin \dcp different from zero, could be an important step in our understanding of the baryon asymmetry in the Universe and will be the main goal of the long-baseline accelerator experiments in the coming years. As of today, indications from T2K show a weak preference for $\dcp\sim-\pi/2$, excluding at more than 3$\sigma$ values of \dcp close to \pipi/2 and excluding CP conservation at 90\% CL~\cite{t2k2016}. 

In addition to the measurement of \dcp, other questions still have to be addressed before having a complete picture of neutrino oscillations: in particular, the mass ordering between the mass eigenstate $\nu_3$ and the other two is still unknown. The ordering can be either normal ($m_3 > m_2 > m_1$) or inverted ($m_2 > m_1 > m_3$), and knowing it will have important consequences for further experimental studies of the neutrinos, and in particular for mass measurements (see Ref.~\cite{capozzi-masses} for a discussion of the current experimental situation).
Different scenarios are open for the neutrino mass spectrum~\cite{pdg}, depending on the mass ordering and on the lightest neutrino mass. In the normal hierarchical scenario, $m_1 \ll m_2 < m_3$ with $m_2 \simeq \sqrt{\Delta m^2_{21}} \simeq 8.6 ~{\rm meV}$ and  $m_3 \simeq \sqrt{\Delta m^2_{31}} \simeq 50 ~{\rm meV}$. The inverted hierarchical scenario is characterized by
$m_3 \ll m_1 \simeq m_2$, with $m_{1,2} \simeq \sqrt{\Delta m^2_{32}} \simeq 50 ~{\rm meV}$.
It is also possible that there is no strong hierarchy between the lightest and the next-to-lightest neutrino masses,
i.e. $m_1 \lesssim m_2 < m_3$ (normal ordering case),
 or $m_3  \lesssim m_1 \simeq m_2$ (inverted ordering).
Finally, in the quasi-degenerate scenario, one has $m_1 \simeq m_2 \simeq m_3$, with $m_1 \gtrsim 100~{\rm meV}$
and any of the two mass orderings.

The experiments searching for neutrinoless double beta decay (see e.g. Ref.~\cite{Gomez-Cadenas:2015twa} for a review) are sensitive to the effective Majorana mass $|m_{\beta \beta}|$, which, for inverted mass ordering, is larger than about 15 meV. Current experiments might be close to explore the upper part of the $|m_{\beta \beta}|$ region for inverted ordering, even for a lightest neutrino of zero mass. 
On the contrary, new techniques and much larger detector masses are needed to explore the $|m_{\beta \beta}|$ region corresponding to normal ordering.

Observational cosmology can probe the sum of the neutrino masses (see e.g. Ref.~\cite{Lesgourgues:2012uu} for a review), which for inverted ordering is larger than 100 meV, while the lower bound for normal ordering is about 59 meV. Also in this case, cosmological probes are getting closer to be sensitive to inverted ordering~\cite{Ade:2015xua}. The absolute neutrino mass scale will also be probed in the near future by the 
Karlsruhe Tritium Neutrino Experiment
(KATRIN), which will measure precisely the end-point of the tritium beta spectrum~\cite{katrin-dr}.
The knowledge of the ordering is also important for model builders. In fact, most Grand Unified models predict normal ordering~\cite{albright}.

The mass ordering has also a direct impact on \dcp. As mentioned in Section~\ref{sec:questdelta}, degeneracies exist in the \nue appearance probability in long-baseline experiments between \dcp and the mass ordering. If the latter can be determined, a major source of degeneracy will be removed and the measurement of \dcp will be easier.

Another important question to be addressed is the precise measurement of \thatm. Currently, the value of this mixing angle is compatible with $\pi/4$, i.e. with maximal disappearance, since the \num disappearance probability is proportional to \sttt.  Values of \thatm of $\pi/4+X$ and $\pi/4-X$ give the same disappearance probability. This is known as the octant degeneracy. Nevertheless, precise measurements of different oscillation channels allow to disentangle among the two solutions and determine the octant of \thatm if $X$ is large enough. A value of \thatm exactly equal to $\pi/4$ is particularly interesting because it might indicate the existence of an underlying symmetry.

This section is organized as follows. We first review the sensitivity of the running long-baseline experiments T2K and \nova, which will give some indications of CP violation, the mass ordering and the \thatm octant in the coming years. We then review the non-accelerator based experiments JUNO, PINGU, ORCA and INO, which have some sensitivity to the mass ordering, and finally, we describe the future long-baseline accelerator experiments Hyper-Kamiokande and DUNE, whose main goal is to observe CP violation in the lepton sector, performing a precise measurement of \dcp.

\subsection{Goals of the running experiments: T2K and \nova}
\label{sec:futuret2knova}

In the next 10 years there will be no new long-baseline experiment; however, the running experiments T2K and \nova will greatly increase their statistics, collecting 10 times more data with respect to what has been collected and analysed so far. 

T2K has presented results based on a total of $1.5 \times10^{21}$ \pot, while the total expected statistics for which the experiment had been approved is $7.8\times10^{21}$ \pot. Since the T2K sensitivity to \dcp will be still statistically limited at $7.8\times10^{21}$ \pot, this experiment will benefit from a larger exposure. 

For this reason, the collaboration has recently submitted a proposal for the T2K phase II~\cite{t2k2prop}. During this phase, which will take place from 2020 to 2026, a total of  $20\times10^{21}$~\pot will be collected. This will allow to exclude CP conservation at more than 3 sigma if \dcp is close to one of the two values that violate CP maximally (namely, $\dcp=-\pi/2$ and $+\pi/2$).

The other running experiment, \nova, has so far presented results based on $6\times10^{20}$ \pot while the experiment is expected to collect $36\times10^{20}$ \pot by 2020. 

The two experiments are highly complementary thanks to their different baselines. The longer \nova baseline allows this experiment to have a better sensitivity to matter effects and therefore to the mass ordering, while for T2K matter effects are small and the experiment is mostly sensitive to \dcp. In the case of \nova, in fact, the two mass ordering solutions lead to a difference in the \nue appearance probability of $\pm19\%$, while the two cases of maximal CP violation, $\dcp=-\pi/2$ and $+\pi/2$, lead to a difference in the \nue appearance probability of $\pm22\%$. In the case of T2K, instead, the mass ordering has an effect of $\pm10\%$ on the appearance probability, while the two maximal CP violation cases lead to a difference of $\pm29\%$. 

\begin{figure} [htbp!]
\begin{center}
\includegraphics[width=8cm]{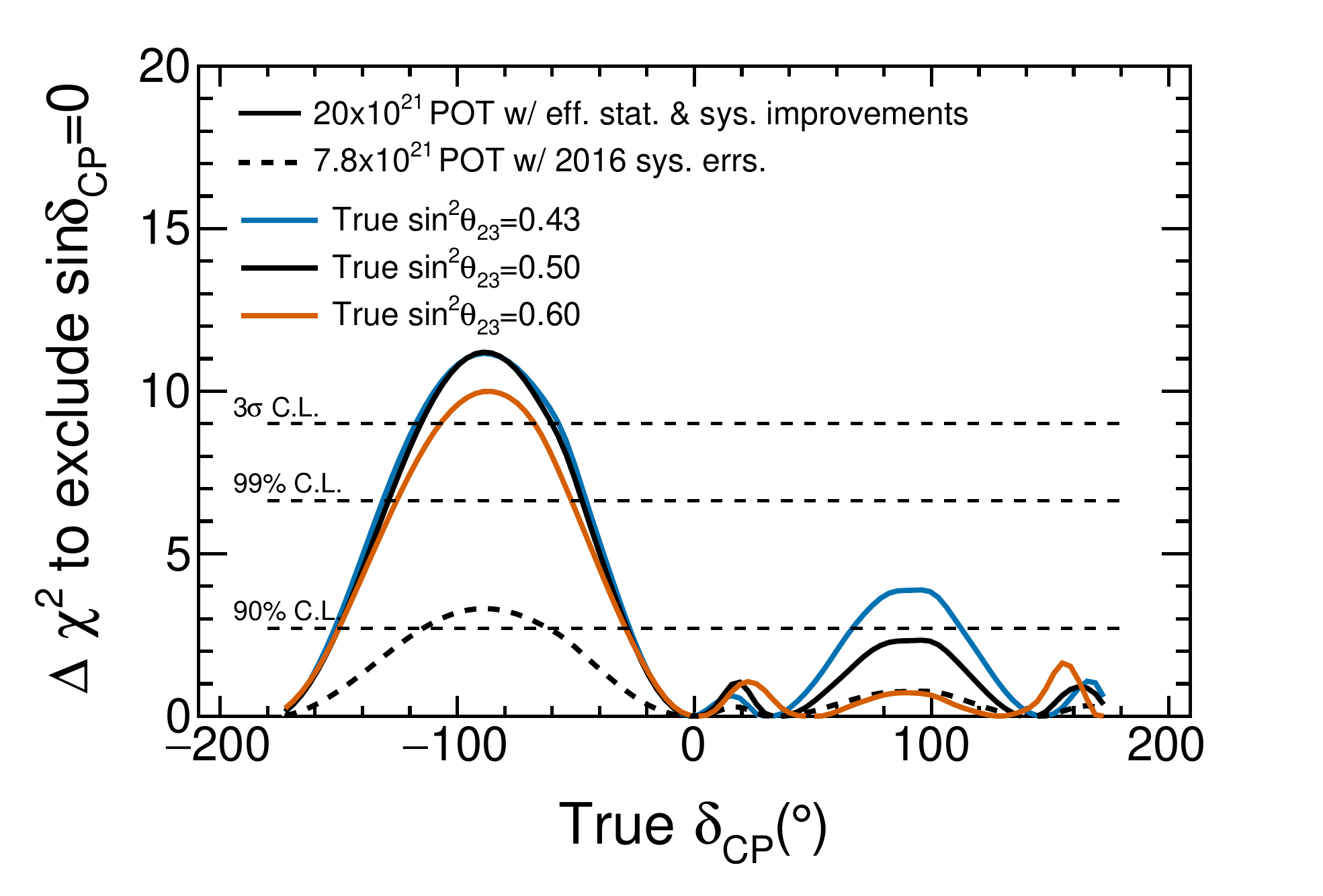}
\includegraphics[width=8cm]{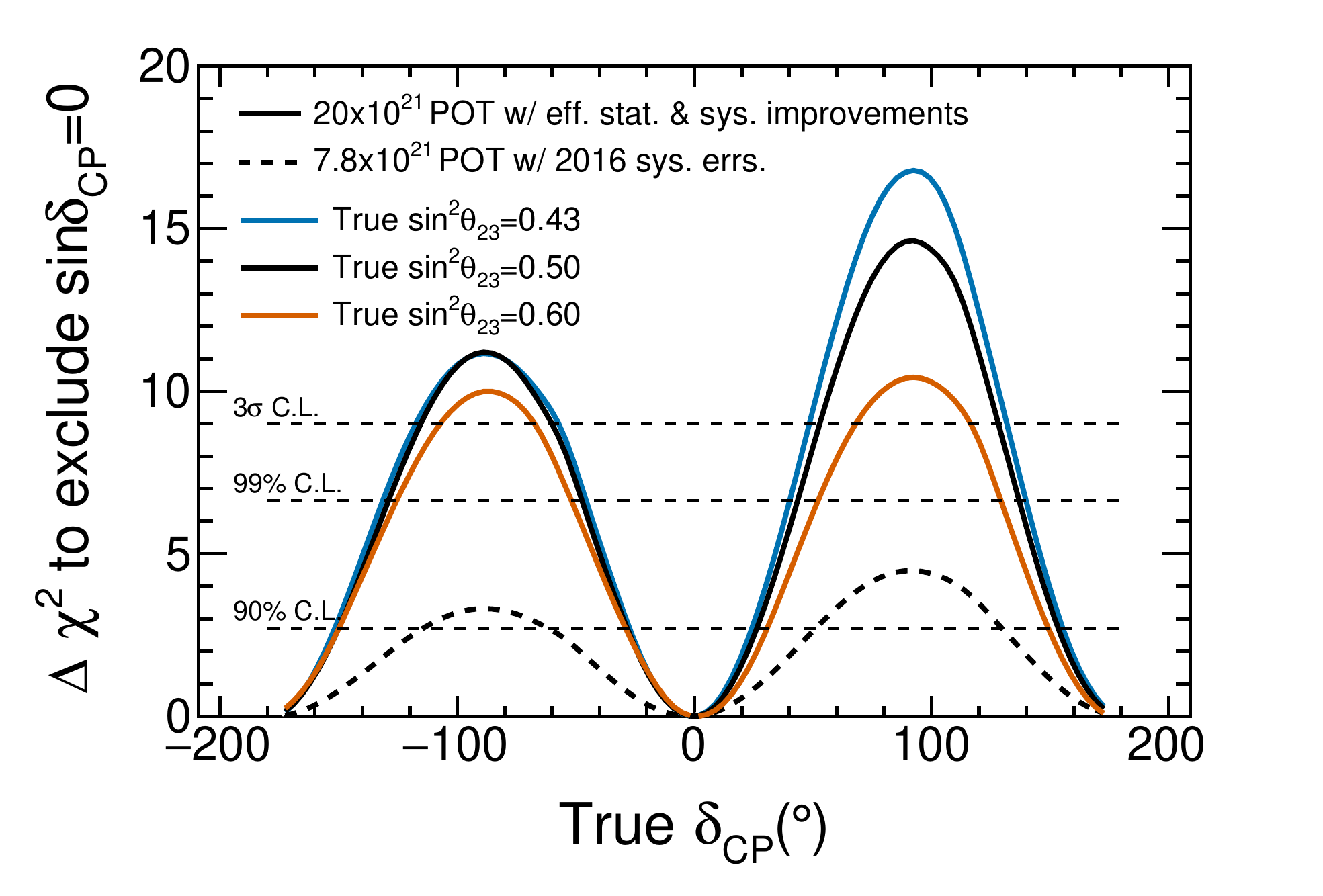}
\caption{\label{fig:t2k2sensi} Sensitivity to CP violation as a function of the true
\dcp for three values of \stt (0.43, 0.50, 0.60) and normal hierarchy, for the full T2K-II exposure of \twopott POT and a reduction of the systematic error to 2/3 of the 2016 T2K uncertainties. On the left plot the mass ordering is considered unknown, while on the right plot it is considered known~\cite{Abe:2016tez}. Courtesy of the T2K collaboration.}
\end{center}
\end{figure}

With $20\times10^{21}$ \pot equally split into neutrino and anti-neutrino mode, and some realistic improvements in the analyses, T2K will observe $559$ ($116$) e-like candidates in neutrino (anti-neutrino) mode and $2735$ ($1284$) \mmu-like candidates for normal ordering and $\dcp=-\pi/2$. The sensitivity to \dcp with such statistics is shown in Fig.~\ref{fig:t2k2sensi}, in the two cases of known and unknown mass ordering~\cite{Abe:2016tez}.

The knowledge of the mass ordering would greatly enhance the sensitivity to positive values of \dcp (assuming the ordering is normal). This effect is due to the degeneracy between normal ordering and positive values of \dcp, since the former feature tends to increase the number of \nue appearance candidates, while the latter tends to reduce it. For negative values of \dcp, instead, there is no degeneracy since normal ordering and negative values of \dcp both tend to increase the \nue appearance probability, hence the sensitivity to the measurement of CP violation is not affected by the prior knowledge of the ordering.

\begin{figure} [htbp!]
\begin{center}
\includegraphics[width=8cm]{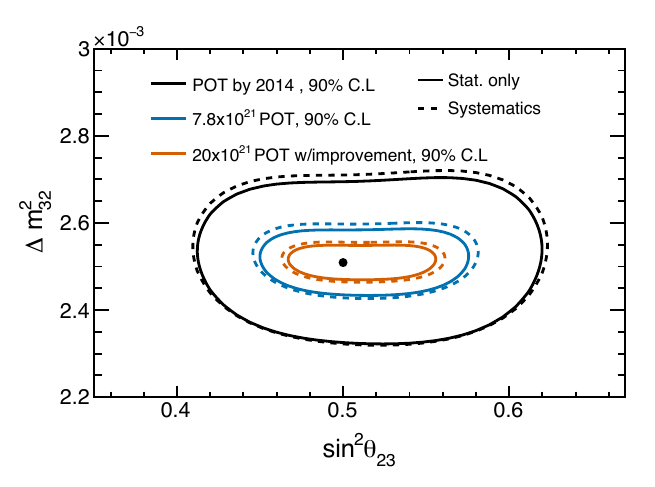}
\includegraphics[width=8cm]{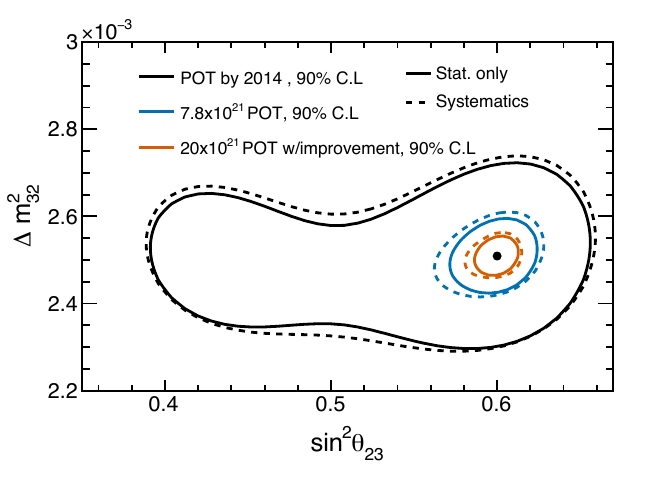}
\caption{\label{fig:t2k2th23} Expected improvements of the 90\% C.L. sensitivity to $\Delta m^2_{32}$ and $\sin^2\theta_{23}$
of T2K and T2K-II, assuming \stt=0.5 (left) and \stt=0.6 (right)~\cite{Abe:2016tez}. Courtesy of the T2K collaboration.}
\end{center}
\end{figure}

In this sensitivity study, an improvement of the systematic uncertainties in the T2K analyses is also assumed. To reach this goal, an upgrade of the Near Detector complex has been launched by the collaboration. Assuming that the mass ordering has been determined by another experiment and that the systematic uncertainties are reduced, the fraction of values of \dcp for which CP conservation (i.e. $\sin \delta_{CP}=0$) can be excluded at the 99\% (3\(\sigma\)) C.L. is 49\% (36\%). Without improvements on the systematic uncertainties, the corresponding fraction is 42$\%$ (21$\%$). These results  depend mildly on the value of \stt, the sensitivity being slightly better for the lower octant solution.
With this statistics, T2K will also be able to perform a greatly improved measurement of the \stt mixing angle, as shown in Fig.~\ref{fig:t2k2th23}.

\begin{figure}[htbp]
\begin{minipage}[c]{.46\linewidth}
   	      \includegraphics[width=0.9\linewidth]{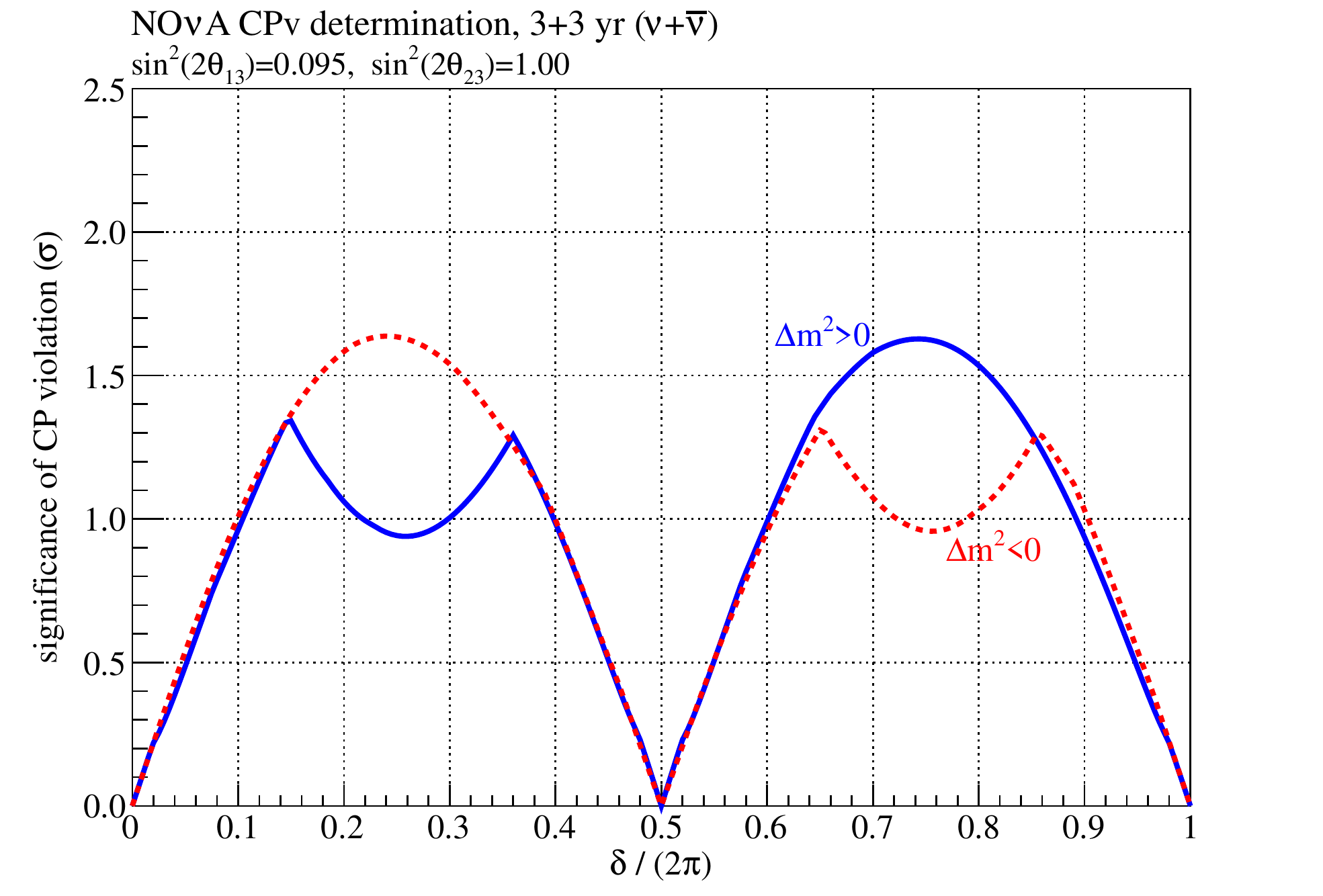}
   \end{minipage} \hfill
   \begin{minipage}{.46\linewidth}
      \includegraphics[width=0.9\linewidth]{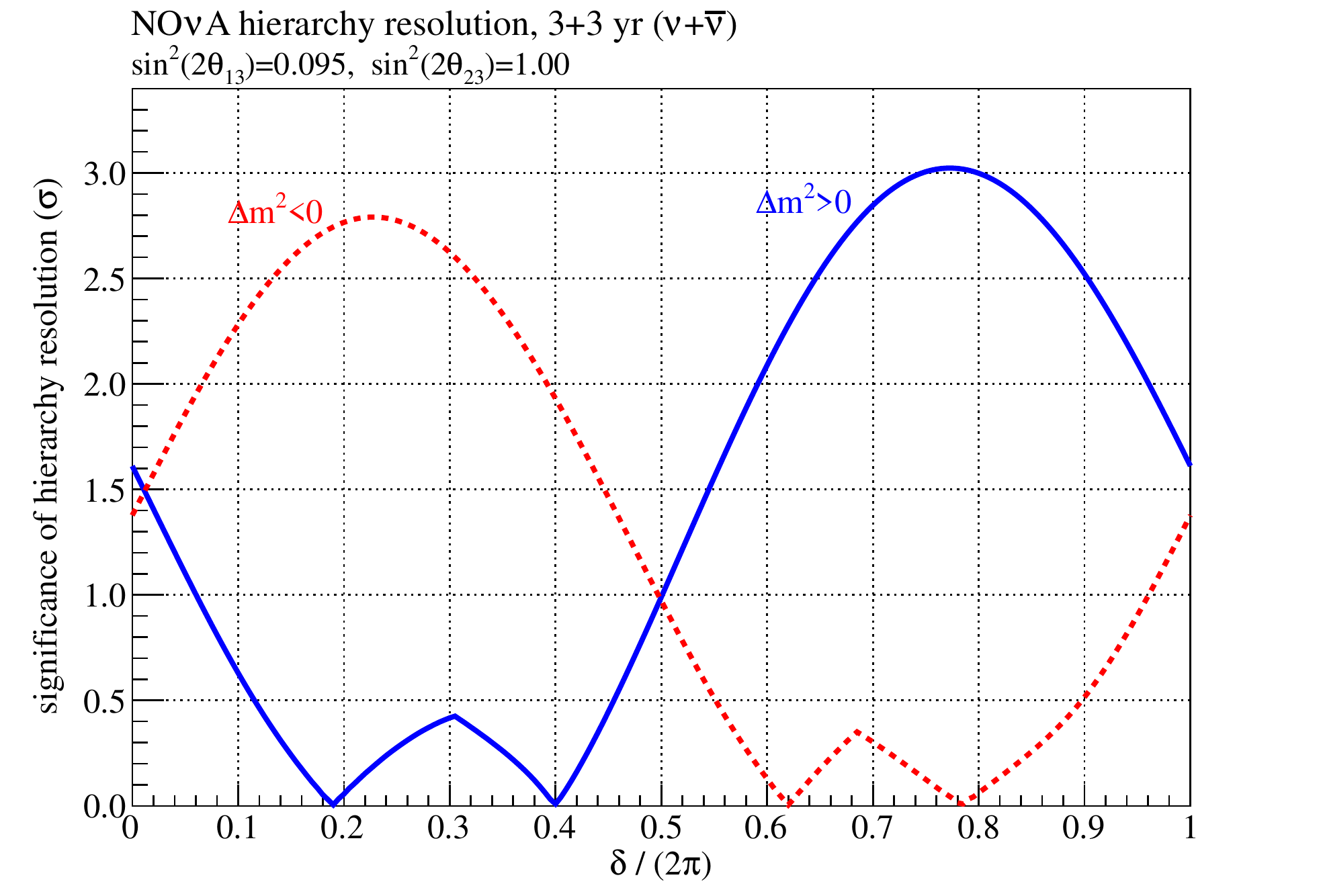}
   \end{minipage}
    \caption{
\nova sensitivity to \dcp (left) and to the mass ordering (right) as a function of \dcp with $36\times10^{20}$~\pot~\cite{messier2016}. The blue solid (red dashed) curve shows the sensitivity assuming normal (inverted) mass ordering. Courtesy of the \nova collaboration.
}
\label{fig:novasensidcp}
\end{figure}

In the coming years, \nova will take advantage  of its longer baseline to further constrain or determine the mass ordering. 
%
%
Also in the case of \nova, the capability of determining the mass ordering can be strongly limited by the combination of matter and CP violation effects. This is true for any experiment in which the baseline is not long enough to fully resolve the degeneracies due to the interplay between matter effects and CP violation. 
For example, if the mass ordering is normal (NO), \nova will have a chance to determine it if \dcp is close to $-\pi/2$, while if \dcp is close to $+\pi/2$, the degeneracy with the solution having inverted mass ordering and $\dcp=-\pi/2$ will make it impossible to establish the mass ordering. In the case of inverted mass ordering (IO), the situation is opposite: it is possible to determine the mass ordering if \dcp= $+\pi/2$, while there is no sensitivity to it for \dcp= $-\pi/2$. This is clearly reflected in the expected sensitivity of \nova to the mass ordering with $36\times10^{20}$~\pot, shown in Fig.~\ref{fig:novasensidcp}.

In the two optimal cases of $\dcp=-\pi/2$ and NO or $\dcp= +\pi/2$ and IO, the mass ordering will be determined at about $3\sigma$ by \nova, while in the other cases the degeneracies among oscillation parameters will reduce the sensitivity of the experiment to the mass ordering. The combination with T2K, which has a different baseline, will help at some level to resolve the degeneracies. The sensitivity of \nova to \dcp with  $36\times10^{20}$~\pot is shown in Fig.~\ref{fig:novasensidcp}.

%
%




\subsection{Next generation of non-accelerator-based experiments}

While the study of CP violation in the lepton sector requires neutrino beams, several different techniques can be used for the measurement of the mass ordering.
It is generally the case that, in order to be able to determine the mass ordering, the experiment must be sensitive either to three-flavour effects or to matter effects. Indeed, the two-flavour vacuum oscillation probability takes the form $P(\nu_\alpha \rightarrow \nu_\beta) = \sin^2 2 \theta \sin^2 (\frac{\Delta m^2 L}{4E})$, which is insensitive to the sign of $\Delta m^2$. 

As we have seen in section~\ref{sec:futuret2knova}, the mass ordering can be determined by exploiting matter effects in long-baseline experiments like \nova, and can be measured for any value of \dcp by doing experiments with a longer baseline such as the proposed DUNE experiment, which will be described in the next section. Other possibilities to measure the mass ordering are provided by matter effects in atmospheric neutrino oscillations~\cite{razzaque}, or by testing the interference between \dmsq and $\Delta m^2_{31}$~\cite{petcov}. 

The latter idea is exploited by the Jiangmen Underground Neutrino Observatory (JUNO) experiment~\cite{An:2015jdp} 
using nuclear reactors.
Antineutrinos produced in nuclear reactors are detected with energies in the range  $(1.8 - 8)$~MeV. For E = 4 MeV, the first oscillation maximum driven by $\Delta m^2_{21}$ occurs at a distance of roughly 65~km from the reactor core, while the oscillation maximum driven by \dmsq occurs at a distance of roughly 2~km. As we have seen in the previous sections, the first effect is exploited by KamLAND to measure \thsol and $\Delta m^2_{21}$, while the latter is exploited by Daya Bay, RENO and Double Chooz to measure \thint. 

JUNO~\cite{An:2015jdp} is a multi-purpose neutrino experiment located in China, whose main goal is to determine the mass ordering. Its 20 kton liquid scintillator detector will be installed at a distance of 53~km from two sites hosting nuclear plants (a total of ten reactor cores will be in operation by 2020). The detector will be located underground with an overburden of more than 700~m of rock. It will consist in a liquid scintillator spherical detector inside a water volume instrumented with PMTs serving as an active shield for natural radioactivity. On the top of the water pool, a tracking detector will also be installed to tag cosmic ray muons.


The \nueb survival probability is given by Eq~(\ref{eq:nue_nue}):

\begin{equation}
P(\bar{\nu}_e \rightarrow \bar{\nu}_e) = 1 - \cos^4 \theta_{13} \sin^2 2 \theta_{12}  \sin ^2 \Delta_{21}
- \sin^2 2 \theta_{13} (\cos^2  \theta_{12} \sin^2 \Delta_{31} + \sin^2  \theta_{12}  \sin^2 \Delta_{32} ),
\end{equation} 
where $\Delta_{ji}= \Delta m^2_{ji} L/(4E)$.

\begin{figure} [htbp!]
\begin{center}
\includegraphics[width=8cm]{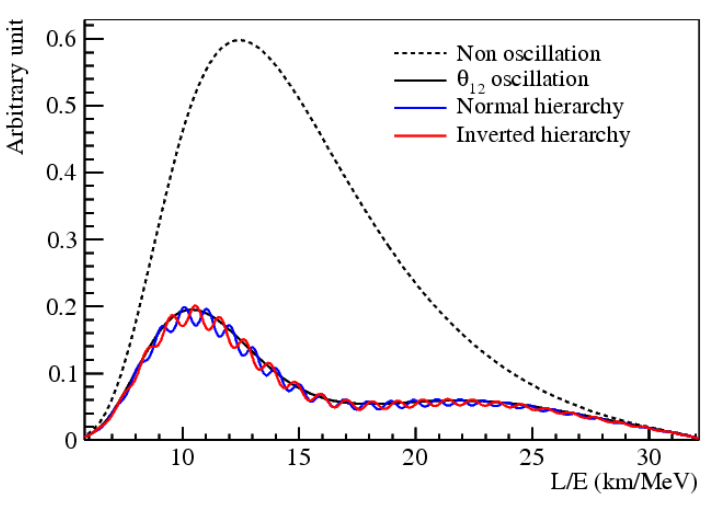}
\caption{\label{fig:junospectrum} Reactor antineutrino signal as a function of L/E for the JUNO detector~\cite{junoplot}. 
The blue and red curves correspond to the normal and inverted mass orderings, respectively. Courtesy of L. Zhan {\it et al.}. Reprinted figure with permission from L. Zhan {\it et al.}, Phys. Rev. D, 78, 111103, 2008. Copyright 2008 by the American Physical Society.}
\end{center}
\end{figure}

The detected neutrino energy spectrum is strongly distorted by oscillations: a large deficit of \nueb is expected, dominated by the parameters \thsol and $\Delta m^2_{21}$, as shown in Fig.~\ref{fig:junospectrum}. The effect of the $\Delta m^2_{32}$-driven oscillations is also visible, manifesting itself as multiple cycles (fast oscillations). 
The period and the phase of these fast oscillations depends on the mass ordering, as discussed in Ref.~\cite{petcov}. 

In a liquid scintillator detector such as JUNO, antineutrinos emitted by nuclear reactors are observed via the inverse beta decay process described in section~\ref{subsec:reactorflux}. 
In order to extract the information on mass ordering from the rapid spectral distortions, an excellent energy resolution ($3\%/ \sqrt{E}$ with $E$ in MeV), a good understanding of the energy response (better than 1\%) and a large statistics ($>100,000$ inverse beta decay events) are required~\cite{An:2015jdp}.
 
This statistics will be obtained thanks to the large size of the JUNO detector, while the needed precision on the energy resolution and on the energy response will be more difficult to reach, and requires important improvements with respect to what has been achieved in the Daya Bay experiment. In particular, it is required to have a PMT photocathode coverage larger than 75\%, a PMT quantum efficiency larger than 35\% and an attenuation length of the liquid scintillator larger than 20~m at 430~nm. If these requirements are met, JUNO will reach a median sensitivity for the determination of the mass ordering of $3\sigma$ after six years of data taking~\cite{An:2015jdp}. 

The sensitivity can be improved to $\sim4\sigma$ if the precision in the measurement of the mass-squared splitting $\Delta m^2_{32}$ in long-baseline experiments reaches the 1\% level. In fact, it can be shown that reactor experiments are sensitive to the parameter~\cite{nunokawa,parkemee}
$\Delta m^2_{ee} = \cos^2 \theta_{12} \Delta m^2_{31} + \sin^2 \theta_{12} \Delta m^2_{32}$,
which can be viewed as 
the ``$\nu_e$-weighted average'' of $\Delta m^2_{31}$ and $\Delta m^2_{32}$, since the $\nu_e$ components in $\nu_1$ and $\nu_2$ are in the ratio $\cos^2 \theta_{12}\, :\, \sin^2 \theta_{12}$ (see Section~\ref{subsec:3-flavour} for details).
On the other hand, long-baseline experiments are sensitive to 
$\Delta m^2_{\mu \mu} \simeq \sin^2 \theta_{12} \Delta m^2_{31} + \cos^2 \theta_{12} \Delta m^2_{32}$, plus a term depending on $\cos \dcp$. The two quantities differ by a few \% for the two mass orderings, and their precise measurement might provide an additional handle on the determination of the mass ordering. However, as these measurements depend on the absolute energy scale of the experiments, this method is difficult and subject to systematic uncertainties. 



Besides the determination of the mass ordering, JUNO will perform several precise measurements of the neutrino mixing parameters. While the precision on \thint will still be dominated by Daya Bay, thanks to its shorter baseline, the huge number of events collected by JUNO will allow to greatly reduce the uncertainties on \thsol and $\Delta m^2_{21}$. As we have seen in Section~\ref{sec:solar}, these parameters are currently measured by KamLAND (which dominates the $\Delta m^2_{21}$ measurement) and solar neutrino experiments (which dominate the \thsol measurement) with uncertainties of the order of 2-4\%. JUNO will be able to reduce these uncertainties well below 1\%, allowing precise tests of the unitarity of the lepton mixing matrix.

In addition, JUNO will also be able to observe the neutrinos emitted by core-collapse supernovae (SN), collecting $\sim5,000$ inverse beta decays, 2,000 elastic neutrino-proton scatterings and 300 elastic neutrino-electron scatterings for a SN explosion at a distance of 10 kpc. The time evolution, energy spectra and flavour content of SN neutrinos can be used to investigate the explosion mechanism (see e.g. Ref.~\cite{Mirizzi:2015eza} for a review). Also, neutrinos from the diffuse supernova neutrino background might be observed by JUNO. Finally, JUNO will be able to perform precise measurements of the flux of the $^8$B and $^7$Be solar neutrinos thanks to its high light yield, exceptional energy resolution and  low threshold. The high statistics samples collected by JUNO will allow to shed light on the solar abundance problem, and probe the transition region between the vacuum oscillation-dominated and MSW-dominated regions of the solar neutrino spectrum. 

\begin{figure} [htbp!]
\begin{center}
\includegraphics[width=12cm]{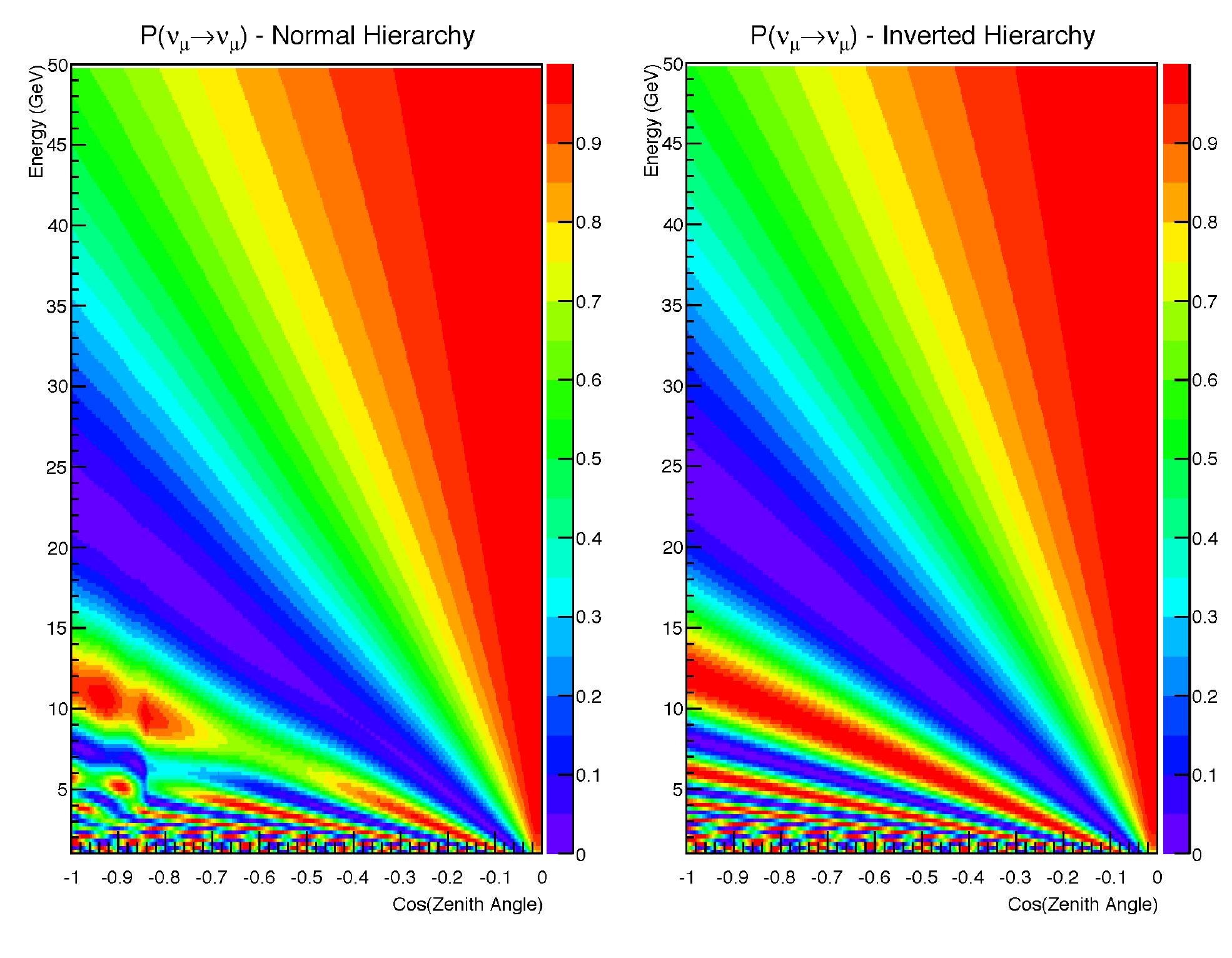}
\caption{ \label{fig:orcaoscprob} 
Muon neutrino survival probability after traveling through the earth, binned in both neutrino
energy and cosine of the zenith angle~\cite{pinguprop}. The survival probabilities for antineutrinos in a given mass hierarchy are essentially the same as
those for neutrinos under the opposite hierarchy. Courtesy of the IceCube collaboration.
}
\end{center}
\end{figure}

An alternative technique for the determination of the mass ordering using matter effects on the propagation of atmospheric neutrinos has been proposed~\cite{razzaque}.
The idea is to exploit the difference in the oscillation probabilities of neutrinos and antineutrinos that are induced by matter effects when they cross the Earth. The oscillation probabilities, or {\it oscillograms}, for the two mass orderings are shown in Fig.~\ref{fig:orcaoscprob} as a function of the energy and the cosine of the zenith angle $\theta_z$.
The qualitative features of these oscillograms can be understood as follows, modelling the Earth as made of two layers, the mantle with an almost constant density of about 5 g/cm$^3$ above a radius of 3483 km, and the core with a density of about 11 g/cm$^3$ below that radius.
The most prominent features of these oscillograms are the $\nu_\mu \rightarrow \nu_e$ MSW resonance in the mantle for E $\simeq$ 7 GeV, visible at $\cos \theta_z \simeq -0.8$, and  the MSW resonance in the core for 
E $\simeq$ 3 GeV and $\cos \theta_z < -0.8$. As already explained, the resonance is present either in the neutrino channel for NO or in the antineutrino channel for IO. This resonance has a spectacular effect
in the $\nu_\mu \rightarrow \nu_e$ channel, 
and results in smaller but significant differences in the oscillograms for 
$\nu_\mu \rightarrow \nu_\mu$ between the NO and IO cases (Fig.~\ref{fig:orcaoscprob}).
The finer structure seen in the $\cos \theta_z < -0.8$ region is due to resonance effects related to the interference between propagation in the mantle and in the core~\cite{petcov-res-1,petcov-res-2}.
  
\begin{figure}[htbp]
\begin{minipage}[c]{.46\linewidth}
   	      \includegraphics[width=0.9\linewidth]{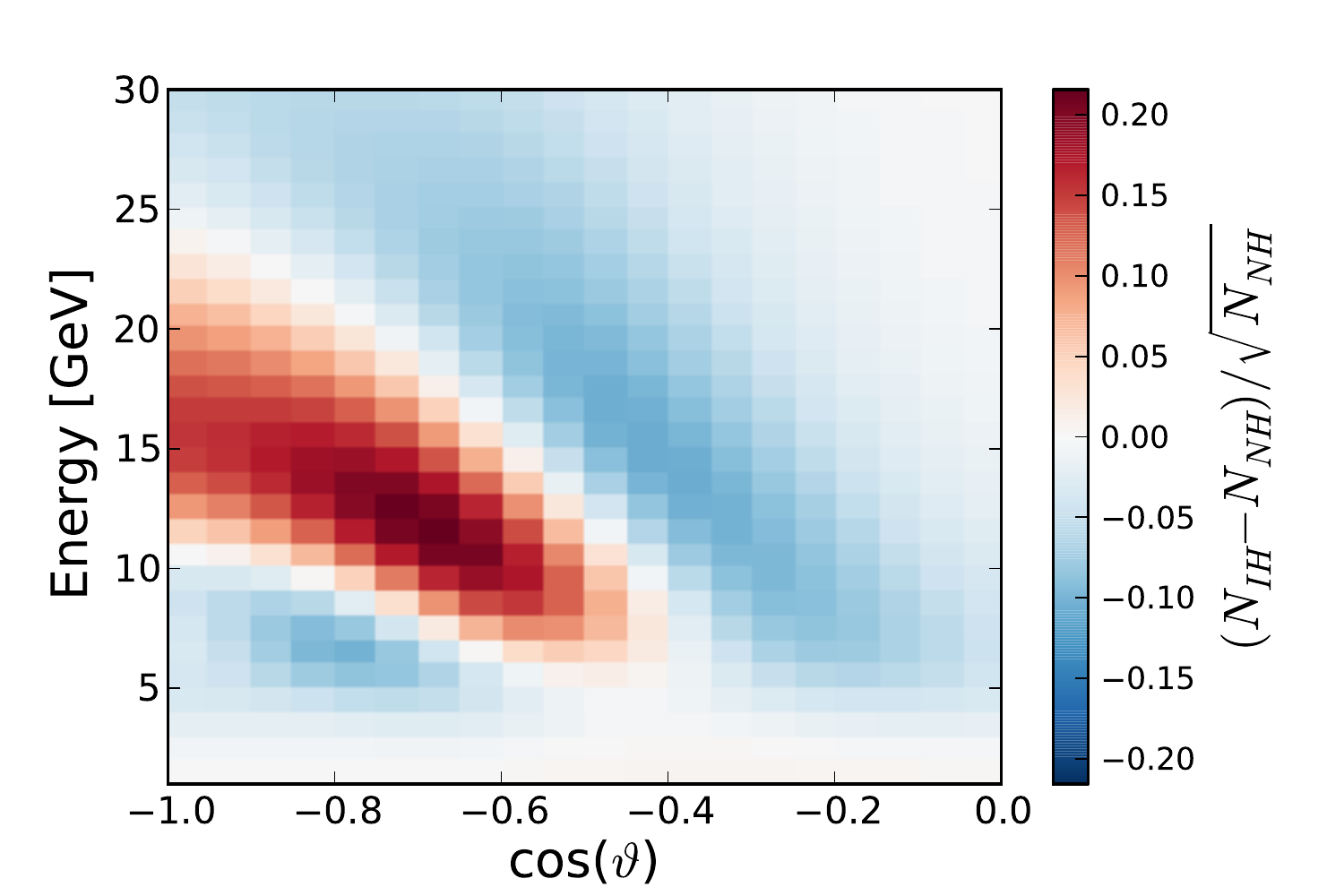}
   \end{minipage} \hfill
   \begin{minipage}{.46\linewidth}
      \includegraphics[width=0.9\linewidth]{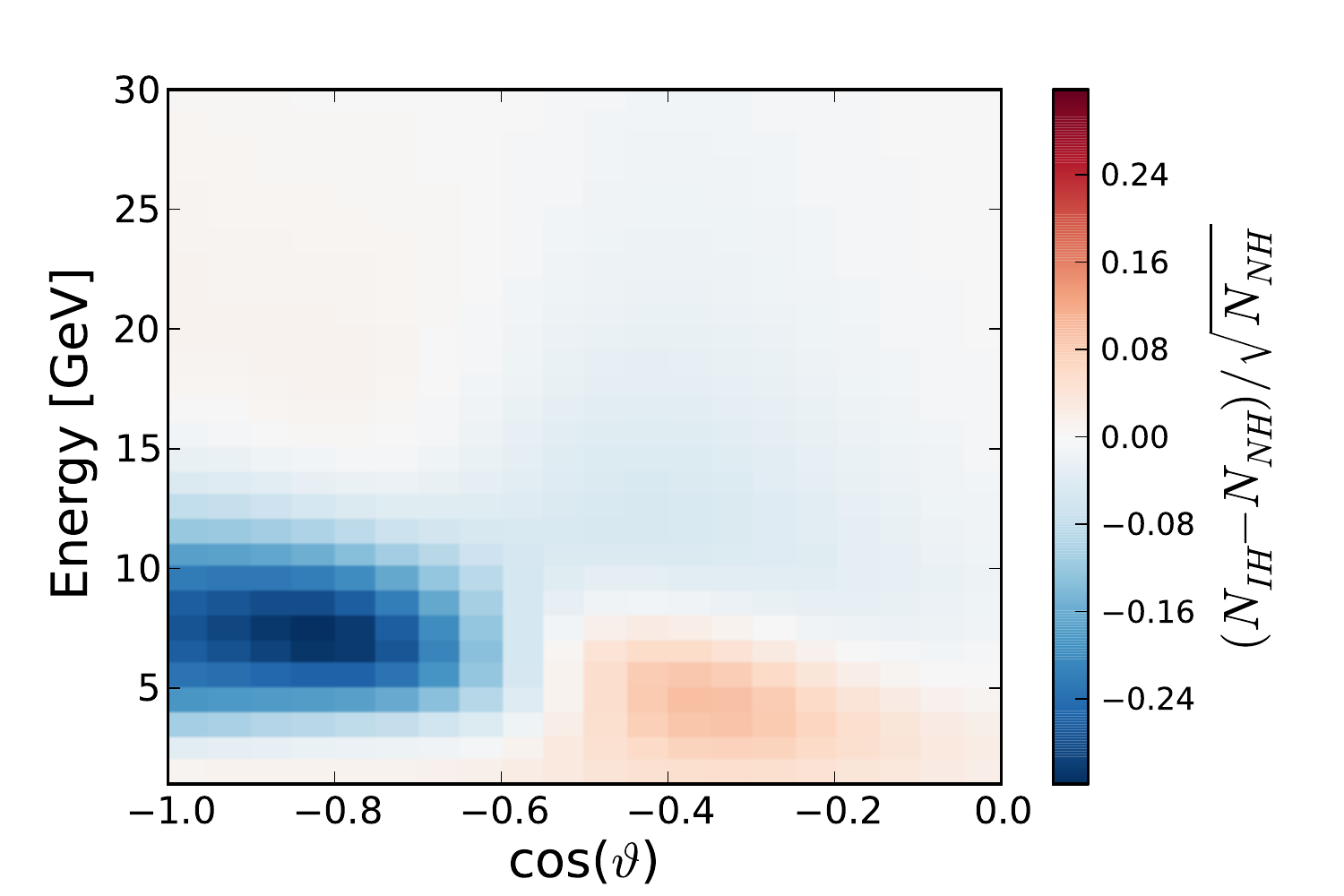}
   \end{minipage}
    \caption{  
    Difference in the event rates for the normal and inverted orderings, normalized by the statistical uncertainty, for one year of simulated PINGU data with
reconstruction and particle identification applied~\cite{pinguprop}. The left panel shows track-like events (mostly due to 
$\nu_\mu$ CC interactions), while the right panel shows cascade-like events (mostly $\nu_e$ and $\nu_\tau$ CC events, as well as NC events from
any neutrino flavors). Courtesy of the IceCube collaboration.
    }
 \label{fig:orcaasym}
\end{figure}

The detectors based on the Cherenkov technique like Hyper-Kamiokande~\cite{hkdr}, ORCA~\cite{orcaprop} or PINGU~\cite{pinguprop}  do not measure the charge of the lepton event-by-event but, due to the different interaction probabilities of neutrinos and antineutrinos, a net asymmetry in the combined ($\nu+\nub$) event rate can be observed in the \nue and \num channels. However, the effect is strongly diluted. Smearing of the reconstructed energy and angle dilutes this effect even more. Once the atmospheric neutrino fluxes and the neutrino-nucleon cross-section are taken into account, bidimensional plots of event rates as a function of the neutrino energy and zenith angle can be built, as shown in Fig.~\ref{fig:orcaasym}. Although the asymmetry is partially washed out, there is still a certain sensitivity to the mass ordering. 

The ORCA and PINGU experiments intend to instrument large volumes of ice in the Antarctica (PINGU) or water in the Mediterranean sea (ORCA) with a dense array of PMTs in order to observe, through the Cherenkov effect, the muon tracks or the electromagnetic showers produced by \num and \nue interactions in the medium. The technique is similar to the one used by IceCube~\cite{Aartsen2016161} and ANTARES~\cite{antares} to observe high-energy neutrinos produced by cosmic rays, but, in order to improve the energy and direction reconstruction, a much denser arrays of PMTs must be built, at the price of instrumenting a smaller volume. Indeed, one of the challenges of these experiments is to reach an energy threshold around 5 GeV, which is considerably lower than the 15 GeV threshold of the IceCube DeepCore denser instrumented volume. This requires to arrange the PMTs at close distance, around 10~m. Moreover, another challenge is to control the systematic uncertainties related to the reconstruction of the energy and the momentum.   

In PINGU and ORCA, two categories of events can be reconstructed: tracks produced by muons in \num charged current interactions, and showers produced by neutral current interactions or \nue. Both contribute to the determination of the mass ordering. The sensitivity  depends mainly on the capability to reconstruct the energy and the angle of the incident neutrino from the observation of a shower or a track on the PMTs. The resolution depends on the spacing between the PMTs and, once the total number of PMTs is fixed, the optimal vertical spacing for ORCA is 9~m. 

The sensitivity to the mass ordering computed by ORCA, assuming this configuration, depends on other oscillation parameters, in particular \thatm and \dcp, but in general a sensitivity of $\sim3\sigma$ can be obtained in three years. Similar results will be reached by PINGU after four years of running.


%

Another experiment aimed at determining the mass ordering is the Iron Calorimeter (ICAL)~\cite{Ahmed:2015jtv} detector to be built at the India-based Neutrino Observatory (INO). ICAL will be a 50 kton magnetized detector that will be mainly sensitive to 1-10 GeV atmospheric neutrinos: 5.6 cm thick iron layers are interleaved with Resistive Plate Chambers (RPC) used for tracking the charged particles. Thanks to the 1.5 Tesla magnetic field, ICAL will be capable of reconstructing the charge of the muons, hence separating the interactions induced by muon neutrinos from the ones induced by muon antineutrinos.
In order to determine the mass ordering, \num and \numb events will be binned in momentum and angle, exploiting the matter effects in a similar way as proposed by ORCA and PINGU. In the case of INO, only muon tracks are used and a sensitivity of 3$\sigma$ to the mass ordering can be reached after 10 years of data taking.

\subsection{Next generation long-baseline experiments}

The next generation of long-baseline experiments will be built in the coming years and operational around the second half of the years 2020, with the main goal of providing precise measurements of the neutrino mixing parameters that are accessible to long-baseline experiments, namely \thatm, \thint, \dmsq and \dcp. 
The two projects for the next generation of long-baseline experiments are the Deep Underground Neutrino Experiment (DUNE)~\cite{Acciarri:2015uup} in the US and the Hyper-Kamiokande experiment~\cite{hkdr} in Japan. Both experiments will be multi-purpose detectors, searching for proton decay and neutrinos produced in core-collapse supernovae explosions, and measuring solar and atmospheric neutrinos. In the context of this review, we focus on the measurements  that are important for a better understanding of the PMNS matrix.

DUNE~\cite{Acciarri:2015uup} will use an innovative technology with Liquid Argon as target for neutrino interactions, performing a calorimetric measurement of the particles produced in neutrino interactions. Neutrinos will be produced at FNAL using protons from the Main Ring, and will be sent to the Sanford Underground Research Facility (SURF) in South Dakota, 1,300 km away from FNAL.

DUNE will build four large (10 kton) Liquid Argon time projection chambers as target for neutrino interactions. When neutrinos interact with the Argon, charged particles are produced that cross the medium, ionizing and exciting Argon nuclei. This produces both ionization and scintillation signals that are detected by a readout system. The careful collection of the ionization electrons allows for a three-dimensional reconstruction of the charged tracks and showers, and hence of the properties of the neutrino responsible for the interaction. Two options are contemplated for the readout of the ionization electrons: a single-phase readout, where the electrons are collected using wire planes in the liquid argon volume, and a dual-phase approach, where the electrons are amplified and detected in a gaseous argon region above the liquid. The single-phase readout is based on the technology used for the ICARUS detector~\cite{CANCI20121257}, while for the dual-phase readout several small-scale prototypes have been built in the last years~\cite{double}. An extended R\&D program is ongoing to demonstrate that both technologies are suitable to build a 10 kton module. Two prototypes, one for the single phase (ProtoDUNE-SP)~\cite{protodune} and one for the dual phase (ProtoDUNE-DP/WA105)~\cite{wa105} of roughly 300 tons, scalable to larger masses, are being built at CERN and will be exposed to a charged-particle beam in 2018.  

\begin{figure} [htbp!]
\begin{center}
\includegraphics[width=10cm]{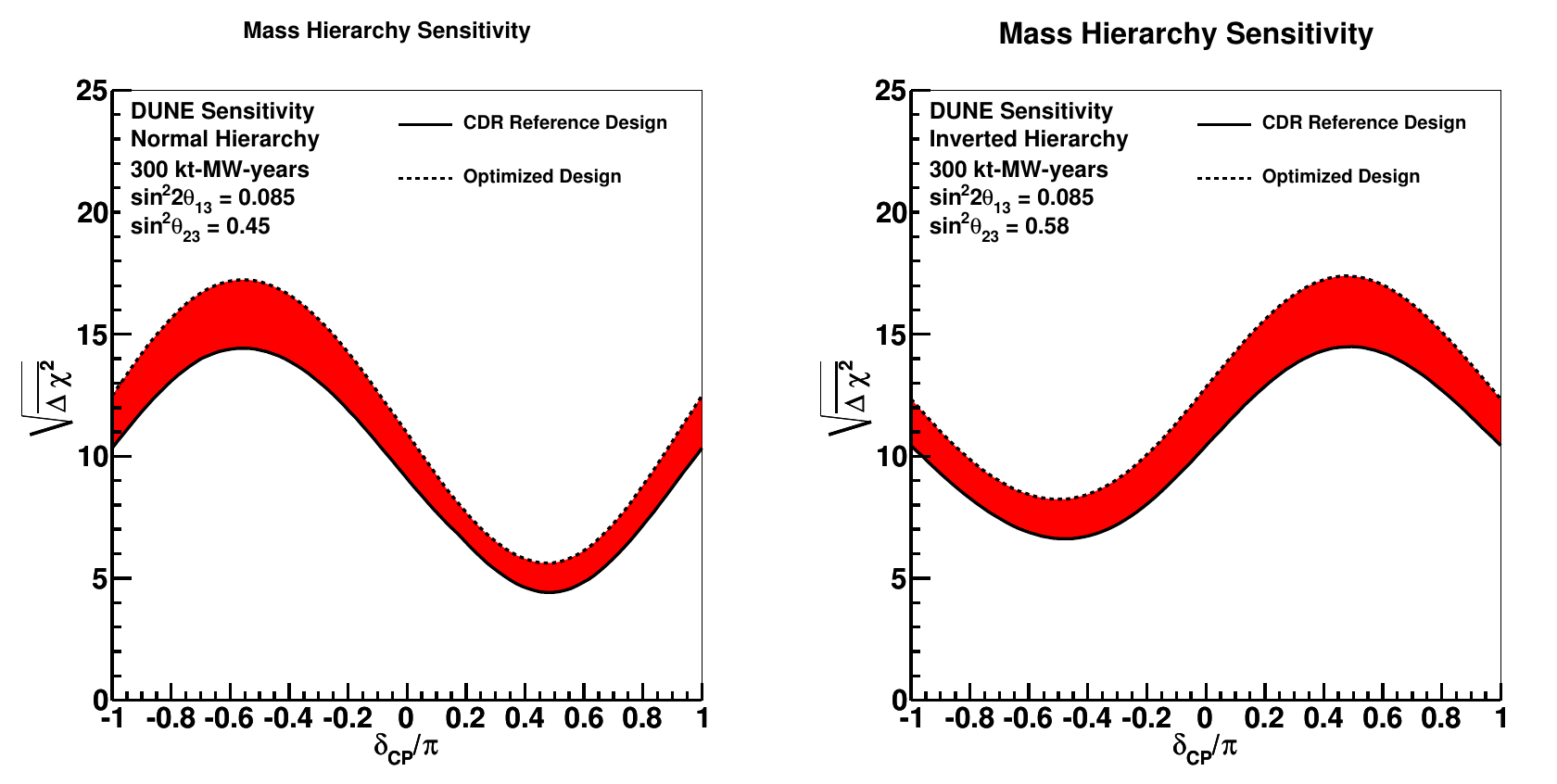}
\includegraphics[width=10cm]{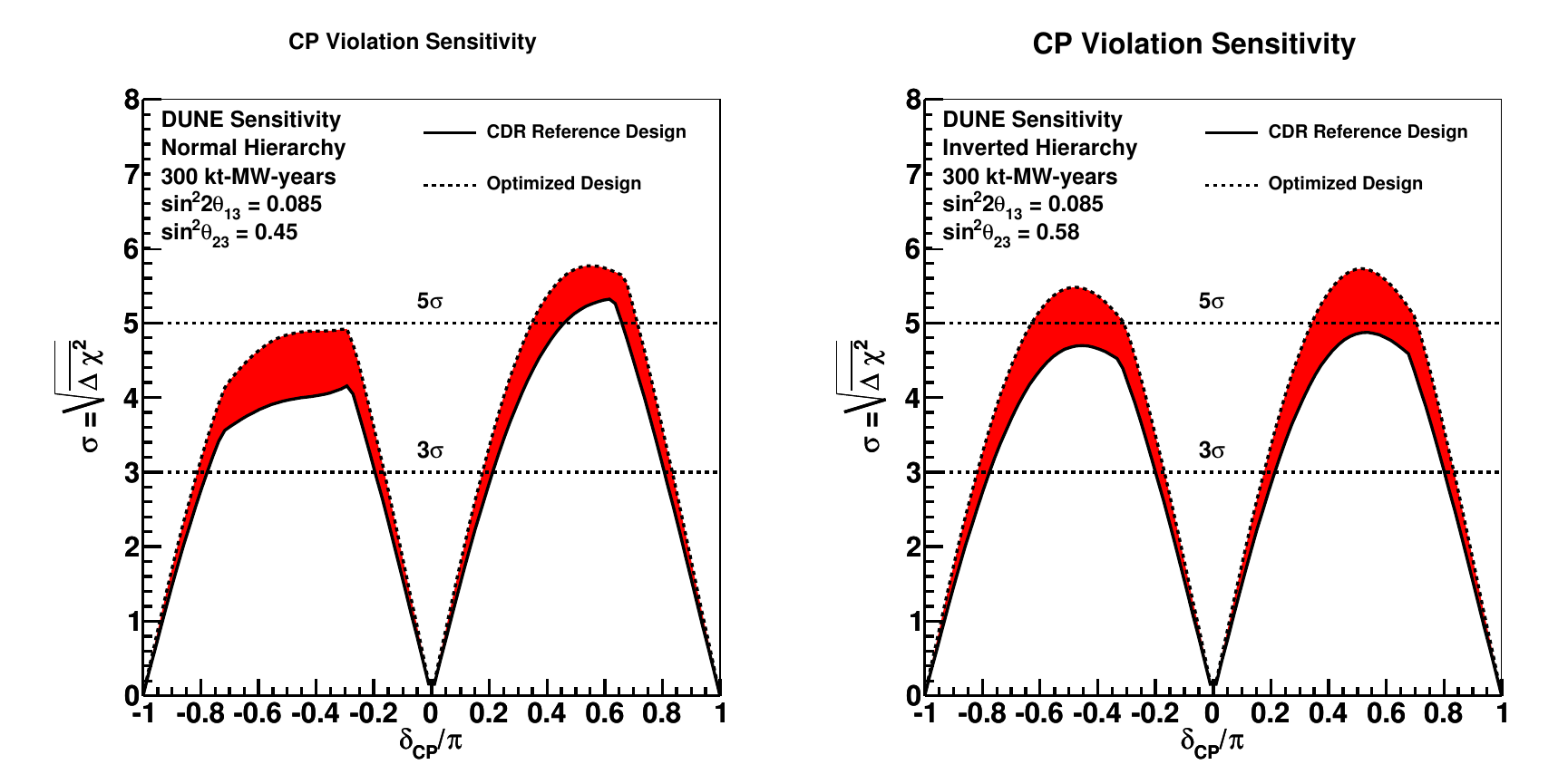}
\caption{\label{fig:dunesensi} DUNE expected sensitivities to the mass ordering (top) and CP-violating phase (bottom) for normal (left) and inverted (right) orderings, assuming an exposure of 300~kt$\times$MW$\times$year~\cite{Acciarri:2015uup}. Courtesy of the DUNE collaboration.}
\end{center}
\end{figure}

The neutrino beam used for DUNE will be a broad band beam produced at the Long Baseline Neutrino Facility (LBNF) at FNAL. The foreseen beam power is 1.07~MW. The neutrino flux will cover both the first and the second oscillation maxima, which given the 1,300 km baseline correspond to energies of 2.5~GeV and 0.8~GeV, respectively. 
At energy above 2~GeV, multiple tracks are produced in charged current neutrino interactions and high efficiency is required to reconstruct all the final state particles. The high granularity of a liquid Argon detector is therefore better suited at these energies than Water Cherenkov detectors, for which most of the charged particles are below the Cherenkov threshold and cannot be reconstructed.
Thanks to the very long-baseline of DUNE, the effects of the mass ordering and of CP violation are decoupled, as shown in Fig.~\ref{fig:novaellipse}. The strategy is to collect neutrino and anti-neutrino data by changing the direction of the current in the magnetic horns, and to measure \num and \numb disappearance and \nue and \nueb appearance probabilities.  
The combination of the four samples allows a clean measurement of the mass ordering for any value of \dcp and in both ordering hypotheses, as shown in Fig.~\ref{fig:dunesensi}. As far as the measurement of CP violation is concerned, the sensitivity depends on the value of \dcp. The requirement for the experiment is to have a sensitivity larger than $5\sigma$ for 50\% of the values of \dcp, and larger than $3\sigma$ for 75\% of the values of \dcp. Such sensitivities will be reached after 7 years of running.

The other proposed long-baseline experiment is Hyper-Kamiokande (HK)~\cite{hkdr}. 
Hyper-Kamiokande builds up on the experience with the K2K and T2K experiments and the Super-Kamiokande detector. It will be composed of two large water Cherenkov detectors, with a total mass 10 times larger than SK, and it will be exposed to the off-axis neutrino beam produced at JPARC.

The baseline design of HK is to have a staged construction of two 187~kton fiducial volume mass modules near the current Super-Kamiokande site, at a distance of 295~km and 2.5 degrees off-axis from the J-PARC site where the neutrinos are produced. The neutrino beam is the same as the one currently used by the T2K experiment but, thanks to an upgrade of the JPARC Main Ring power supplies, it will be able to operate at more than 1.3~MW  by 2025 (to be compared with a beam power of 470~kW in 2017).

An alternative design is being pursued to install the second module in South Korea~\cite{Abe:2016ero}. This second detector would be operated at a baseline of roughly 1,100~km and a smaller off-axis angle. In this configuration, HK would be sensitive to the first and second oscillation maxima, with the first one taking place at $\sim2$~GeV and the second one at $0.6~$GeV. In this configuration, the first oscillation maximum would give sensitivity to the mass ordering, while the second one would allow to exploit the larger CP asymmetry, improving the measurement of \dcp.  If both detectors are installed at Kamioka, external measurements of the mass ordering will be used in the search for \dcp. If no measurement of mass ordering is available by the time HK is collecting data, the mass ordering will be determined at more than $3\sigma$ by measuring matter effects in the large atmospheric neutrino sample collected by HK.    

\begin{figure} [htbp!]
\begin{center}
\includegraphics[width=15cm]{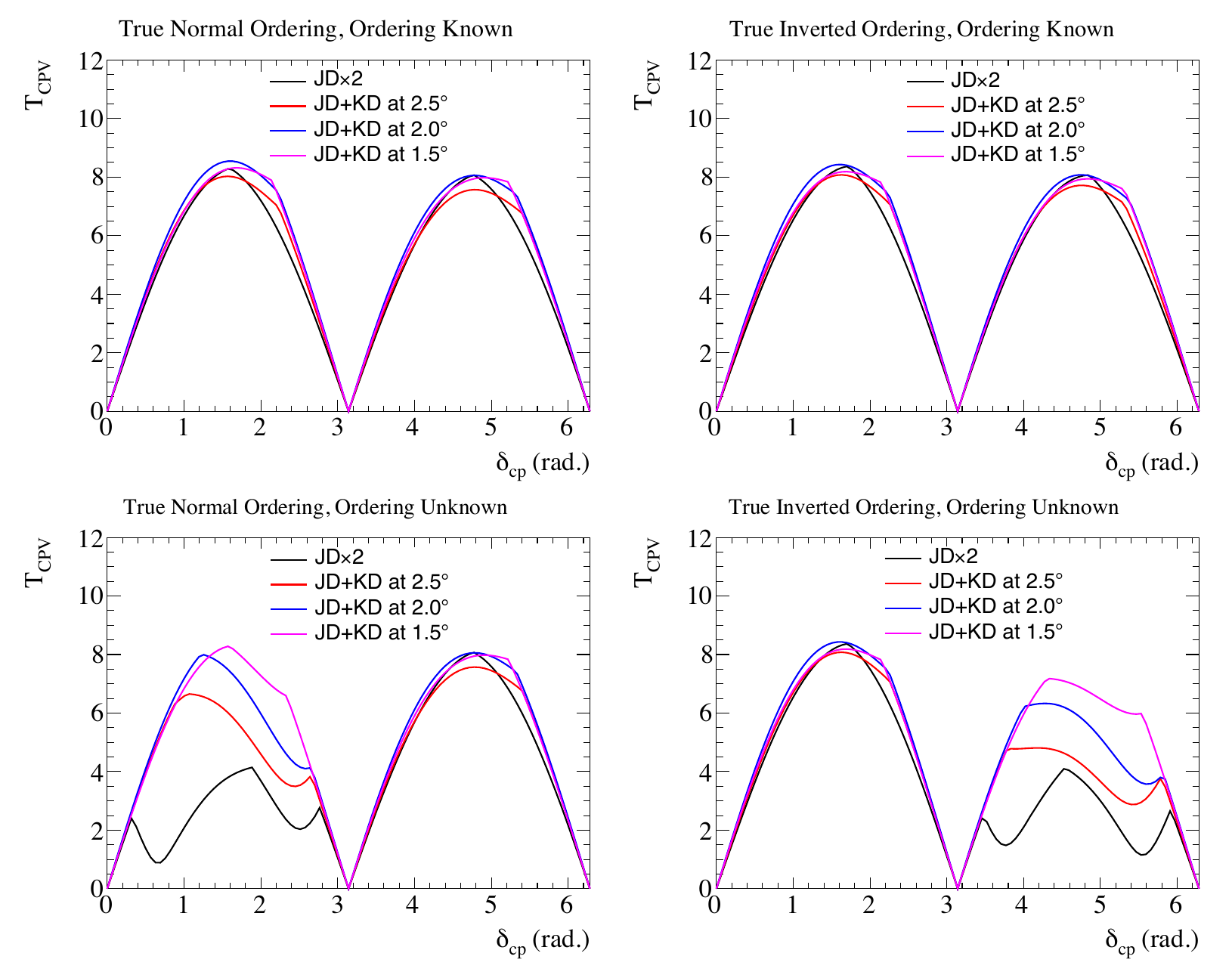}
\caption{\label{fig:hkcpv} Hyper-Kamiokande significance (in number of standard deviations) to reject CP conservation as a function of \dcp for normal ordering (left) and inverted ordering (right) and different hypotheses for the location of the second tank. In the top row, the mass ordering is assumed to be known, while in the bottom row it is assumed to be unknown~\cite{Abe:2016ero}. The black lines (labeled JD$\times$2) corresponds to two detectors in Japan, while the colored lines labeled KD correspond to one detector in Japan and one in Korea, with different off-axis angles. Courtesy of the Hyper-Kamiokande collaboration.}
\end{center}
\end{figure}

After 10 years of data taking, HK will collect $\sim1,000$ ($\sim130$) \nue and \nueb signal events per tank at 295 km (1,100 km and 2.0 degrees off-axis) assuming a 3:1 ratio of antineutrino mode to neutrino mode operation. The sensitivity of the experiment for different options for the location of the second tank is shown in Fig.~\ref{fig:hkcpv}. The figure shows the benefit of having the second detector in Korea, if no measurement of the mass ordering is available by 2025. On average, the second detector in Korea also allows for a more precise measurement of the value of \dcp, thanks to the measurement at the second maximum. A precision between 6 and 13 degrees can be obtained, depending on the value of \dcp. Finally, the sensitivity in Korea will be limited by statistical uncertainties even after 10 years of data taking, while in Kamioka the measurements will be limited by systematic uncertainties.

In conclusions, the  DUNE and Hyper-Kamiokande projects are largely complementary. Hyper-Kamiokande is based on a known technology and will offer large $\nu_e$ samples. DUNE will help develop the Liquid Argon technology and will offer more detailed information on the particles in the final state. Complementarity between the two projects is also present for other physics goals like SN neutrinos and the search for proton decay.  


\section{Conclusions}
\label{sec:conc}

The discovery of neutrino oscillations is a major milestone in particle physics. It is so far the only experimental evidence of nonzero neutrino masses, and the first positive indication of physics beyond the Standard Model.
The field of neutrino oscillations and its findings therefore deserve great attention
and require reliable and complementary top-quality measurements to assure a steady progress of our knowledge. 

Neutrino oscillation physics benefited from rapid progress in the last two decades. After the establishment of neutrino oscillations as the mechanism behind the anomalies observed in the study of solar and atmospheric neutrinos, a large corpus of experimental data has emerged. This has been obtained with solar, atmospheric, reactor and accelerator neutrinos, in a variety of experimental configurations, detection methods and neutrino energies, spanning the full range from eV to TeV.

Neutrino oscillations have been studied and confirmed by several independent experiments in all sectors: solar and reactor experiments for the 1-2 sector, atmospheric and accelerator experiments for the 2-3 sector, and reactor and accelerator experiments for the 1-3 sector. The latest major discovery is indeed related to the 1-3 sector, with the precise measurement of the $\theta_{13}$ mixing angle by Daya Bay in 2012 and its confirmation by RENO and Double-Chooz.
This overall picture, which has been obtained mainly through disappearance measurements, has been confirmed recently by the direct observation of the appearance of new flavours: T2K has observed $\nu_\mu \rightarrow \nu_e$ appearance, later confirmed by NOvA, and OPERA has established $\nu_\mu \rightarrow \nu_\tau$ appearance, with supporting indications from Super-Kamiokande.  

It is remarkable that this extensive corpus of data can be interpreted with only five parameters: the three mixing angles of the PMNS matrix $\theta_{12}$, $\theta_{23}$ and $\theta_{13}$, and two independent squared-mass differences, $\Delta m^2_{21}$ and $\Delta m^2_{31}$. This has led to the emergence of the PMNS paradigm of three active neutrino oscillations, which has been the subject of this review. At present the experimental data is only weakly sensitive to the CP-violating phase \dcp, but the inclusion of this parameter will become necessary to provide an accurate description of future oscillation results.

The PMNS paradigm is now entering the precision period, as can be seen from the results
of a global fit~\cite{nufit} to all present oscillation measurements:
\begin{eqnarray}
\theta_{12}\, ({\rm deg}) &=& 33.56\:^{+0.77}_{-0.75} \\
\theta_{23}\, ({\rm deg}) &=& 41.6\:^{+1.5}_{-1.2}\ (50.0\:^{+1.1}_{-1.4})\\
\theta_{13}\, ({\rm deg}) &=& 8.46\:^{+0.15}_{-0.15} \\
\Delta m^2_{21}\, (10^{-5} {\rm eV}^2) &=& 7.50\:^{+0.19}_{-0.17} \\
\Delta m^2_{3l}\, (10^{-3} {\rm eV}^2)&=&  2.524\:^{+0.039}_{-0.040}\ (-2.514\:^{+0.038}_{-0.041})
\end{eqnarray}
where the mixing angles are given in degrees, and $\Delta m^2_{3l}$ is equal to $\Delta m^2_{31}$
for normal ordering and to $\Delta m^2_{32}$ for inverted ordering. The values in parenthesis are
relative to inverted ordering.

A few outstanding anomalies, most notably the LSND excess, cannot be interpreted in the PMNS framework. Recent experimental results have not confirmed these anomalies, which are still the object of intense scrutiny and debate. A rich experimental program is under way and will provide more sensitive tests of these anomalies
in the next few years. One should therefore know in the near future whether they must be removed
from the list of open questions or not. An unambiguous confirmation of some of the existing anomalies
would represent a major discovery.

Now that the experimental precision on the squared-mass differences and on most of the parameters of the PMNS matrix
is approaching the percent level, great attention needs to be devoted to the experimental systematic uncertainties to make further progress. A robust and precise control of the neutrino flux is needed for future experiments devoted to the improvement of the experimental accuracy. This calls for a full-fledged program of auxiliary experiments and measurements, for instance experiments like NA61/SHINE in the field of hadro-production. A renewed program to establish a reliable model for the neutrino-nucleus cross-section is also needed, in which new phenomenological investigations and high-precision data will be major ingredients. The goal is to reach a control at the few percent level, to be compared with the current uncertainty of 10\% (or worse).

A rich experimental program is under preparation to answer the remaining open questions: the determination
of the octant of the $\theta_{23}$ mixing angle and the precise measurement of its deviation
from the maximal value $\pi/4$, the determination of the mass ordering and the measurement of the CP-violating phase \dcp. 
These are fundamental questions, and measurements in the field of neutrino oscillations are likely to play
a major role on the experimental particle physics scene in the coming years. Partial answers to these questions will come from the running long-baseline experiments T2K and \nova.
The question of mass ordering will be addressed to a certain extent by non-accelerator experiments using either atmospheric neutrinos (INO, PINGU, ORCA) or nuclear reactors (JUNO). 
Two major long-baseline projects,
DUNE and Hyper-Kamiokande,
will perform high-precision measurements in the next decade, with the aim to determine the neutrino
mass ordering and to establish CP violation in the lepton sector.

\section*{Acknowledgements}

We thank Davide Franco, Lucio Ludovici, Marco Martini, and Boris Popov for reading the manuscript and providing useful comments.







\bibliography{biblio}

\end{document}